\documentclass[fleqn,usenatbib]{mnras}

\usepackage{newtxtext,newtxmath}
\usepackage[T1]{fontenc}

\DeclareRobustCommand{\VAN}[3]{#2}
\let\VANthebibliography\thebibliography
\def\thebibliography{\DeclareRobustCommand{\VAN}[3]{##3}\VANthebibliography}

\usepackage{graphicx}	
\usepackage{amsmath}	
\usepackage{tikz}
\usepackage{lscape}
\usepackage{longtable}
\usepackage{supertabular,booktabs}
\usepackage{multirow}
\usepackage{ragged2e}

\newcommand{\HP}{\phantom{Ajjjj\,}}

\title[CAvity DEtection Tool (CADET)]{CAvity DEtection Tool (CADET): Pipeline for automatic detection of X-ray cavities in hot galactic and cluster atmospheres}

\author[T. Plšek et al.]{
T. Plšek,$^{1}$\thanks{E-mail: plsek@physics.muni.cz}
N. Werner$^{1}$
M. Topinka,$^{1,2}$
A. Simionescu$^{3,4,5}$
\\
$^{1}$ Department of Theoretical Physics and Astrophysics, Masaryk University, Brno, Czech Republic\\
$^{2}$ INAF-Istituto di Astrofisica Spaziale e Fisica Cosmica, Via A. Corti 12, I-20133 Milano, Italy\\
$^{3}$ SRON Netherlands Institute for Space Research, Niels Bohrweg 4, 2333 CA Leiden, The Netherlands
\\
$^{4}$ Leiden Observatory, Leiden University, PO Box 9513, 2300 RA Leiden, The Netherlands\\
$^{5}$ Kavli Institute for the Physics and Mathematics of the Universe, The University of Tokyo, Kashiwa, Chiba 277-8583, Japan\\
}

\date{Accepted XXX. Received YYY; in original form ZZZ}

\pubyear{2023}

\begin{document}
\label{firstpage}
\pagerange{\pageref{firstpage}--\pageref{lastpage}}
\maketitle

\begin{abstract}
The study of jet-inflated X-ray cavities provides a powerful insight into the energetics of hot galactic atmospheres and radio-mechanical AGN feedback. By estimating the volumes of X-ray cavities, the total energy and thus also the corresponding mechanical jet power required for their inflation can be derived. Properly estimating their total extent is, however, non-trivial, prone to biases, nearly impossible for poor-quality data, and so far has been done manually by scientists. 
We present a novel and automated machine-learning pipeline called \textit{Cavity Detection Tool} (CADET), developed to detect and estimate the sizes of X-ray cavities from raw \textit{Chandra} images. The pipeline consists of a convolutional neural network trained for producing pixel-wise cavity predictions and a DBSCAN clustering algorithm, which decomposes the predictions into individual cavities. The convolutional network was trained using mock observations of early-type galaxies simulated to resemble real noisy Chandra-like images. 
The network's performance has been tested on simulated data obtaining an average cavity volume error of 14\% at an 89\% true-positive rate. For simulated images without any X-ray cavities inserted, we obtain a 5\% false-positive rate. When applied to real \textit{Chandra} images, the pipeline recovered 91 out of 100 previously known X-ray cavities in nearby early-type galaxies and all 14 cavities in chosen galaxy clusters. Besides that, the CADET pipeline discovered 8 new cavity pairs in atmospheres of early-type galaxies and galaxy clusters (IC\,4765, NGC\,533, NGC\,2300, NGC\,3091, NGC\,4073, NGC\,4125, NGC\,4472, NGC\,5129) and a number of potential cavity candidates.
\end{abstract}



\section{Introduction}

Studies of hot atmospheres of galaxies, groups and clusters of galaxies underwent substantial progress at the turn of the millennium mainly due to the launch of three X-ray telescopes: ROSAT, \textit{Chandra} X-ray Observatory, and \textit{XMM-Newton}. Among many other important discoveries, their observations have revealed the existence of prominent surface brightness depressions, so-called {\it X-ray cavities}, in the hot atmospheres of clusters, groups, and massive early-type galaxies \citep{Bohringer1993,Huang1998,McNamara2000b,Fabian2000,Blanton2001}. Subsequent multiwavelength data (radio observations, \citealp{McNamara2001,Fabian2002,Birzan2004}; Sunyaev-Zeldovich effect observations, \citealp{Abdulla2019}) of these systems have shown that the cavities are filled with nonthermal relativistic electrons producing radio emission via synchrotron radiation. In the following two decades, several comprehensive studies of X-ray cavities and radio lobes in the atmospheres of nearby giant ellipticals \citep{Dunn2010}, distant galaxy clusters \citep{Hlavacek2012,Hlavacek2015} or both nearby and distant systems \citep{Birzan2004,Rafferty2006,Diehl2008,Dong2010,McNamara2011,Panagoulia2014,Shin2016} were performed, tens of new cavities were discovered, and their underlying attributes were inferred.

The presence of extended radio emission suggests that the cavities are products of the interaction between the relativistic jets emanating from the active galactic nucleus (AGN) and the hot intergalactic medium \citep{McNamara2000b}. Signs of previous AGN activity can thus be observed either in the radio band as extended radio lobes or in the X-ray band as brightness depressions (X-ray cavities). For some of the cavities and especially for older cavity generations, however, the extended radio emission might be significantly misaligned \citep{Gitti2006} or it can be completely missing \citep[`ghost' cavities;][]{McNamara2000,Birzan2004}. In such cases, X-ray cavities represent the only remnant of the previous AGN activity. The latest radio observations at MHz frequencies (GMRT, \citealp{Giacintucci2011}; LOFAR, \citealp{Birzan2020,Capetti2022}) have, however, revealed the existence of extended radio emission filling some of the X-ray cavities previously classified as `ghost'.

The process of cavity inflation is expected to involve a significant amount of energy released by the central supermassive black hole (SMBH) and therefore play an important role in the energetics of the whole galactic atmosphere and the AGN feedback. For nearby radio galaxies, this type of radio-mechanical feedback is believed to be dominant over the energy expelled by electromagnetic radiation. Furthermore, the mechanical power of the AGN jet is, for radio-mechanical feedback, within an order of magnitude consistent with the expected accretion power \citep{Allen2006, Plsek2022}. The enormous amount of energy deposited in cavities and in the relativistic plasma is then on timescales of $10^7-10^8$ years being dissipated back into the hot atmosphere through shocks, sound waves and turbulent flows \citep{Churazov2002,Werner2019}. A detailed description of such processes is therefore crucial for understanding the evolution and energetic balance of the entire galaxy. 

The comparison of basic cavity parameters and their corresponding mechanical jet powers with other underlying properties of their host galaxies has shown interesting implications. \cite{Birzan2004} reported that mechanical jet powers derived from the X-ray cavities correlate with the radio luminosity at 1.4$\:$GHz and the X-ray luminosity of the host atmospheres. Similarly, \cite{Rafferty2006} found a relation between the mechanical jet powers and luminosity of the X-ray emitting gas from within a cooling radius (cooling luminosity). \cite{Allen2006} found a tight correlation between the mechanical jet powers determined from the combination of observations of radio lobes and X-ray cavities and approximate SMBH accretion powers estimated from the assumption of the spherical Bondi accretion, which has been confirmed by \cite{Plsek2022}. 

\subsection{Properties of X-ray cavities}

X-ray cavities, just like radio lobes, typically come in pairs and originate in a single relativistic outflow. Their observations show that cavities are discrete separate bubbles rather than continuous funnel-like structures and for many galactic systems even multiple generations of cavities are observed (e.g. NGC$\,$5813; \citealp{Randall2015}). However, to this day it is not clear whether the discrete nature of X-ray cavities is a result of occasional episodic outbursts or whether they are caused by the fragmentation of relatively continuous outflows.

Many X-ray cavities are surrounded by shell-like or arm-like features of enhanced brightness. These bright rims correspond to regions of shocked or piled-up gas created by the inflation of cavities. Although shock fronts caused by the highly supersonic motion of material were expected, no significant temperature jumps connected with strong shocks were detected and some of the bright rims were even observed to be cooler than the surrounding gas \citep{Fabian2000}. Instead, mostly only mildly supersonic weak shock fronts are observed with Mach numbers between 1.2 and 1.7 (NGC\,4636, \citealp{Jones2002}; NGC\,5813, \citealp{Randall2015}), which are close to being in pressure balance with the ambient medium \citep{McNamara2007}. Nonetheless, these weak shocks, when present, can also deposit a significant amount of energy in the X-ray gas \citep[e.g. NGC\,5813,][]{Randall2015}.

As long as the inflation of cavities is ongoing, cavities are marked as `attached' and due to the weak nature of the surrounding shocks, they are assumed to be inflated approximately at the speed of sound \citep{Birzan2004}. Once they detach from the central AGN, they rise buoyantly and increase their volumes at altitudes with lower ambient pressure. As they rise, part of their energy is extracted and, for an M87-like pressure profile \citep{Bohringer2001}, after reaching a 20 kpc distance, nearly half of the original energy is lost \citep{Churazov2002}. Moreover, rising cavities can drag the central low entropy gas, uplift it and thus dilute the hot atmospheres of host galaxies (e.g. M87, \citealp{Werner2010}).

Although X-ray cavities come in various shapes, they are most commonly approximated as prolate or oblate spheroids with rotational symmetry along the semi-axis closer to the direction towards the centre of the galaxy \citep[see][]{Birzan2004,Allen2006}. We note, however, that many real cavities are far from being ideally ellipsoidal and their structure is much more complex, which is supported also by current idealized simulations of radio-mechanical AGN feedback \citep{Bruggen2009,Mendygral2011,Mendygral2012,Guo2015}.

As cavities rise into the thinner lower-pressure environment, they increase their volumes to maintain pressure equilibrium with the surrounding medium, which leads to an approximately constant projected area filing factor. As a result, we can observe a correlation between cavity areas (or radii) and distances \citep{Diehl2008,Dong2010,Shin2016}. X-ray cavities tend to have their semi-major axis either aligned or perpendicular to the galactocentric direction \citep{Shin2016}. Furthermore, cavities are preferentially elongated towards the centre for attached or recently detached outflows and flattened for previously separated cavities \citep{Churazov2001, Guo2020}.

Individual cavities vary significantly in their size and also in the amount of displaced material ranging from 0.5\:kpc and $10^{8}\;M_{\odot}$ for NGC\,4636 up to hundreds of kiloparsecs and more than $10^{12}\;M_{\odot}$ for Hydra~A \citep{McNamara2007}. The mechanical jet powers required to inflate typical X-ray cavities are of the order of $10^{41}-10^{44}$ erg s$^{-1}$, however, for the largest X-ray cavities typically found in brightest cluster galaxies in massive clusters (Hercules A, \citealp{Nulsen2005}; MS$\,$0735.6+7421, \citealp{Vantyghem2014}) it can be up to $10^{46}$ erg s$^{-1}$, which is comparable to the energy output of a typical powerful quasar.

\subsection{Motivation}

The total energy deposited in X-ray cavities and also the corresponding mechanical jet power required for their inflation can be measured by estimating the volumes of X-ray cavities \cite[e.g.][]{Birzan2004}. The accurate identification of X-ray cavities and estimation of their total extent is, however, non-trivial, prone to biases and nearly impossible for poor-quality data. Additionally, most of the current detection and size-estimation methods (unsharp masking, $\beta$-modelling), widely used in the previous studies of X-ray cavities \citep[e.g.][]{Birzan2004,Diehl2008,Dong2010,Panagoulia2014,Hlavacek2015,Shin2016}, are all ultimately based on visual inspection and manual estimation of cavity sizes. The total extent of thereby identified and size-estimated cavities are, also due to over-simplifying assumptions (ellipsoidal shape, rotational symmetry, projection effects), therefore rather uncertain, and difficult to reproduce among different studies performed by different teams.

The utilization of machine learning techniques has registered significant progress in the field of observational astronomy and astrophysics. It has great potential in the automation of tasks that would otherwise require human or even expert insight into the problematics.
 
Neural networks and other machine learning techniques are already being widely used in various astronomical fields for classification tasks: point source vs extended source classification \citep{Alhassan2018}, galaxy morphology classification \citep{Dieleman2015,Hausen2020}, radio galaxy morphology classification \citep{Wu2019}, detection of galaxy clusters using multiwavelength observations \citep{Kosiba2020}, distinguishing between astrophysical gamma-ray photons and cosmic-ray induced events in satellite observations \citep{Shilon2019,Wilkins2022} or for distinguishing between single and multi-temperature plasma from X-ray spectra \citep{Ichinohe2019}; for regression tasks: photometric redshift estimation \citep{Disanto2018} and pre-fitting of X-ray spectral properties \citep{Ichinohe2018}; and also for more complex problems: predicting the cosmological structure formation \citep{He2019} or producing hydrodynamical simulations by learning the basic physical laws \citep{Dai2021}. In the past, there were also attempts aiming for cavity detection and size estimation using either granular convolutional neural networks \citep{Ma2017} or Inception-like convolutional neural networks \citep{Fort2017}, the latter served as an inspiration for this work.

For these reasons, we have decided to tackle the problem of searching for and estimating the sizes of X-ray cavities on \textit{Chandra} images using the power of machine learning techniques and modern computer technology. Using a set of artificially generated images, we have trained a detection pipeline composed of a convolutional neural network (CNN) and a clustering algorithm, which we have called the \textit{CAvity DEtection Tool}\footnote{\url{https://github.com/tomasplsek/CADET}} (CADET).

The paper is organised as follows. In Section \ref{section:artificial_data}, we describe how the set of artificial training images was generated and discuss our main assumptions. The architecture of the CADET pipeline as well as the process of training and subsequent testing of the pipeline are described in Section \ref{secion:CADET}. In Section \ref{section:real_data}, we present results obtained by applying the CADET pipeline to a sample of 70 nearby early-type galaxies and 7 more distant galaxy clusters. We discuss the precision and accuracy of CADET predictions in Section \ref{section:discussion} and we conclude in the last Section. In the Supplementary online material, we share the resulting CADET predictions for the whole sample of 70 nearby early-type galaxies and 7 distant galaxy clusters.

\section{Artificial data}
\label{section:artificial_data}

The CADET pipeline was trained on artificially generated mock images processed to resemble real X-ray images as observed by the \textit{Chandra X-ray Observatory}. The real data was not used for the training for two main reasons: the number of known galactic systems with well-defined X-ray cavities is low, and even if the number of real images with cavities was sufficiently larger ($> 10^3-10^4$ images), all cavities would need to be manually size-estimated, which would bring a systematical human bias into the training process.

We, therefore, produced a large set of simplified 3D models of early-type galaxies and randomly inserted pairs of ellipsoidal cavities into them. The 3D models were then reprojected onto the 2D plane by simply summing the values across one axis -- we used this approximation assuming that atmospheres of early-type galaxies are optically thin.
Thereby obtained brightness maps were noised using Poisson statistics to resemble real low-count X-ray images.

The gas distribution of the 3D galaxy models was generated based on observed surface brightness profiles of nearby early-type galaxies. 
To approximate the gas distribution in both real and simulated galaxies, we used either single or double $\beta$-models \citep{Cavaliere1967}:
\begin{equation}
\label{eq:beta}
     \rho_{\text{gas}}(r) = \rho_0 \left[ 1 + \left(\frac{r}{r_0} \right)^2 \right]^{-3\beta/2},
\end{equation}
where $r_0$ is the core radius, $\rho_0$ is the central density and $\beta$ is the beta parameter, which describes the logarithmic slope of the distribution at larger radii. The resulting surface brightness of the simulated images was obtained as an emissivity of the gas $\epsilon(r) = \Lambda(T) \, n^2_{\text{p}}(r)$, where $n_{\text{p}} = \rho_{\text{gas}} / (2.21 \, m_{\text{p}})$, and $\Lambda(T)$ is the cooling function of the X-ray emitting gas (see \citealp{Schure2009}).

We note that the gas distribution in real galaxies is often not ideally smooth and symmetric but is disrupted by sloshing effects and cold fronts due to past mergers (NGC\,1275, \citealp{Fabian2006}; NGC\,4696, \citealp{Sanders2016}), ram pressure striping (NGC\,1404, \citealp{Machacek2005}; NGC\,4552, \citealp{Kraft2017}), shock fronts (NGC\,4552, \citealp{Machacek2006}; NGC\,5813, \citealp{Randall2015}), and also by a low entropy gas uplifted by the inflation of X-ray cavities (e.g. NGC\,4486, NGC\,4649, NGC\,4778; see also \citealp{Churazov2001,Simionescu2008,Werner2010}). However, realistically simulating such inner structure is beyond the scope of this work and it would be reachable only using detailed hydrodynamical simulations of radio-mechanical feedback in early-type galaxies. Instead, we tried to reproduce these irregularities by generating simple geometrical features tailored to resemble real observed structures on a purely empirical basis.

To imitate the non-spherical perturbation of gas distribution brought on by gas sloshing, we generated a 2D grid with an anti-symmetric spiral-like pattern and, before applying Poisson noise, we multiplied the surface brightness maps with this grid. Basic parameters of the spiral pattern (e.g. periodicity and depth) were generated according to real galaxies with prominent sloshing patterns (e.g. NGC\,507, NGC\,1275, NGC\,4696, NGC\,7618) and are based on basic assumptions further discussed in Section \ref{subsection:distrib}.

Real images of galaxies also contain bright point sources such as background AGNs, distant galaxies, and stellar sources located in the host galaxy: Low Mass X-ray Binaries (LMXB), Coronally Active Binaries (CAB) or Cataclysmic Variables (CV) \citep{Irwin2003}. Besides that, the central parts of many galaxies are often dominated by point-like non-thermal emission of their AGNs, and some of them might even contain prominent jets (e.g. NGC\,315, NGC\,383, NGC\,4261, NGC\,4486). However, we did not take these point sources into account while generating the artificial dataset. Instead, before applying the CADET pipeline to real \textit{Chandra} images, we detected these regions and replaced them with their average surrounding background (see Section \ref{subsection:data_analysis}).

To account for X-ray cavities, we randomly generated pairs of ellipsoidal masks and cut off the corresponding regions from our 3D models. The gas inside cavity regions has been removed completely assuming that they are filled purely with relativistic plasma not producing any X-ray emission \citep{McNamara2012}, which is supported also by Sunyaev-Zeldovich observations (e.g. MS\,0735, \citealp{Abdulla2019}, \citealp{Orlowski2022}). X-ray cavities were generated at various galactocentric distances, radii and shapes (ellipticities) -- their parameters were sampled from distributions based on measurements of real X-ray cavities (Section~\ref{subsection:distrib}). When generating the cavities, we tried to both entirely remove the gas from the corresponding regions as well as displace part of it to the edges of cavities to resemble bright rims caused by shocks. The rims were generated as ellipsoidal shells of the same ellipticity as the corresponding cavities (see Section \ref{subsection:distrib}).

\subsection{Data analysis}
\label{subsection:data_analysis}

When generating 3D $\beta$-models of artificial galaxies, their parameters were sampled from distributions derived from $\beta$-modelling analysis of real X-ray observations. For this purpose, we fitted a sample of 70 nearby early-type galaxies with a single or double $\beta$-model and determined the parameters and uncertainties describing the best-fitting models.

Observations of all galaxies were processed using standard CIAO 4.14 procedures \citep{Fruscione2006} and current calibration files (CALDB 4.9.8). For galaxies with multiple \textit{Chandra} observations, individual OBSIDs were reprojected and merged. All observations were deflared using the \texttt{lc\_clean} algorithm within the \texttt{deflare} routine. Most objects were observed using the ACIS-S chip, however, for some galaxies, we included also ACIS-I observations.

The images were generated from reprojected and cleaned observations using the \texttt{flux\_obs} procedure with a binsize of 1 pixel (0.492 arcsec) in the broad energy band ($0.5-7.0\;$keV). Point sources were found using the \texttt{wavdetect} tool and filled with an average surrounding background using the \texttt{dmfilth} procedure. Before applying the \texttt{dmfilth} script, the exposure-corrected flux images were converted back to units of counts (scaled by the lowest pixel value) to enable the use of Poisson statistics.

Individual processed and source-filled images were, based on the scale of interest of each galaxy, cropped to the size of an integer multiple of 128 \textit{Chandra} pixels, mostly to 512 pixels or 640 pixels. For galaxies with prominent cavities, the scale of interest was chosen to encompass the outer edge of cavities (oldest cavities if more were present). For galaxies without any visible cavities, the scale of interest was visually chosen to enclose most of the galactic emission or set to the edge of the chip for very extended sources (e.g. NGC\,4406).

The resulting cropped images were fitted with a 2D representation of a single or double $\beta$-model -- for the purpose of modelling the surface brightness distribution of X-ray images, a classical 3D $\beta$-model \citep{Cavaliere1967} has been projected onto a 2D plane under the assumption of isothermality and thus constant emissivity \citep{Ettori2000}:
\begin{equation}
\label{eq:beta_proj}
     S(r) = n_0^2 \, \Lambda(T) \, r_0 \, B(3 \beta - 0.5, 0.5) \left[ 1 + \left(\frac{r}{r_0} \right)^2 \right]^{0.5-3\beta},
\end{equation}
where $n_{0}$ is the central particle concentration, $\Lambda(T)$\footnote{For simplicity, the temperature and thus also the cooling function was assumed to be the same for all the analyzed as well as generated galaxies and the cooling function was therefore omitted from further calculations. We instead accounted a scatter of $0.3$ dex in the distribution of the $\beta$-model amplitudes, which corresponds to the difference of cooling functions at $0.5$ and $2.5$ keV at solar abundance.} 
is the cooling function, $B$ is the beta function, $r_0$ is the core radius and $\beta$ is the beta parameter of the $\beta$-model. When using composite $\beta$-models, individual $\beta$-components were simply summed together. For all galaxies, we also added a constant background model ($c_0$).

The fitting was performed in the \textsc{Sherpa 4.14} package \citep{Freeman2001} using Cash statistics \citep{Cash1979} and Levenberg-Marquardt optimization method \citep{Levenberg1944}. Within the \textsc{Sherpa 4.14} package, each $\beta$-model is described by a set of 7 parameters: centre coordinates $x_0$ and $y_0$, amplitude $A$, core radius $r_0$, power-law slope $\alpha$, ellipticity $e$ and rotational angle $\varphi$; and following equations\footnote{For simplicity, rotations of the XY grid by the angle $\varphi$ were omitted.}:
\begin{equation}
\label{eq:sherpa_r}
     r^2 = \frac{(x_0 - x)^2 (1 - e)^2 + (y_0 - y)^2}{r_0^2 (1 - e)^2},
\end{equation}
\begin{equation}
\label{eq:sherpa_beta}
     S(r) = A \left(1 + r^2 \right)^{-\alpha}.
\end{equation}

For simplifying the fitting process, an interactive Python tool\footnote{\url{https://github.com/tomasplsek/Beta-modelling}} based on Jupyter Ipywidgets was developed. Using this tool, the galaxies were first fitted using a single $\beta$-model and for galaxies for which it introduced a significant improvement in the fit statistics, we included an additional beta component. For double $\beta$-models, both ellipticities and for most objects also central coordinates of individual components were linked together during the fitting. In several cases (mainly for images with a low number of counts but a complex radial profile), we even tied the $\beta$ parameters of individual beta components in order to reduce the number of free parameters. After finding the best-fitting $\beta$-models, median values and uncertainties of individual parameters were derived from their posterior distributions obtained from Markov-Chain-Monte-Carlo (MCMC) simulations \citep[pyBLoCXS algorithm;][]{Siemiginowska2011}. The fitted parameters with their uncertainties are listed in Table \hyperlink{tableD1}{D1}. Plots with distributions of parameters for the whole sample of galaxies and correlations between them can also be found in Figures \hyperlink{figD1}{D1}, \hyperlink{figD2}{D2}.

The best-fitting $\beta$-models were subtracted from the original images and smoothed versions of thus obtained residual images were used to visually detect X-ray cavities. Individual detected cavities were size-estimated by manually overlaying ellipses using the \textit{SAOImageDS9} software \citep{Joye2003}. Significances of estimated cavities were probed using azimuthal surface brightness profiles \citep[see][]{Hlavacek2015,Ubertosi2021} and we only chose cavities with depth lower than or equal to $3\sigma$ below the surrounding background. Combining this method with known cavities from the literature \citep{Dunn2010,Dong2010,Giacintucci2011,Panagoulia2014,Shin2016,Birzan2020}, we detected in total 100 cavities in 33 galaxies. 

For describing a pair of X-ray cavities, we introduced 10 parameters (see Figure \ref{fig:schema}) -- parameters describing the first cavity: distance from the galactic centre~$d$, positional angle $\phi_{\text{cav}}$, semi-major axis $R$, ellipticity~$e_{\text{cav}}$, rotational angle $\varphi_{\text{cav}}$, and parameters describing the second cavity with respect to the first one: relative change in distance $r_{d_{\text{cav}}}$, the difference in positional angle $d \phi_{\text{cav}}$, the relative change in semi-major axis $r_{R_{\text{cav}}}$, the relative change in ellipticity $r_{e_{\text{cav}}}$ and difference in rotational angle~$d \varphi_{\text{cav}}$. Properties of all significant cavities were derived and are stated in Table \hyperlink{tableD2}{D2} (see also Figures \hyperlink{figD3}{D3}, \hyperlink{figD4}{D4}). 

\subsection{Parameter distributions}
\label{subsection:distrib}

\begin{figure}
    \centering
    \includegraphics[width=\columnwidth]{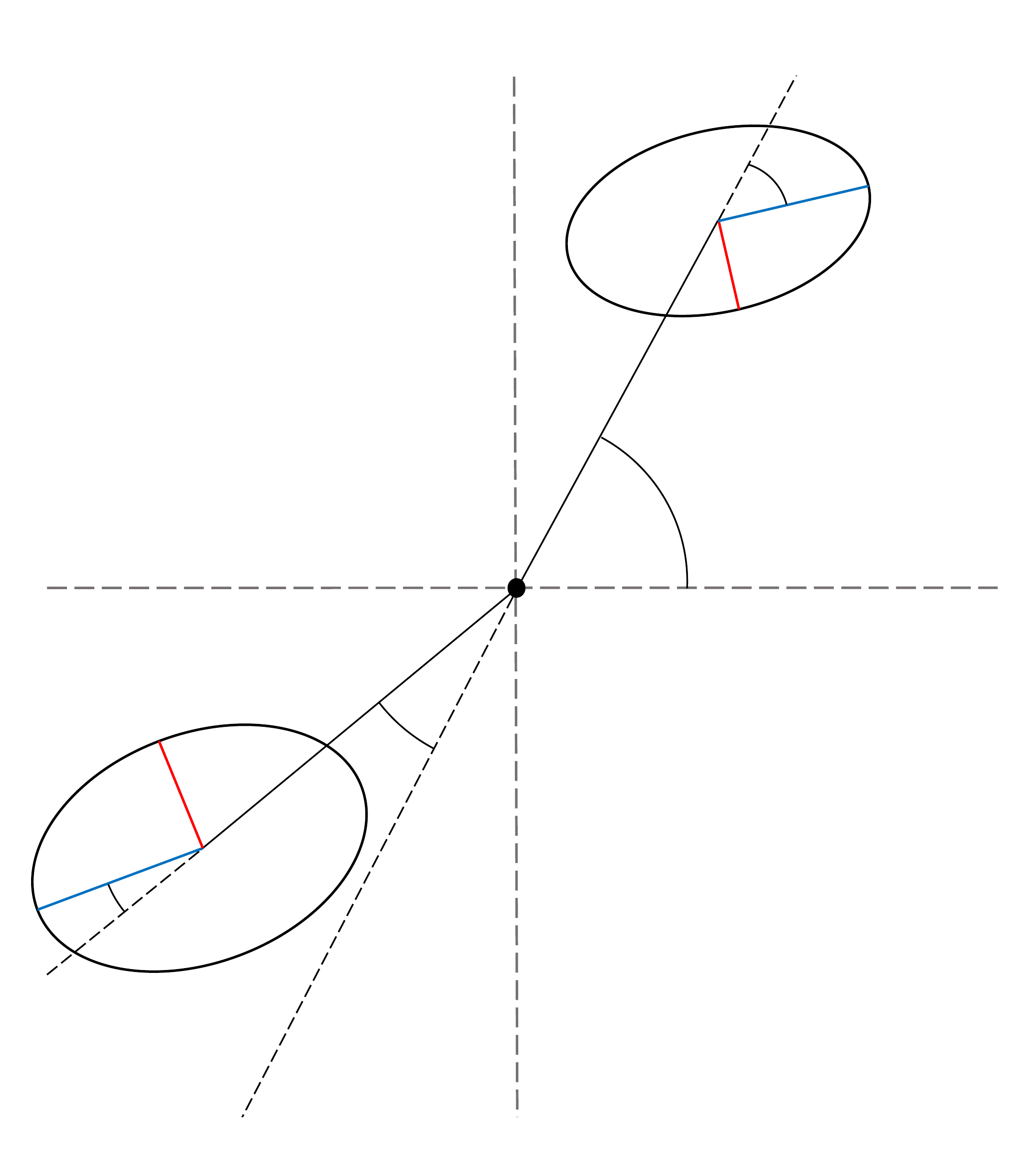}
    \begin{tikzpicture}[overlay]
    {\normalsize
    \draw (2.07,8.25) node[right] {$a_1$};
    \draw (1.35,7.73) node[right] {$b_1$};
    \draw (0.23,6.76) node[right] {$d_1$};
    \draw (1.25,6.20) node[right] {$\varphi_1$};
    \draw (2.38,9.17) node[right] {$\phi_1$};    
    \draw (-3.60,8.85) node[right] {$a_1 = R_{\text{cav}}$};
    \draw (-3.60,8.25) node[right] {$b_1 = R_{\text{cav}} \, (1 - e_{\text{cav}})$};
    \draw (-3.60,7.65) node[right] {$d_1 = d_{\text{cav}}$};
    \draw (-3.60,7.05) node[right] {$\varphi_1 = \varphi_{\text{cav}}$};
    \draw (-3.60,6.45) node[right] {$\theta_1 = \theta_{\text{cav}}$};
    \draw (-3.60,5.85) node[right] {$\phi_1 = \phi_{\text{cav}}$};
    
    \draw (-3.85,3.00) node[right] {$a_2$};
    \draw (-2.81,3.60) node[right] {$b_2$};
    \draw (-1.70,4.55) node[right] {$d_2$};
    \draw (-3.55,1.60) node[right] {$d\varphi_{\text{cav}}$};
    \draw (-4.66,2.18) node[right] {$\phi_2$};
    \draw (0.21,4.35) node[right] {$a_2 = R_{\text{cav}} \cdot r_{R_{\text{cav}}}$};
    \draw (0.21,3.75) node[right] {$b_2 = R_{\text{cav}} \cdot r_{R_{\text{cav}}} \, (1 - e_{\text{cav}} \cdot r_{e_{\text{cav}}})$};
    \draw (0.21,3.15) node[right] {$d_2 = d_{\text{cav}} \cdot r_{d_{\text{cav}}}$};
    \draw (0.21,2.55) node[right] {$\varphi_2 = \varphi_{\text{cav}} + 180^{\circ} + d\varphi_{\text{cav}}$};
    \draw (0.21,1.95) node[right] {$\theta_2 = -\theta_{\text{cav}} + d\theta_{\text{cav}}$};
    \draw (0.21,1.35) node[right] {$\phi_2 = \phi_{\text{cav}} + d\phi_{\text{cav}}$};
    }
    \end{tikzpicture}
    \vspace{-4mm}
    \caption{
    Parametric model of a pair of X-ray cavities -- each cavity is described by a set of 6 parameters: distance from the galactic centre $d$, semi-major axis $R_{\text{cav}}$, ellipticity $e_{\text{cav}}$, positional angle $\varphi_{\text{cav}}$, positional angle with respect to the line-of-sight $\theta_{\text{cav}}$, and rotational angle $\phi_{\text{cav}}$. Due to the parameters of both cavities being closely correlated, parameters of the secondary cavity ($d_2$, $a_2$, $b_2$, $\varphi_2$, $\theta_2$, $\phi_2$) were expressed with respect to the primary cavity using relative ($r_{d_{\text{cav}}}$, $r_{R_{\text{cav}}}$, $r_{e_{\text{cav}}}$) and differential factors ($d\varphi_{\text{cav}}$, $d\theta_{\text{cav}}$, $d\phi_{\text{cav}}$.) Positional angles with respect to the line-of-sight direction were for simplicity omitted from the scheme.}
    \label{fig:schema}
\end{figure}
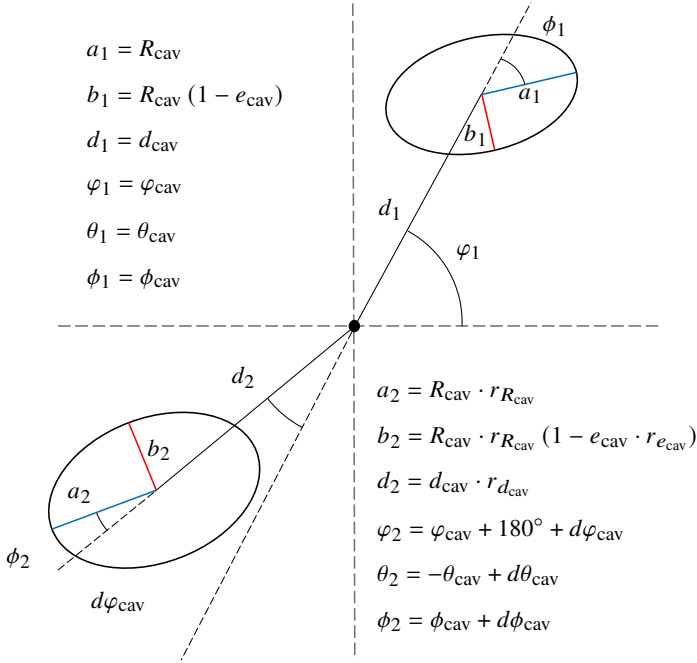

Some of the parameters describing $\beta$-models and X-ray cavities can occur according to certain distributions and may even correlate with others (e.g. cavity sizes and distances), while others are naturally totally random (e.g. rotational angle of $\beta$-model).
In order to properly reproduce real observations, we, therefore, derived the distributions from our $\beta$-modelling and cavity analyses and we sampled the parameters of simulated data from approximations of those distributions\footnote{We note that the measured distributions may suffer from selection effects and biases and are representative only of the given sample of galaxies. For the purpose of this work, however, we sampled the parameters of simulated galaxies directly from measured distributions and thus we `optimized' the CADET pipeline for the given sample of 70 early-type galaxies.} (further described in Section \ref{subsection:datagen}).

Although most of the parameters were uncorrelated and therefore could have been generated purely from their measured marginal distributions, some of them seem to correlate strongly with others. We tried to uncover these relations and take them into account while generating the artificial dataset. Linear correlations were identified using the Pearson product-moment correlation coefficient ($r$) \citep{fisher1944}. We also investigated monotonous but nonlinear correlations using Kendall's Tau coefficient ($\tau$) \citep{Kendall1938}. Corresponding correlation coefficients are stated above plots of individual quantity pairs in Figures~\hyperlink{figD1}{D1}, \hyperlink{figD2}{D2}, \hyperlink{figD3}{D3}, and \hyperlink{figD4}{D4}.

\begin{figure}
    \includegraphics[width=\columnwidth]{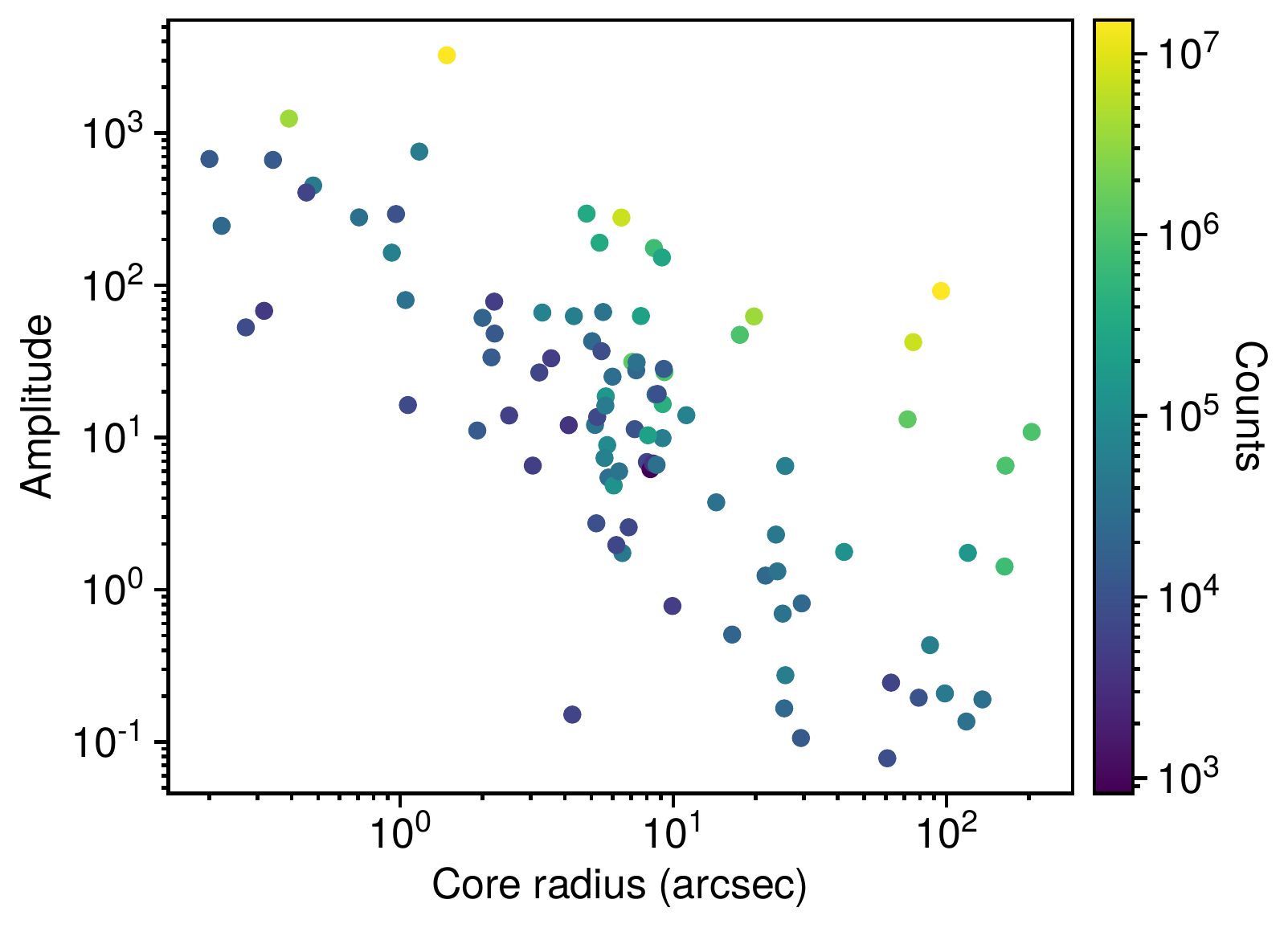}
    \caption{Colour-coded number of counts in $512\times512$ images for a sample of 70 nearby early-type galaxies as a function of the core radius and amplitude of the primary $\beta$-model. The plot shows that the number of counts is, with certain scatter given by different values of $\beta$ parameter and level of background ($c_0$), proportional to the product of core radius and amplitude of their best-fitting $\beta$-model.}
    \label{fig:counts_ampl_r0}
\end{figure}

Correlations were found between individual parameters of the primary $\beta$-component as well as between primary and secondary $\beta$-components. Parameters of secondary $\beta$-components were therefore generated with respect to the primary component. Among other key correlations, we found the correlation between $\beta$ parameters, core radii and amplitudes of the primary $\beta$-component especially important because these parameters define the total number of counts of the resulting images (see Figure \ref{fig:counts_ampl_r0}). It is, therefore, necessary to simulate these parameters according to the correlated distributions, because populating uniformly the whole parameter space would result in generating a big number of images with either an extremely low ($< 10^3$) or unrealistically high ($> 10^7$) number of counts, both of which are not suitable for training the network.

Parameters of individual cavities within the cavity pair were expected to be correlated from the beginning of the analysis and parameters of secondary cavities were therefore expressed directly via relative and differential parameters with respect to primary cavities (see Figure \ref{fig:schema}). Among other correlated primary cavity parameters, a very strong correlation ($r = 0.94$) is observed between galactocentric distances of cavities and their radii\footnote{The existence of this correlation is probably a combination of physical (cavities grow as they rise into decreasing pressure) and selection effects (detectability of small cavities drastically decreases with their distance).}. One of the consequences of sampling from these correlated distributions is that simulated cavities will not have bigger radii than their separation from the centre and therefore they will not intersect each other or the galactic centre. Sampling from such a correlated distribution will also not produce hardly detectable, very small cavities located far from the centre.

\subsection{Dataset generation}
\label{subsection:datagen}

\setlength{\tabcolsep}{5.0pt}
\renewcommand{\arraystretch}{1.3}
\begin{table}
    \centering
    \caption{Parameter ranges and distributions used for sampling parameters of training and testing data. For positional angles of $\beta$-models ($\varphi$, $\varphi_2$) and X-ray cavities ($\varphi_{\text{cav}}$), a uniform distribution between 0 and 360 degrees was used since both should emerge at random orientations. Distributions of parameters describing bright rims and sloshing effect were generated uniformly randomly and are based on basic assumptions further discussed in Section~\ref{subsection:distrib}. The rest of the parameters that could have been determined from observed properties of real galaxies (from data) were derived from $\beta$-modelling or cavity analyses.}
    \begin{tabular}{c c c}
        \hline
        Parameter & Distribution & Range \\
        \hline
         & primary beta model ($100\,\%$) & \\
        \hline
        $dx_0$ & from data & $\pm 1$ pixels\\
        $dy_0$ & from data & $\pm 1$ pixels\\
        $r_0$ & from data & $0.1-20$ pixels\\
        $\alpha$ & from data & $0.45-1.40$\\
        $A$ & from data & $1-1000$\\
        $e$ & from data & $0-0.3$\\ \vspace{-0.5mm}
        $\varphi$ & uniform & $0^{\circ}-360^{\circ}$\\
        $c_{0}$ & from data & $0.001-5$\\
        \hline
         & secondary beta model ($50\,\%$) & \\
        \hline
        $r_{c,2} \, / \, r_{c}$ & from data & $5-55$\\
        $A_2 \, / \, A$ & from data & $0.005-1$\\
        $\alpha_2 \, / \, \alpha$ & from data & $1-6$\\
        $\varphi_2$ & uniform & $0^{\circ}-360^{\circ}$\\
        \hline
         & primary cavity parameters ($50$, $90$ or $100\,\%$) & \\
        \hline
        $d_{\text{cav}}$ & from data & $5-60$ pixels\\
        $R_{\text{cav}}$ & from data & $5-30$ pixels\\
        $e_{\text{cav}}$ & from data & $0 - 0.65$\\
        $\varphi_{\text{cav}}$ & uniform & $0^{\circ}-360^{\circ}$\\
        $\phi_{\text{cav}}$ & from data & $\pm90^{\circ}$\\ \vspace{-0.5mm}
        $\theta_{\text{cav}}$ & normal & $\pm25^{\circ}$\\
        \hline
         & secondary cavity parameters & \\
        \hline
        $r_{d_{\text{cav}}}$ & from data & $0.4-1$\\
        $r_{R_{\text{cav}}}$ & from data & $0.5-1$\\
        $r_{e_{\text{cav}}}$ & from data & $0 - 1$\\
        $d\varphi_{\text{cav}}$ & from data & $\pm65^{\circ}$\\
        $d\phi_{\text{cav}}$ & from data & $\pm90^{\circ}$\\ \vspace{-0.5mm}
        $d\theta_{\text{cav}}$ & normal & $\pm10^{\circ}$\\
        \hline
         & cavity rims ($20\,\%$ of cavities)& \\
        \hline
        width & uniform & $0.05 - 0.6$\\
        height & uniform & $0 - 0.25$\\ \vspace{-0.5mm}
        type & binary & \{I, II\}\\
        \hline
         & sloshing effect ($33\,\%$)& \\
        \hline
        periodicity & uniform & $0.5-2.5$\\
        depth & uniform & $0-0.15$\\
        angle & uniform & $0^{\circ} - 360^{\circ}$\\ \vspace{-0.5mm}
        direction & binary & \{-1, +1\}\\
        \hline
    \end{tabular}
    \label{tab:distribution}
\end{table}

For generating artificial images, we used single and double $\beta$-models. Their relative frequencies were estimated from our $\beta$-modelling analysis described in Section \ref{subsection:data_analysis} to be nearly 50\,\% and 50\,\% for single and double $\beta$-models, respectively. Into each galaxy model, we inserted either one or zero cavity pairs in order for the network to learn the possibility that there can also be no cavities present on the image. Other features (bright cavity rims, sloshing effects) were generated based on their probabilities stated in Table \ref{tab:distribution} and randomly combined.

Central coordinates of $\beta$-models were set to be equal to the centre of the image with random Gaussian variation in both axes with a 1-pixel standard deviation. For double $\beta$-models, the central coordinates of individual components were not linked together but generated independently from the same distribution. Positions of X-ray cavities were generated with respect to the coordinates of the centre of the $\beta$-model (the most compact one for multiple beta components). 

Except for ellipticities, which are for most galaxies clustered at zero and follow an approximately exponential distribution, and rotational angles, which by nature should be distributed uniformly randomly between 0 and 360 degrees, marginal distributions of all other parameters were close to normal (Gaussian). All other parameters of single $\beta$-models were therefore randomly sampled from approximations of their measured distributions. For double $\beta$-models, parameters ($r_0$, $\beta$ and $A$) of secondary components were generated with respect to the parameters of the primary component. Rotational angles of individual beta components were generated independently and their ellipticities were all tied with respect to the primary component.

Similarly, as for $\beta$-models, plane-of-the-sky positional angles of primary cavities were expected to be uniformly random in the range from $0$ to $360$ degrees. All other cavity parameters were generated based on their measured distributions. Among all the distributions, strong deviations from normality are observed only in the distribution of rotational angles (angle between cavity semi-major axis and galactocentric direction) -- cavities tend to have their semi-major axis either aligned ($\phi_{\text{cav}} \approx 0^{\circ}$; prolate shape) or perpendicular ($\phi_{\text{cav}} \approx 90^{\circ}$; oblate shape) to the galactocentric direction, which is in a good agreement with results of \cite{Shin2016}. Although positional angles along the line-of-sight should also be naturally random, it is not possible to estimate them from X-ray observations and usually X-ray cavities are assumed to lie directly in the plane of the sky. As discussed later (see Section \ref{section:angular_dependence}), this assumption is good enough for most cavities, since their detectability decreases steeply at higher launching angles. When generating artificial images we have, however, varied the line-of-sight positional angles by drawing them from a normal distribution with zero mean and standard deviation of 20 degrees\footnote{In the preliminary training phase, we also tried other standard deviations (0$^{\circ}$, 10$^{\circ}$, 30$^{\circ}$, 40$^{\circ}$), however, the best performance was obtained using the standard deviation of 20 degrees.}. Rotational angles along the line of sight were for simplicity omitted.

Since the CADET pipeline can only input $128\times128$ images, real images will usually be probed on multiple scales by binning variously cropped images into $128\times128$ pixels (see Section \ref{section:real_data}). For most of the galaxies, images will maximally be cropped to 512 pixels (4 binning) and minimally to 64 pixels (0.5 binning). When simulating artificial galaxy models, core radii of $\beta$-models are, however, generated in units of arcsec based on the measured distribution of core radii in real images (later transformed to \textit{Chandra} pixels; 1 pixel $\approx$ 0.4928 arcsec). In order to sufficiently adapt the training data for the possibility of probing images on different scales, before generating 3D models of galaxies, we rescaled core radii\footnote{Based on Eq. \ref{eq:beta_proj}, we also rescaled amplitudes of $\beta$-models accordingly.} by a random factor in the range from 1 (512 pixels) to 8 (64 pixels). Radii and distances of X-ray cavities have also been rescaled to always be located within the simulated $128 \times 128$ image.

When sampling parameters of simulated $\beta$-models and X-ray cavities, the parameters were drawn from an approximation of measured distributions (see Figures~\hyperlink{figD1}{D1}, \hyperlink{figD2}{D2}, \hyperlink{figD3}{D3}, and \hyperlink{figD4}{D4}). For approximating the measured distributions, we used a Gaussian kernel density estimation (KDE) truncated at data margins\footnote{Distributions of all parameters were not truncated directly at data margins, but instead, we used a 10 per cent overlay where possible in order to properly populate the parameter space even around extreme values.} taking into account also multidimensional correlations. For a single $\beta$-model with a pair of cavities, the parameters were therefore drawn from a 19-dimensional correlated parameter space (7 for the $\beta$-model, 12 for a pair of cavities) and additional 7 parameters describing bright rims and gas sloshing were generated uniformly randomly.

\begin{figure}
    \centering    
    \includegraphics[width=0.9\columnwidth]{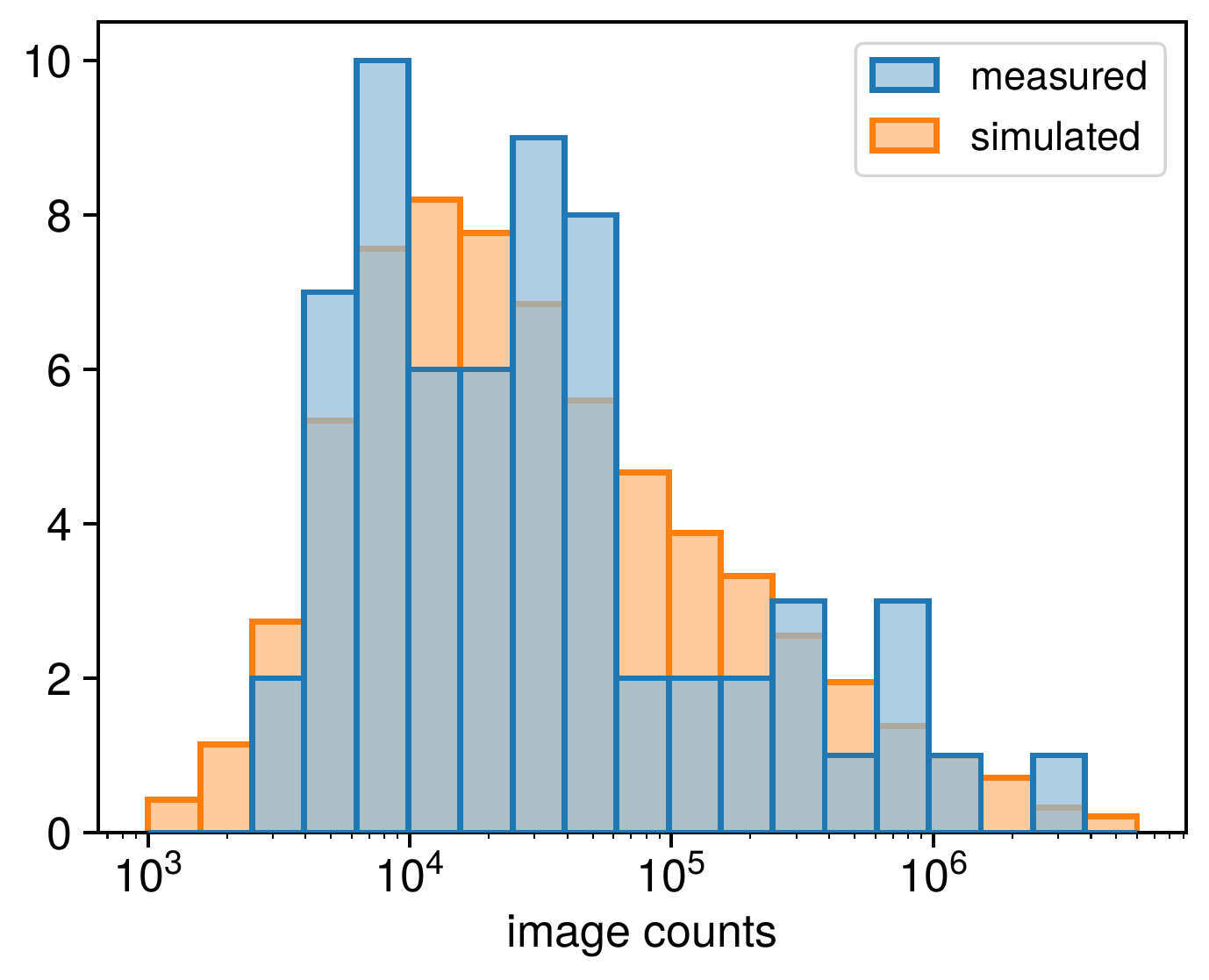}
    \caption{Number-of-counts distributions for real \textit{Chandra} images (blue) and simulated data (orange). The distribution of a number of counts for simulated images was obtained by simulating $14000$ images, the parameters of which were sampled from identical distributions as used for training of the network.}
    \label{fig:counts}
\end{figure}

The bright rims around cavities representing shocked or entrained regions were produced for approximately one-fifth of generated cavities. The rims were generated as ellipsoidal shells of the same ellipticity as corresponding cavities and they can be in general described by many various parameters. For simplicity, we linked the parameters of individual rims within a cavity pair together and assumed only 3 basic parameters: \textit{width}, \textit{height} and \textit{type} of the rim. Values of the \textit{width} of the rim were drawn from a uniform distribution ranging from 0.05 to 0.6 times the radius of the cavity. For the \textit{height} of the rim, we assumed a uniform distribution from 0 to 0.25 times the local surface brightness at the position of the rim. The actual height of the rim was set to be linearly decreasing from its maximal value (\textit{height}) at the inner side of the rim to zero at the outer side, so a sharp boundary of the rim is only produced at its inner edge. Bright rims were produced in two different realizations (Figure \hyperlink{figE1}{E1}) -- type~I: the height of the rim decreases with angular distance from the minor axis, and type~II: the height of the rim decreases with the physical distance from the centre of the cavity. The relative frequency of each rim type is 50$\,\%$.

The effects of gas sloshing were simulated for approximately one-third of galaxies by multiplying the original surface brightness map with a grid of spiral-like pattern (see Figure \hyperlink{figE1}{E1}). The values in the grid are axially anti-symmetric with a mean value of 1, and the whole pattern can be described by four parameters: \textit{periodicity} describing the number of angular periods within the image, \textit{depth} which represents a maximal$\,$/$\,$minimal value of the whole pattern, \textit{starting angle} of the `overdensity$\,$/$\,$under-density' boundary, and \textit{direction} that sets if the pattern is clockwise or anti-clockwise. The possible values of the \textit{periodicity} were taken from a uniform distribution from 0.5 to 2.5, the \textit{depth} of the `overdensity$\,$/$\,$under-density' field was drawn from a uniform distribution between 0 and 0.15, the \textit{starting angle} of the spiral pattern was generated uniformly randomly (0 to 360 degrees) and the \textit{direction} was either clockwise (+1) or anti-clockwise~(-1).

Before the training process, a large number ($\approx 10^6$) of simulated parameters was generated and saved into a CSV file. By comparing generated parameters to their measured distributions, we can see a good agreement between the distributions. We also tested an agreement for numbers of counts (Figure~\ref{fig:counts}) between simulated and real images by simulating a set of $14000$ mock images which later served as validation and testing images.

\section{Cavity Detection Tool}
\label{secion:CADET}

\begin{figure*}
    \includegraphics[width=\textwidth]{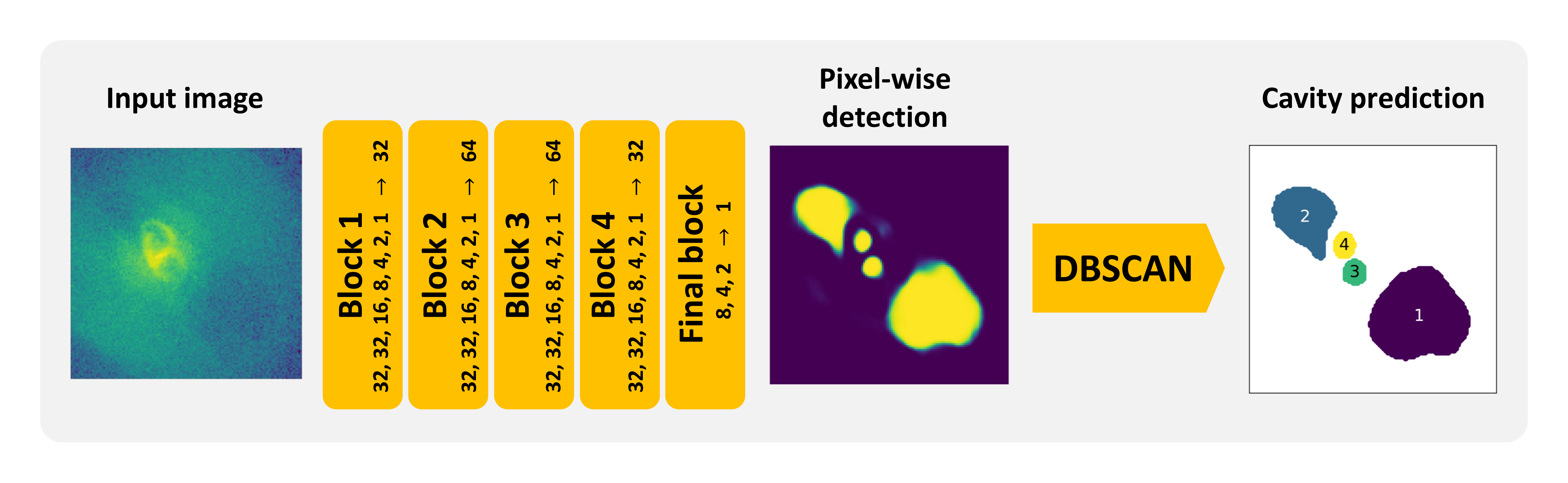}
    \vspace{-2mm}
    \caption{The architecture of the CADET pipeline consisting of a convolutional neural network (composed of five Inception-like convolutional blocks) producing a pixel-wise cavity prediction and a clustering algorithm \texttt{DBSCAN}, which decomposes the prediction into individual cavities. The number of individual filters within the Inception layers and subsequent dimensionality-reducing layers is stated alongside individual blocks. The scheme is inspired by Fig. 3 in \protect\cite{Fort2017}.}
    \label{fig:architecture}
\end{figure*}

The \textit{CAvity DEtection Tool} (CADET) consists of a convolutional neural network (CNN) composed of a series of Inception-like convolutional blocks \citep{Szegedy2015} and a density-based spatial clustering algorithm \citep[DBSCAN;][]{Ester1996}. The CNN part of the pipeline is trained for finding pairs of elliptical surface brightness depressions (X-ray cavities) on noisy \textit{Chandra}-like images of early-type galaxies and for producing pixel-wise predictions of their position and total extent. The clustering algorithm is used for the decomposition of thereby obtained predictions into individual cavities (Figure \ref{fig:architecture}). The creation of the pipeline was inspired by the work of \cite{Fort2017} and \cite{Secka2018}, although we made several changes in the network and mainly also in the process of generating the artificial dataset.

The convolutional neural network (CNN) was implemented using a high-level Python \textit{Keras} library \citep{chollet2015keras} with \textit{Tensorflow} back-end \citep{tensorflow2015-whitepaper}. The CNN was written using a functional \textit{Keras} API which enables saving and loading the model into the Hierarchical Data Format (HDF5) without the need to re-define the model when loading. For the clustering task, we used the DBSCAN implementation in the \textit{Scikit-Learn} library \citep{Pedregosa2011}. For monitoring learning curves, comparing final test statistics and selecting optimal hyper-parameters, we used the \textit{Tensorboard} dash-boarding tool \citep{tensorflow2015-whitepaper}. The final model as well as training and data generating scripts can be found on CADET Github page\footnote{\url{https://github.com/tomasplsek/CADET}}.

Elliptical $\beta$-models and ellipsoidal cavities were generated with the use of the \textit{JAX} library\footnote{\url{https://github.com/google/jax}} \citep[version 0.2.26;][]{jax2018github}. Thanks to the Graphical Processing Unit (GPU) support of the \textit{JAX} library, training images were generated `on the fly` in a vectorized way using the same GPU as was used for training of the network. This dramatically improved the data generation time compared to generating the data using a CPU and also enabled additional tweaking of the parameters of training images between individual training runs (e.g. the fraction of images containing X-ray cavities). To enable the comparison of CADET performance for various values of network hyper-parameters, fixed sets of testing images with X-ray cavities ($10^4$ images), testing images without X-ray cavities ($2000$ images) and validation images ($2000$ images) were generated before the entire training process so the same sets could be used when testing the CADET performance for various values of network hyper-parameters.

\subsection{Network architecture}
\label{section:architecture}

On the input of the CNN, there is a single channel $128 \times 128$ image. Since radial profiles of $\beta$-models in both real and simulated images are rather exponential, we transform the input images by a decimal logarithm\footnote{We tested various approaches to input image normalization including widely used techniques incorporated in \textit{Scikit-Learn} or \textit{Keras} libraries (e.g. zero mean, unit variance), however, the best results were obtained for the above-stated approach.} (the value of each pixel was raised by one to avoid calculating the logarithm of zero). Before the images are processed by the first Inception-like convolutional block, they are further normalized in mini-batches by a batch-normalization layer within the convolutional neural network.

The architecture of the convolutional neural network is similar to that developed by \cite{Fort2017} and is composed of a series of 5 convolutional blocks (Figure \ref{fig:architecture}). Each block resembles an Inception-like layer \citep{Szegedy2015} as it applies a set of multiple parallel 2D convolutions with various kernel sizes and concatenates their outputs (Figure \hyperlink{figF1}{F1}). Inception layers within the first 4 blocks consist of convolutional layers with 32 of $1 \times 1$ filters, 32 of $3 \times 3$ filters, 16 of $5 \times 5$ filters, 8 of $7 \times 7$ filters, 4 of $9 \times 9$ filters, 2 of $11 \times 11$ filters, and one $13 \times 13$ filter. The output of each convolutional layer within the Inception-like layer is activated by Rectified Linear Unit (ReLU; \citealp{Fukushima1975}) activation function, which brings non-linear elements into the network, and then normalized by batch normalization \citep{Ioffe2015}. As opposed to the original Inception block \citep{Ioffe2015}, we omitted the max-pooling layer because it did not bring any performance improvement. Each Inception layer is then followed by a 2D convolutional layer with 32 or 64 of $1 \times 1$ filters, which is introduced mainly due to dimensionality reduction. The output of this convolutional layer is also activated using the ReLU activation function and batch-normalized. The $1 \times 1$ convolutional layers are, in order to prevent overfitting, followed by a dropout layer, where the dropout rate was varied as a hyper-parameter.

The convolutional neural network is ended by a final block, which is as well composed as an Inception-like layer but differs from the previous blocks by the numbers and sizes of individual 2D convolutional filters (8 of $8 \times 8$ filters, 4 of $16 \times 16$ filters, 2 of $32 \times 32$ filters, and one $64 \times 64$ filter) and also by the activation function of the last $1\times1$ convolutional filter. Since the output of the network is intended to be a prediction of whether a corresponding pixel belongs to a cavity (value 1) or not (value 0), the activation function of the final layer was set to be the \textit{sigmoid} function, which outputs real numbers in the range between 0 and 1.

Initial weights of individual 2D convolutional layers were generated using He initialization \cite{He2015_2} and biases were initialized with low but non-zero values (0.01). In total, the network has $563\:146$ trainable parameters and the size of the model is 7.4\,MB. 

On the output of the CNN, there is a pixel-wise prediction of the same shape as the input image with a value in each pixel ranging from 0 to 1, which expresses whether that pixel corresponds to a cavity or not. The pixel-wise prediction is then decomposed into individual X-ray cavities using the DBSCAN clustering algorithm. Before the decomposition, a pair of discrimination thresholds are applied for the pixel-wise prediction excluding low-significance regions and keeping only solid cavity predictions while properly estimating their areas and volumes (see Appendix \ref{section:discrimination_threshold}).

\subsection{Training}
\label{subsection:training}

\setlength{\tabcolsep}{6.0pt}
\renewcommand{\arraystretch}{1.33}
\begin{table*}
    \centering
    \caption{Hyper-parameters and test statistics for individual training runs: fraction of cavities in training data, learning rate, dropout, test loss function, average volume error, false-positive (FP) rate, true-positive (TP) rate, real data performance. The loss function, average cavity volume error and true-positive rate were calculated using the set of testing images with X-ray cavities (10000 images), while the false-positive rate correspond to the dataset of images without any X-ray cavities (2000 images). Optimal hyper-parameters are marked in boldface.}
    \begin{tabular}{c c c c c c c c}
        \hline
        Cavities (\%) & Learning rate & Dropout & Loss function & Average error (\%) & TP rate & FP rate & Real data performance\\
        \hline
        50 & 0.0005 & 0.0 & 0.056 & ${14}^{+11}_{-9}$ & 0.78 & 0.03 & 6/6\\
        50 & 0.0005 & 0.2 & 0.057 & ${15}^{+13}_{-8}$ & 0.79 & 0.02 & 4/6\\
        90 & 0.0005 & 0.0 & 0.051 & ${14}^{+13}_{-7}$ & 0.81 & 0.04 & 2/6\\
        90 & 0.0005 & 0.2 & 0.049 & ${14}^{+14}_{-8}$ & 0.83 & 0.07 & 2/6\\
        100 & 0.0005 & 0.0 & 0.041 & ${16}^{+15}_{-8}$ & 0.87 & 0.08 & 5/6\\
        \textbf{100} & \textbf{0.0005} & \textbf{0.2} & 0.042 & ${14}^{+13}_{-8}$ & 0.89 & 0.05 & 6/6\\
        50 & 0.001 & 0.0 & 0.052 & ${14}^{+12}_{-7}$ & 0.80 & 0.01 & 2/6\\
        50 & 0.001 & 0.2 & 0.057 & ${14}^{+12}_{-7}$ & 0.76 & 0.02 & 3/6\\
        90 & 0.001 & 0.0 & 0.044 & ${15}^{+12}_{-8}$ & 0.84 & 0.04 & 5/6\\
        90 & 0.001 & 0.2 & 0.048 & ${15}^{+12}_{-8}$ & 0.84 & 0.07 & 4/6\\
        100 & 0.001 & 0.0 & 0.051 & ${16}^{+14}_{-9}$ & 0.86 & 0.19 & 1/6\\
        100 & 0.001 & 0.2 & 0.040 & ${15}^{+13}_{-8}$ & 0.86 & 0.07 & 2/6\\
        50 & 0.002 & 0.0 & 0.061 & ${15}^{+11}_{-8}$ & 0.75 & 0.02 & 2/6\\
        50 & 0.002 & 0.2 & 0.063 & ${16}^{+15}_{-8}$ & 0.78 & 0.08 & 3/6\\
        90 & 0.002 & 0.0 & 0.052 & ${15}^{+12}_{-8}$ & 0.82 & 0.07 & 1/6\\
        90 & 0.002 & 0.2 & 0.073 & ${19}^{+18}_{-10}$ & 0.80 & 0.11 & 3/6\\
        100 & 0.002 & 0.0 & 0.044 & ${15}^{+13}_{-8}$ & 0.86 & 0.12 & 1/6\\
        100 & 0.002 & 0.2 & 0.042 & ${15}^{+12}_{-8}$ & 0.86 & 0.07 & 1/6\\
        \hline
    \end{tabular}
    \label{tab:hyperparams}
\end{table*}

During the training, the convolutional neural network was given a set of $128 \times 128$ training images on the input, and as ground truth images on the output we used cavity masks projected into $128 \times 128$ images and binned to contain either ones (inside of the cavity) and zeros (outside of it)\footnote{We also tried not binning label masks to express depths of cavities (with MSE as loss function and ReLU as last layer activation function). However, this drastically worsened the performance of the network.}. For galaxies with no inserted cavities, the ground truth output image was represented by a zero matrix.

The loss function of the network was calculated as the binary cross-entropy of the pixel-wise prediction with respect to the ground truth image: \vspace{2mm}
\begin{equation}
    \mathcal{L} = - \frac{1}{N}\sum^N_i y_{\text{true}, i} \cdot \log  y_{\text{pred}, i} + (1 -  y_{\text{true}, i}) \cdot \log (1 -  y_{\text{pred}, i}),
\end{equation}
where $y_{\text{true}, i}$ is the value in the i-th pixel of the binary cavity mask (ground truth), $y_{\text{pred}, i}$ is the value in the i-th pixel of the pixel-wise prediction, and $N$ is the total number of pixels in each image. The minimization of the loss function was realized using the \textit{Adaptive Moment Estimation} (ADAM) optimizer \citep{Kingma2014}. The learning rate of the ADAM optimizer was varied as a hyper-parameter and the decay of the learning rate was for all runs set to zero.

During the training, we examined various sets of hyper-parameters describing the data and also the network itself: the fraction of images containing X-ray cavities (50\,\%, 90\,\% or 100\,\%), learning rate ($0.0005$, $0.001$, $0.002$), and dropout rate ($0.0$, $0.2$). The batch size was not varied as a hyper-parameter, and for all networks, it was set to 24 images per mini-batch, which corresponds to the maximal possible batch size due to large GPU memory consumption caused by on-the-fly data generation. We tested all combinations of hyper-parameters and produced in total 18 trained networks.

All networks were first trained for 32 epochs, wherein each epoch the network was given 12288 images. Nevertheless, instead of using the same images every epoch, we continuously generated new data by sampling their parameters from identical distributions. During the training, the network was therefore given 393\,216 different images in total. After every epoch, the performance of the network was validated using a fixed set of 2000 validation images. For controlling the learning rate, in order for the network to prevent overfitting, we utilized the \texttt{ReduceLROnPLateau} \textit{Tensorflow} callback, which reduces the learning rate when the validation loss (loss function calculated for the validation dataset) stops improving in a given number of epochs. The training of all networks including on-the-fly data generation was performed on NVIDIA GPU type GeForce RTX 3080 (10$\,$GiB) and lasted approximately 8 hours per single network. The network with the best performance was trained for additional 32 epochs and further tested using the testing dataset.

\subsection{Testing}
\label{section:testing}

To compare the performance of individual networks, we calculated the cross-entropy loss function for the whole testing dataset, true-positive (TP) rate for testing images with X-ray cavities ($10^4$ images) and false-positive (FP) rate for testing images without X-ray cavities ($2000$ images). For correctly recovered X-ray cavities (true-positives), we also compared inferred cavity volumes to their real values known from the simulations and calculated their average differences. We note, however, that the accuracy of CADET predictions depends strongly on the total number of counts of the given images (further discussed in Section \ref{section:cavity_reliability}). For this reason, individual testing statistics were also computed as a function of the number of counts and corresponding dependencies were logged and compared using \textit{Tensorboard} callback.

For qualitative comparison, we picked 3 real images with well-defined and obvious cavities (NGC\,4649, NGC\,4778 and NGC\,5813), and 3 images with hardly detectable cavities due to more complex features or the low number of counts (NGC\,1600, NGC\,4552, NGC\,6166) and we visually compared CADET predictions with presumed cavity extent estimated manually. Real-data performance was based upon visual inspection of CADET predictions for individual galaxies evaluated as sufficient (1 point) or insufficient (0 points). Each network could therefore get between 0 and 6 data points for real-data performance.

For the final testing (see Section \ref{section:discussion}) and application of the network, we chose the combination of hyper-parameters with the best possible test score (lowest possible loss function), lowest FP rate, highest TP rate, best volume prediction accuracy, and best real-data performance. Hyper-parameters of the best-performing network are: $100\,\%$ of training images containing cavities, a learning rate of 0.0005 and a dropout of 0.2. The architecture and saved weights of the best-performing network can be found on the CADET GitHub page\footnote{\url{https://github.com/tomasplsek/CADET}} (\texttt{CADET.hdf5}).

\section{Application on real data}
\label{section:real_data}

The CADET network with the best testing score was further applied to real \textit{Chandra} images of early-type galaxies, groups and clusters of galaxies. Firstly, we applied the CADET pipeline to images of 70 nearby early-type galaxies, the parameters of which were used to generate the training dataset. For galaxies known to contain X-ray cavities (33 sources), we tested whether the CADET pipeline is able to recover all previously detected X-ray cavities and for the rest of the galaxies (37 sources), we looked for new cavity candidates. To test the performance on also out-of-distribution sources, the CADET pipeline was applied to a sample of 7 more distant galaxy clusters with known X-ray cavities.

Before being processed by the CADET pipeline, real \textit{Chandra} images were cropped on various different scales (multiples of 128 pixels) and binned to $128\times128$ pixels. For most galaxies, we used 6 different scales: $64\times64$ pixels (0.5 binning\footnote{For 0.5 binning, we used \textit{Chandra} sub-pixel resolution images.}), $128\times128$ pixels (no binning), $196\times196$ pixels (1.5 binning), $256\times256$ pixels (2 binning), $384\times384$ pixels (3 binning), $512\times512$ pixels (4 binning). For very extended or nearby sources, also larger scales were used (e.g. 640, 728 pixels). The binning of images was performed using the \texttt{dmregrid} routine within the \texttt{CIAO} pipeline.

To partly mitigate possible systematical uncertainties due to different positional angles of X-ray cavities, in the final version of the CADET pipeline, the resulting prediction is averaged over predictions produced for 4 different rotation angles ($0^{\circ}$, $90^{\circ}$, $180^{\circ}$, $270^{\circ}$). To suppress systematic effects connected with improper centring of the input image, CADET predictions are additionally averaged for 25 possible shifts of the central pixel (maximally by $\pm2$ pixels in both coordinates). Both of these approaches are combined, so the final prediction represents an average of 100 possible combinations.

\subsection{Recovering known X-ray cavities}

\begin{figure}
    \centering
    \includegraphics[width=0.95\columnwidth]{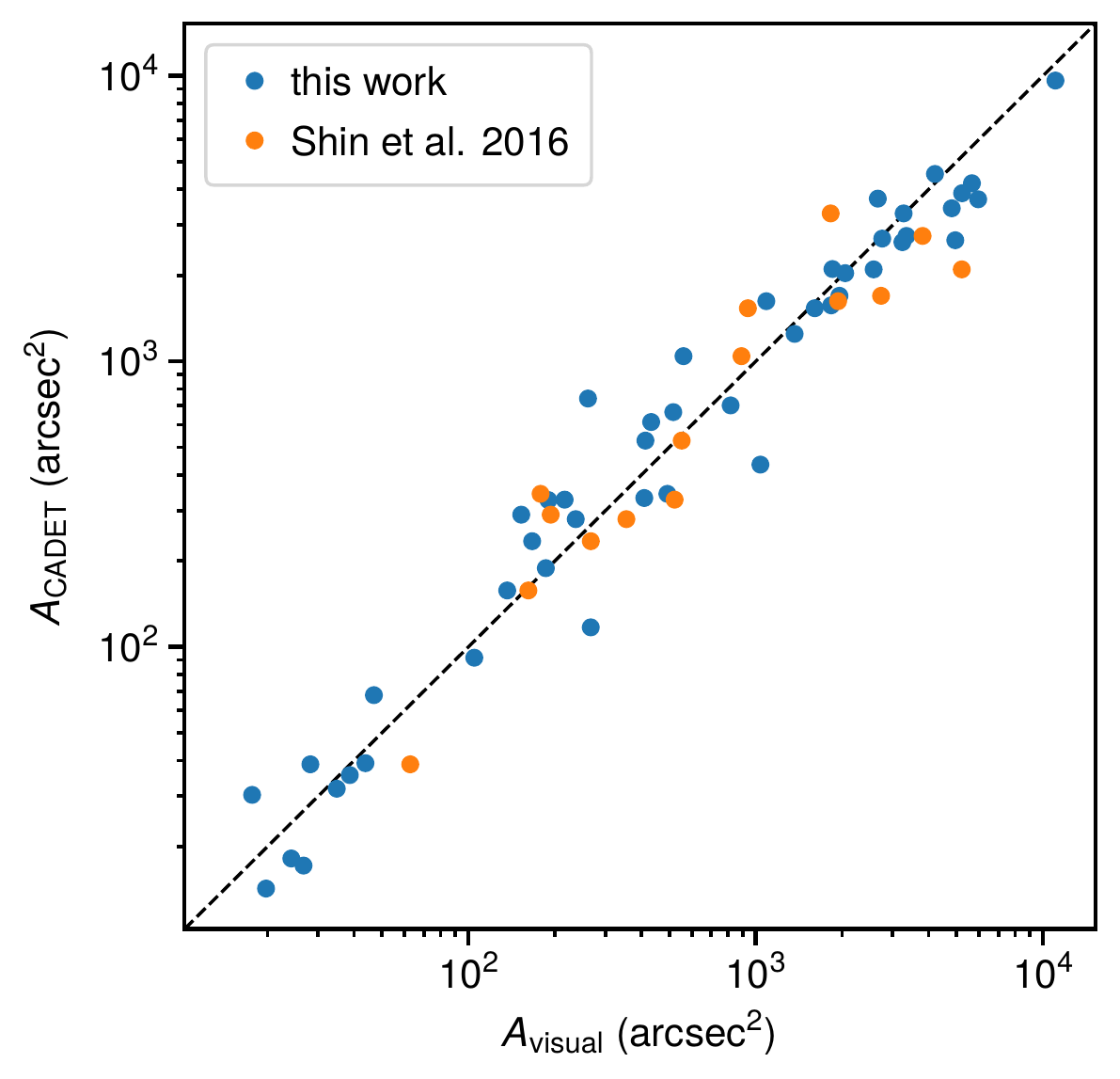}
    \vspace{-3mm}
    \caption{Areas of CADET predictions compared with manually derived cavity sizes for the sample of 33 early-type galaxies with known X-ray cavities estimated in our work and for  15 sources estimated by \protect\cite{Shin2016}.}
    \label{fig:CADET_vs_manual}
\end{figure}

The CADET pipeline was applied on 33 systems previously known to harbour X-ray cavities in their hot atmospheres \citep[reported by][]{Dong2010,Cavagnolo2011,Giacintucci2011,Shin2016}. Resulting CADET predictions were examined and cavities of interest, that were spatially co-aligned (verified visually) with previously known X-ray cavities, were selected and their estimated areas were compared (Figure~\ref{fig:CADET_vs_manual}).

For 33 galaxies known to harbour 50 pairs of X-ray cavities, the CADET pipeline was able to recover 91 out of 100 cavities. We note that this well corresponds to the TP rate ($89$ per cent) when applied to simulated images. Besides that, CADET detected a number of potential cavity candidates (Section \ref{section:new_cavities}) as well as false positive predictions (further discussed in Section \ref{section:discussion}).

A comparison between predictions produced by the CADET pipeline and human-made predictions shows a good agreement (see Figure~\ref{fig:CADET_vs_manual}) for estimated cavity volumes with an average difference of $-0.04_{-0.10}^{+0.19}$ dex and $-0.06_{-0.15}^{+0.26}$ dex for this work and \cite{Shin2016}, respectively.

\subsection{New cavity candidates}
\label{section:new_cavities}

\setlength{\tabcolsep}{3.0pt}
\renewcommand{\arraystretch}{1.45}
\begin{table}
    \centering
    \caption{Significances of newly detected X-ray cavities (upper part) and potential cavity candidates (lower part): galaxy name, cavity side and generation,
    maximal azimuthal significance, maximal radial significance, false positive rate (fraction of non-cavity images with positive cavity prediction with similar properties as given cavities), true positive rate (fraction of images with properly recovered cavities), volume error (relative cavity volume error estimated for true positive cases). For cavities with TP rates smaller than 25\%, average volume errors were omitted.}
    \begin{tabular}{c c c c c c c}
    \hline \vspace{-0.6mm}
    \multirow{2}{*}{Galaxy} & \multirow{2}{*}{Cavity} & Azimuthal & Radial & FP & TP & Volume error\\
     & & ($\sigma$) & ($\sigma$) & rate & rate & (\%)\\
    \hline
    IC\,4765 & S1 & 6.7 & 4.3 & 0.04 & 0.95 & $-0^{+15}_{-14}$\\
    IC\,4765 & N1 & 4.9 & 4.4 & 0.11 & 0.91 & $-7^{+22}_{-17}$\\
    NGC\,533 & SE1 & 7.1 & 7.5 & 0.01 & 0.96 & $2^{+14}_{-15}$\\
    NGC\,533 & NW1 & 10.1 & 11.5 & 0.04 & 0.97 & $17^{+21}_{-19}$\\
    NGC\,2300 & SE1 & 3.8 & 4.1 & 0.0 & 0.94 & $-11^{+15}_{-12}$\\
    NGC\,2300 & NW1 & 5.2 & 6.8 & 0.01 & 0.62 & $-12^{+19}_{-17}$\\
    NGC\,3091 & W1 & 4.6 & 6.5 & 0.02 & 0.93 & $-1^{+15}_{-13}$\\
    NGC\,3091 & NE1 & 7.9 & 5.6 & 0.02 & 0.93 & $-2^{+15}_{-14}$\\
    NGC\,4073 & SW1 & 6.6 & 3.7 & 0.02 & 0.91 & $0^{+17}_{-13}$\\
    NGC\,4073 & SE1 & 1.9 & 2.0 & 0.02 & 0.87 & $-18^{+13}_{-11}$\\
    NGC\,4125 & S1 & 7.4 & 10.2 & 0.06 & 0.98 & $10^{+18}_{-17}$\\
    NGC\,4125 & N1 & 6.0 & 6.3 & 0.08 & 0.97 & $23^{+26}_{-42}$\\
    NGC\,4472 & W1 & 17.1 & 23.7 & 0.0 & 1.0 & $-1^{+2}_{-2}$\\
    NGC\,4472 & E1 & 7.3 & 11.9 & 0.0 & 1.0 & $-0^{+3}_{-2}$\\
    NGC\,5129 & SE1 & 3.4 & 4.2 & 0.01 & 0.75 & $2^{+20}_{-18}$\\
    NGC\,5129 & NW1 & 3.7 & 5.6 & 0.01 & 0.96 & $-6^{+14}_{-10}$\\
    \hline
    IC\,1860 & W1 & 4.1 & 4.0 & 0.11 & 0.09 & -\\
    IC\,1860 & NE1 & 2.4 & 2.1 & 0.08 & 0.2 & -\\
    NGC\,499 & E1 & 2.1 & 2.3 & 0.09 & 0.35 & $-2^{+32}_{-34}$\\
    NGC\,499 & NW1 & 2.3 & 5.9 & 0.06 & 0.39 & $40^{+66}_{-71}$\\
    NGC\,1521 & SE1 & 3.8 & 5.2 & 0.12 & 0.67 & $25^{+34}_{-35}$\\
    NGC\,1521 & W1 & 5.3 & 4.1 & 0.09 & 0.46 & $-3^{+43}_{-28}$\\
    NGC\,1700 & SE1 & 3.5 & 3.6 & 0.24 & 0.39 & $36^{+89}_{-44}$\\
    NGC\,1700 & N1 & 5.2 & 3.3 & 0.23 & 0.44 & $39^{+86}_{-53}$\\
    NGC\,3923 & W1 & 3.1 & 2.5 & 0.07 & 0.01 & -\\
    NGC\,3923 & E1 & 2.8 & 2.1 & 0.1 & 0.03 & -\\
    NGC\,4325 & E1 & 4.9 & 2.1 & 0.05 & 0.26 & $49^{+65}_{-62}$\\
    NGC\,4325 & N1 & 6.2 & 3.3 & 0.09 & 0.84 & $0^{+21}_{-16}$\\
    NGC\,4325 & SE2 & 2.5 & 1.3 & 0.02 & 0.85 & $27^{+31}_{-24}$\\
    NGC\,4325 & N2 & 1.8 & 3.9 & 0.02 & 0.50 & $36^{+49}_{-36}$\\
    NGC\,4526 & SW1 & 8.3 & 9.2 & 0.01 & 0.67 & $-17^{+28}_{-23}$\\
    NGC\,4526 & N1 & 4.7 & 10.9 & 0.01 & 0.54 & $60^{+97}_{-63}$\\
    NGC\,6482 & S1 & 3.0 & 3.0 & 0.0 & 0.05 & -\\
    NGC\,6482 & N1 & 2.1 & 4.1 & 0.01 & 0.37 & $60^{+35}_{-25}$\\
    NGC\,6482 & SW2 & 2.4 & 3.5 & 0.0 & 0.60 & $11^{+24}_{-18}$\\
    NGC\,6482 & NE2 & 2.6 & 3.9 & 0.01 & 0.94 & $21^{+22}_{-18}$\\
    \hline
    \end{tabular}
    \label{tab:new_detections}
\end{table}

Furthermore, we applied the CADET pipeline to images of 37 galaxies without previously detected X-ray cavities and looked for new cavity candidates. For this purpose, we increased the discrimination threshold ($\sim 0.8$) to reduce the possibility of obtaining a false positive detection and to only obtain high-significance brightness drops. Moreover, the predictions were only taken as valid if present on at least two different size scales (verified visually). For subsequent significance testing and cavity volume estimation, we usually used the smaller size scale with better spatial resolution.

\begin{figure*}
    \hspace{-4.0mm}
    \begin{tikzpicture}
    \draw (-6.77, 7) node {\includegraphics[height=0.22\textwidth]{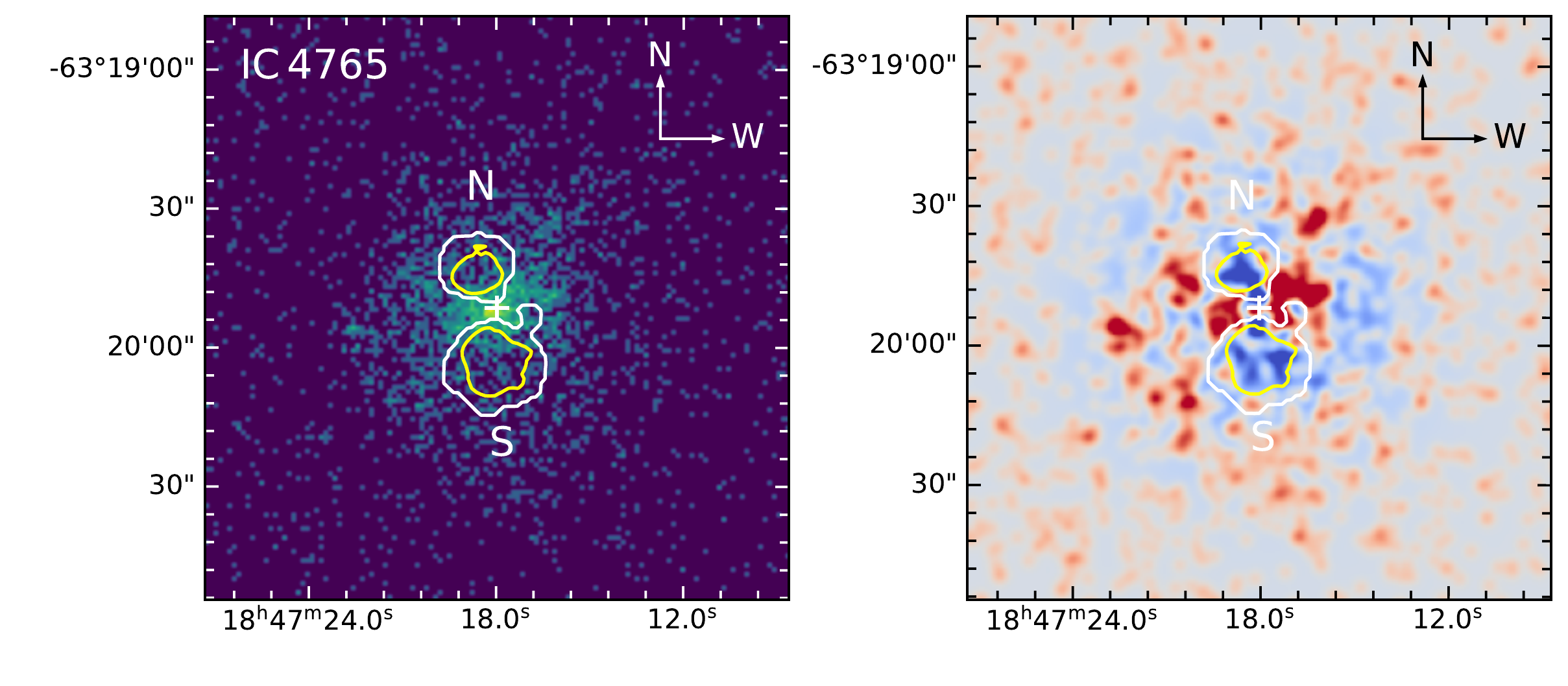}};
    \draw (2.30, 7) node {\includegraphics[height=0.22\textwidth]{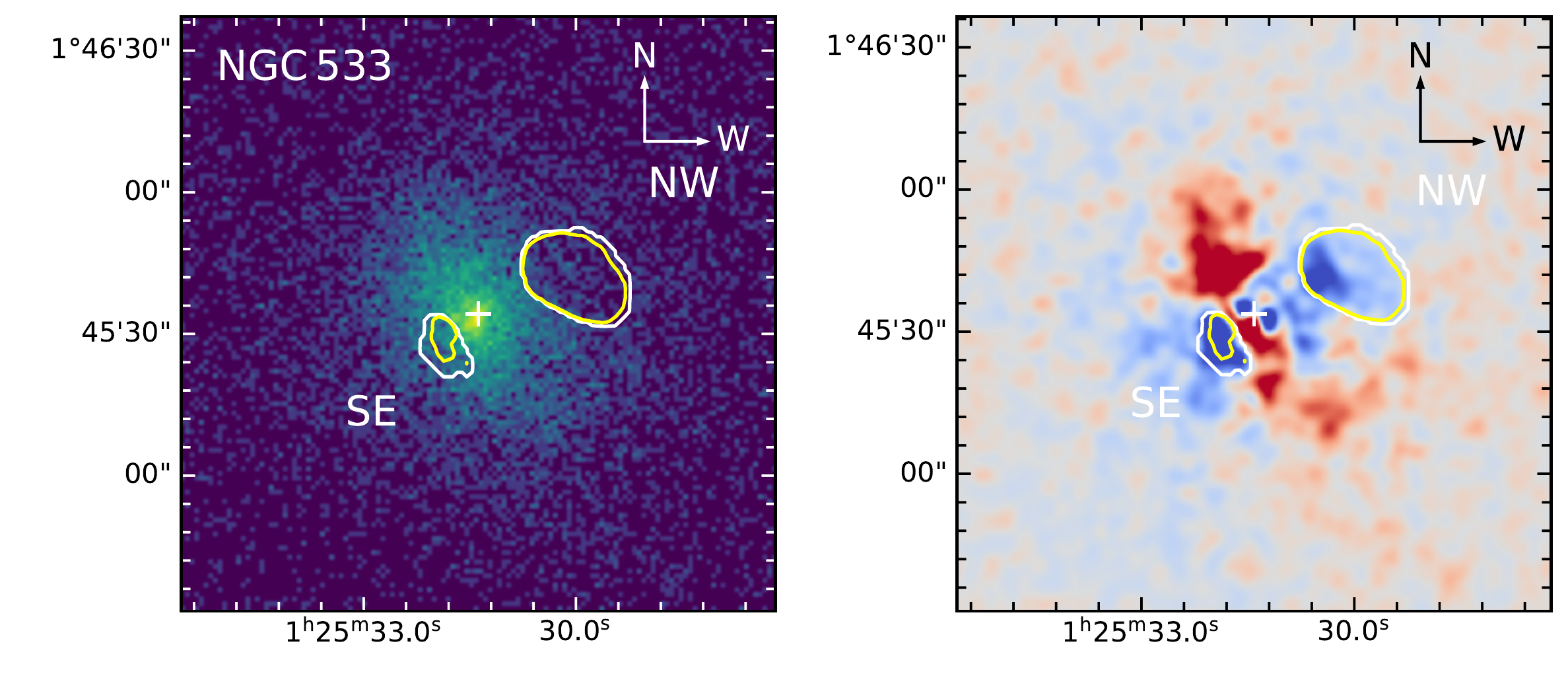}};
    \draw (-6.65, 3.0) node {\includegraphics[height=0.22\textwidth]{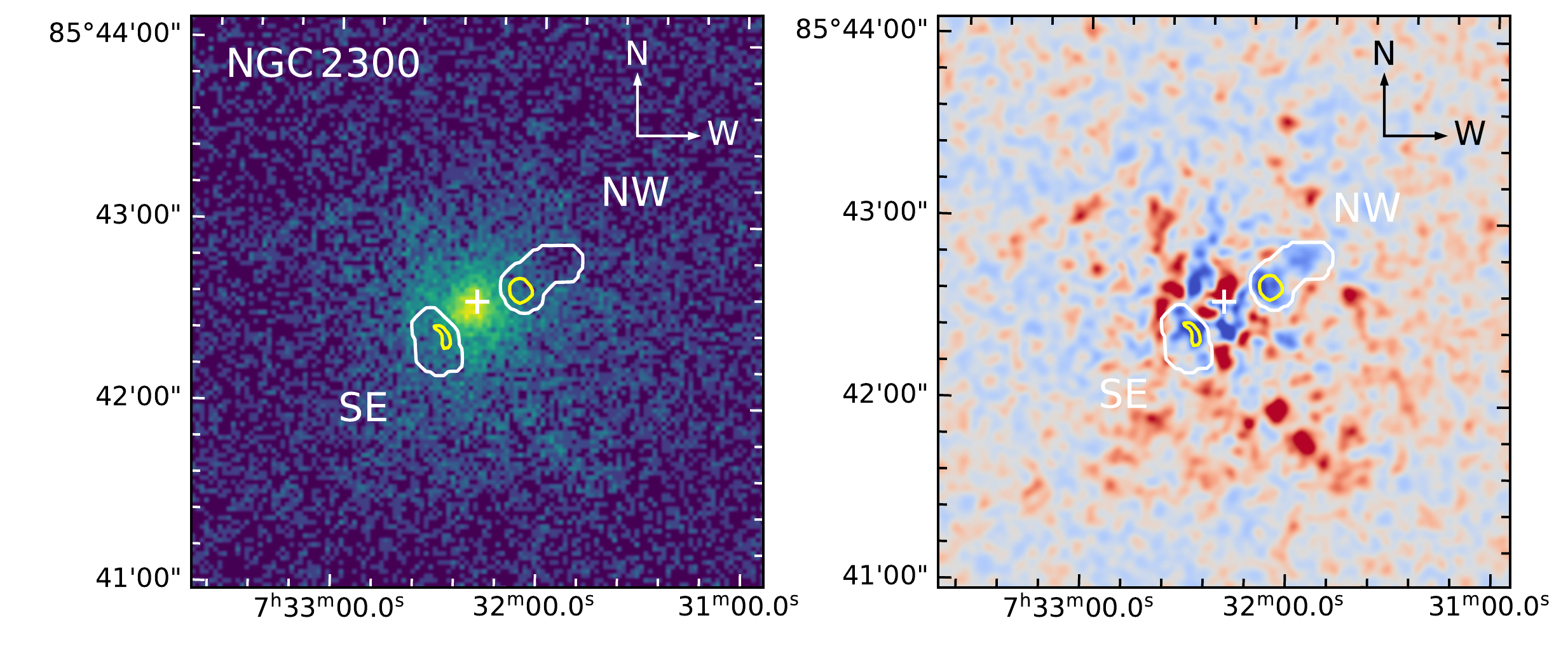}};
    \draw (2.25, 3.0) node {\includegraphics[height=0.22\textwidth]{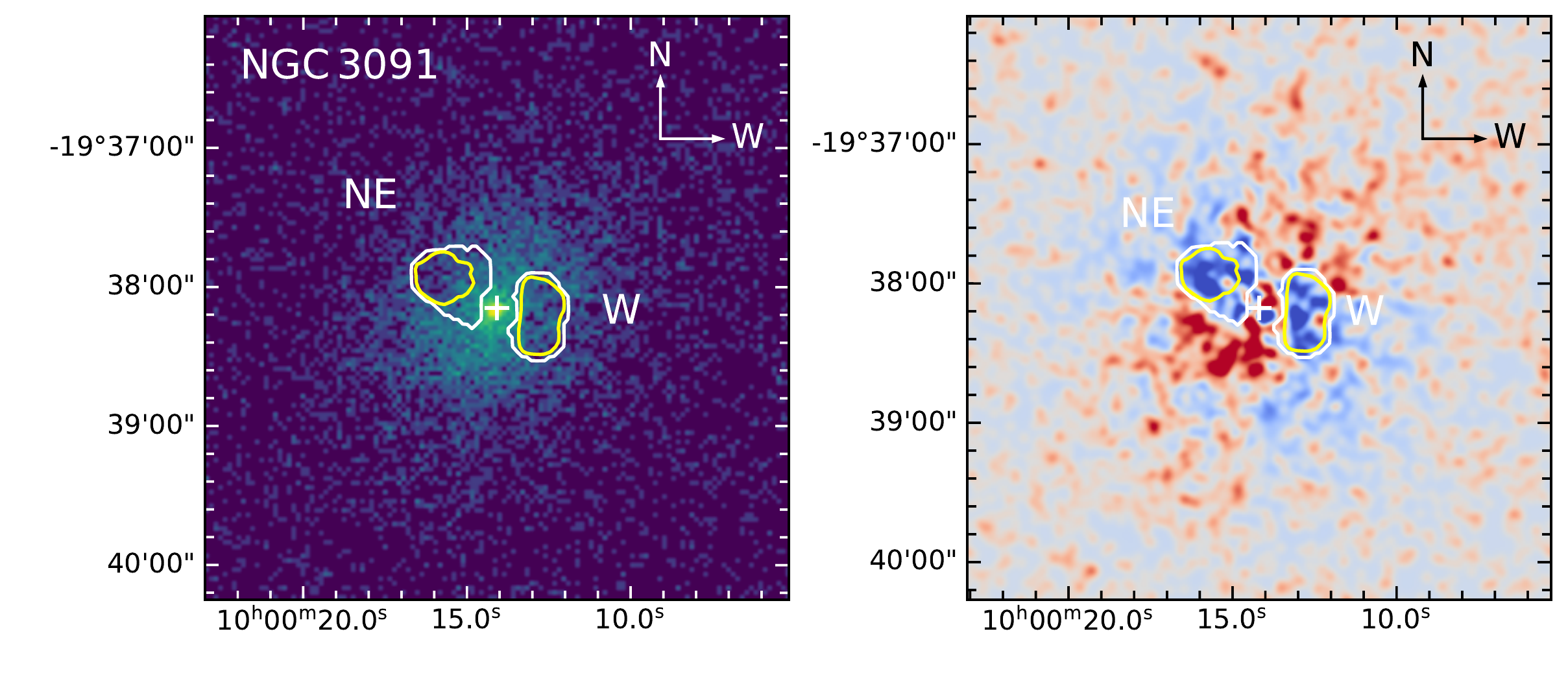}};
    \draw (-6.70, -1.1) node {\includegraphics[height=0.22\textwidth]{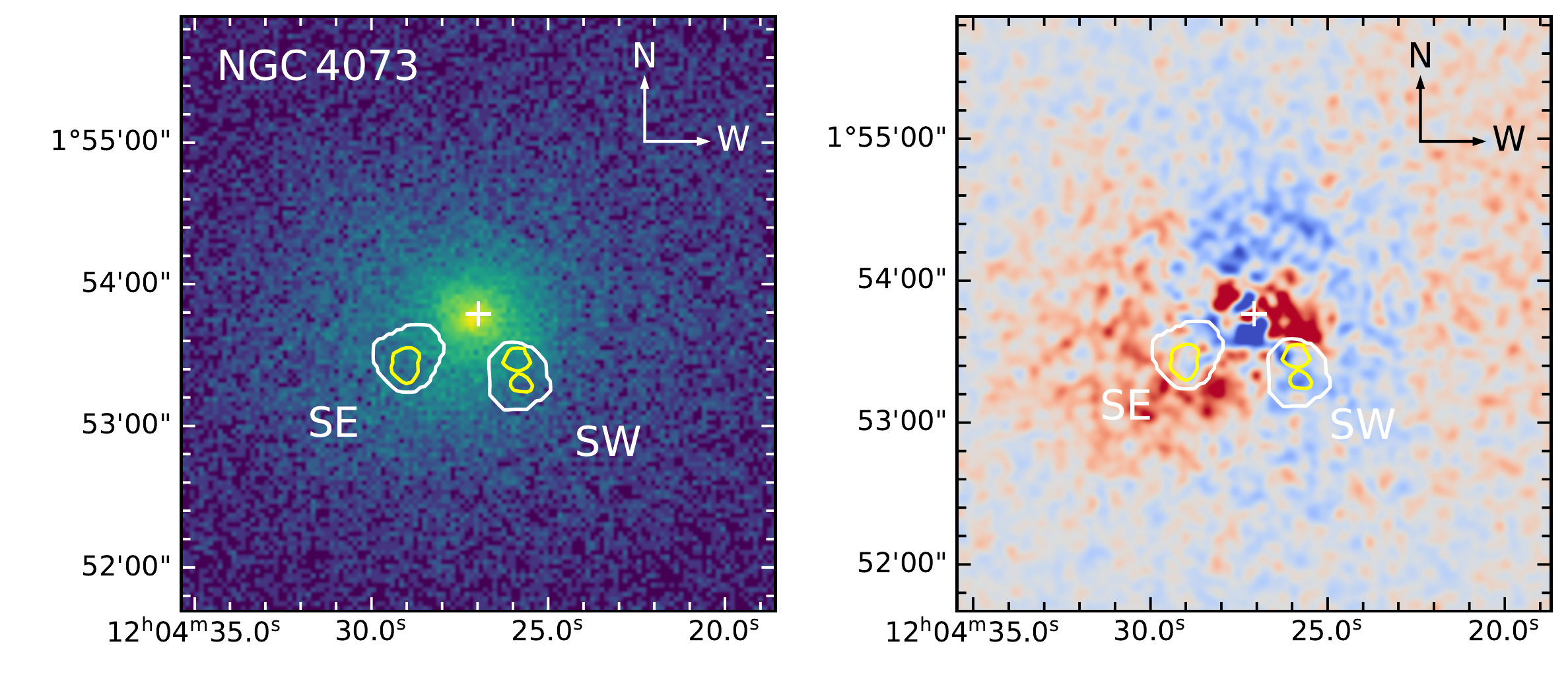}};
    \draw (2.20, -1.1) node {\includegraphics[height=0.22\textwidth]{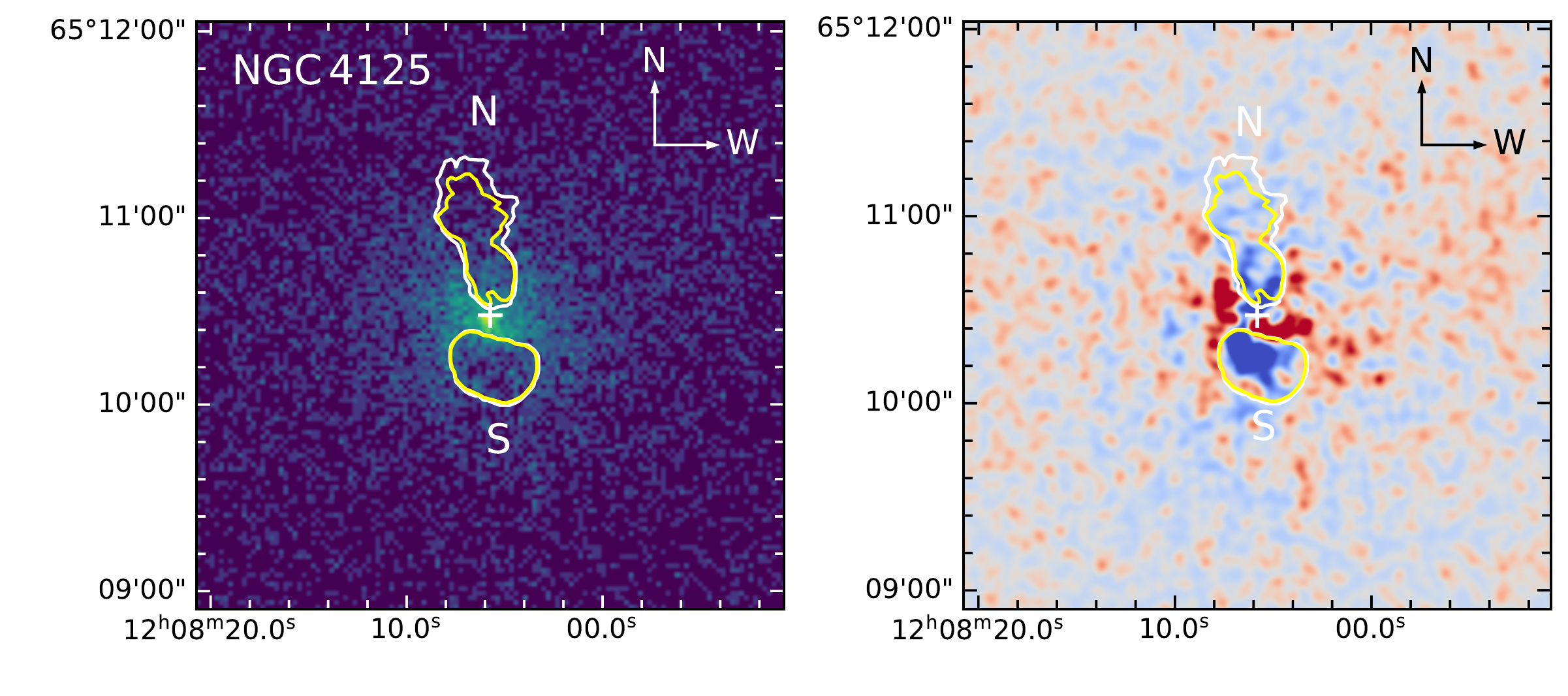}};
    \draw (-6.70, -5.2) node {\includegraphics[height=0.22\textwidth]{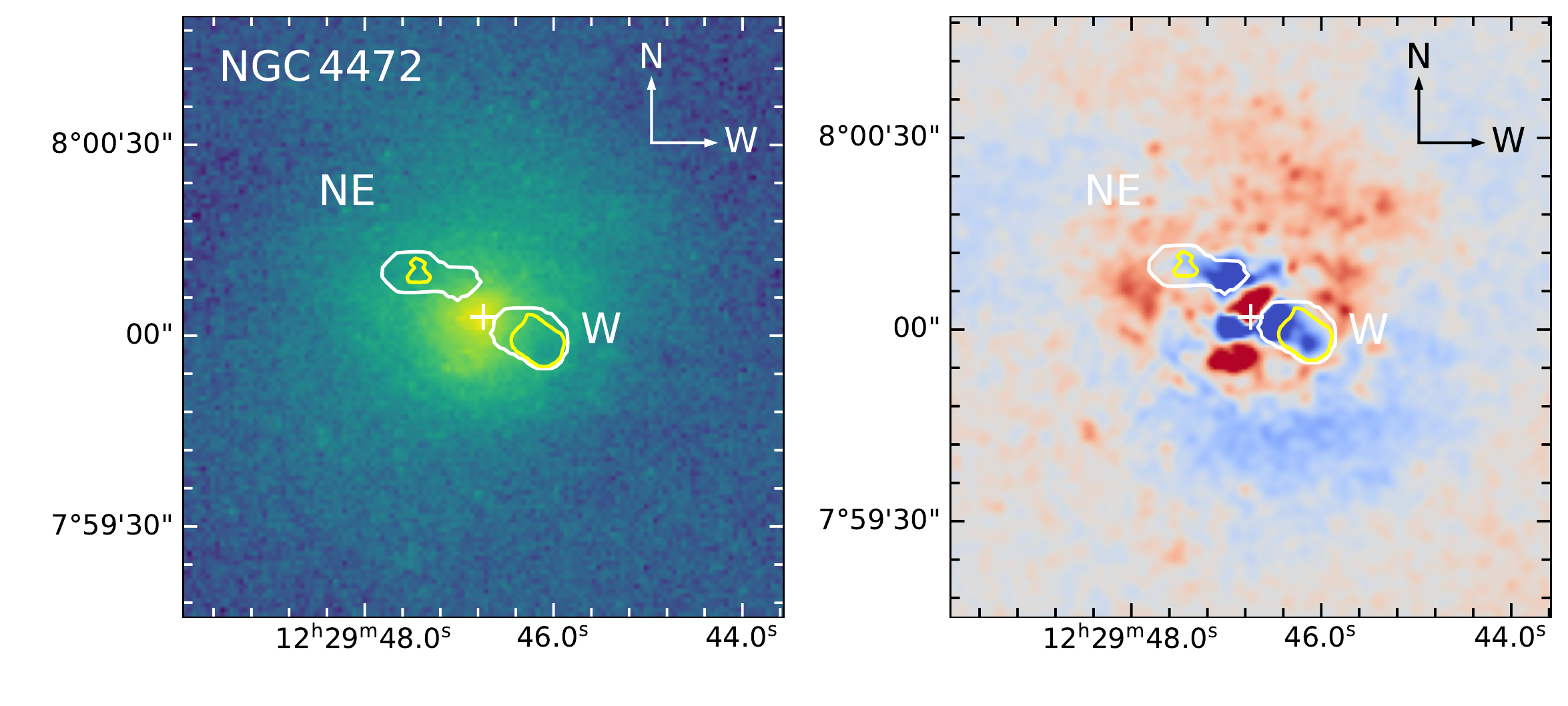}};
    \draw (2.20, -5.2) node {\includegraphics[height=0.22\textwidth]{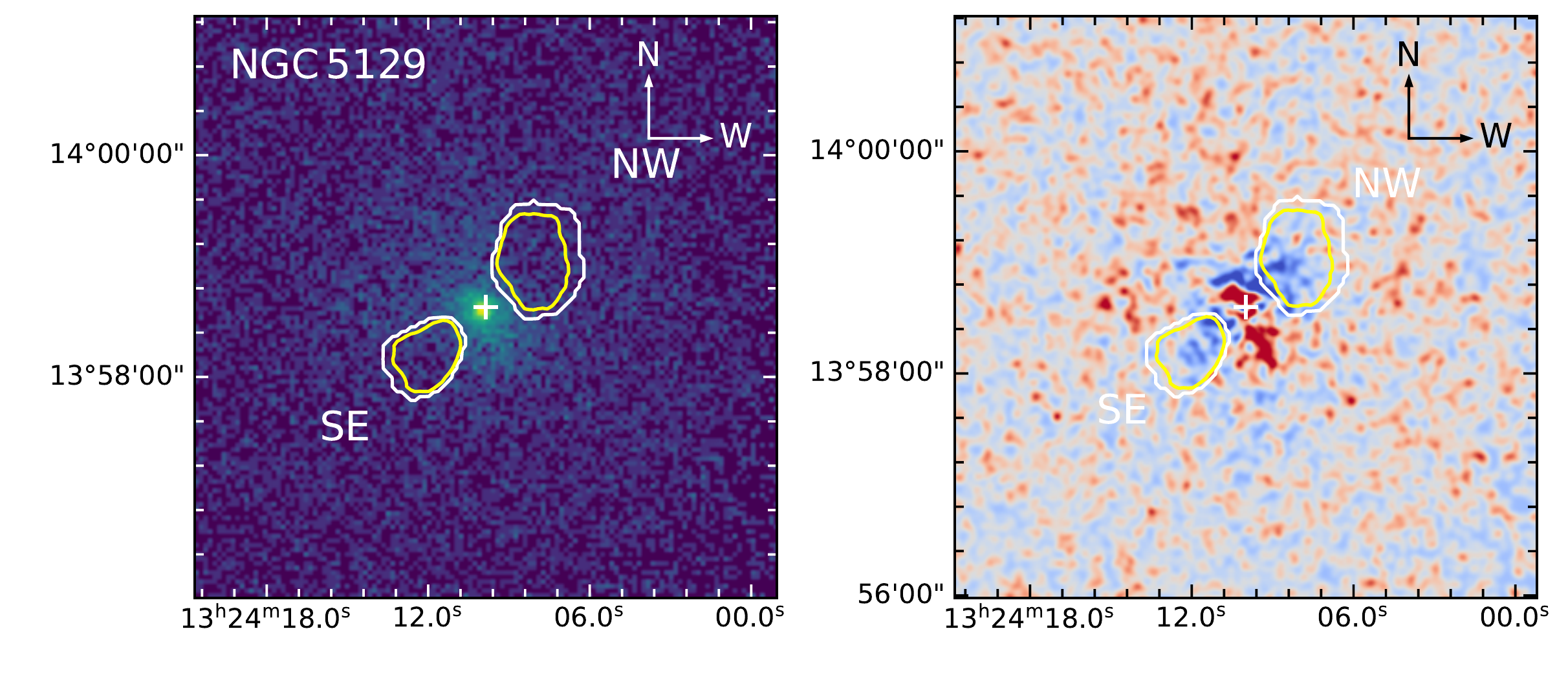}};
    \end{tikzpicture}
    \caption{Newly detected and confirmed X-ray cavities: IC\,4765, NGC\,533, NGC\,2300, NGC\,3091, NGC\,4073, NGC\,4125, NGC\,4472. \textbf{Left subplots}: original X-ray images with filled point sources overlaid by CADET prediction contours, \textbf{right subplots}: smoothed residual images, obtained by dividing the image by its best-fit beta-model, overlaid by the CADET prediction contours.}
    \label{fig:new_cavities}
\end{figure*}

\begin{figure*}
    \hspace{-5.0mm}
    \begin{tikzpicture}
    \draw (0, 9.6) node {\phantom{a}};
    \draw (-6.70, 7) node {\includegraphics[height=0.22\textwidth]{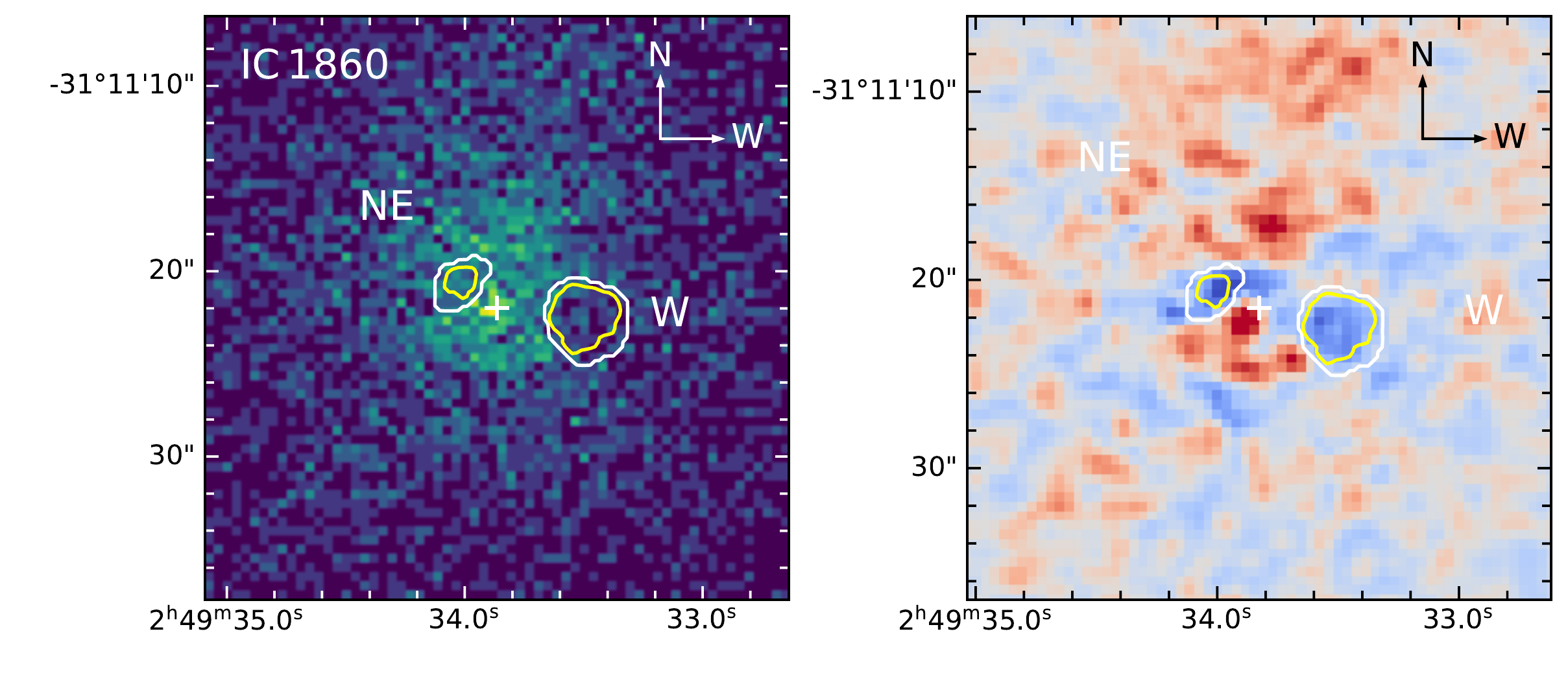}};
    \draw (2.20, 7) node {\includegraphics[height=0.22\textwidth]{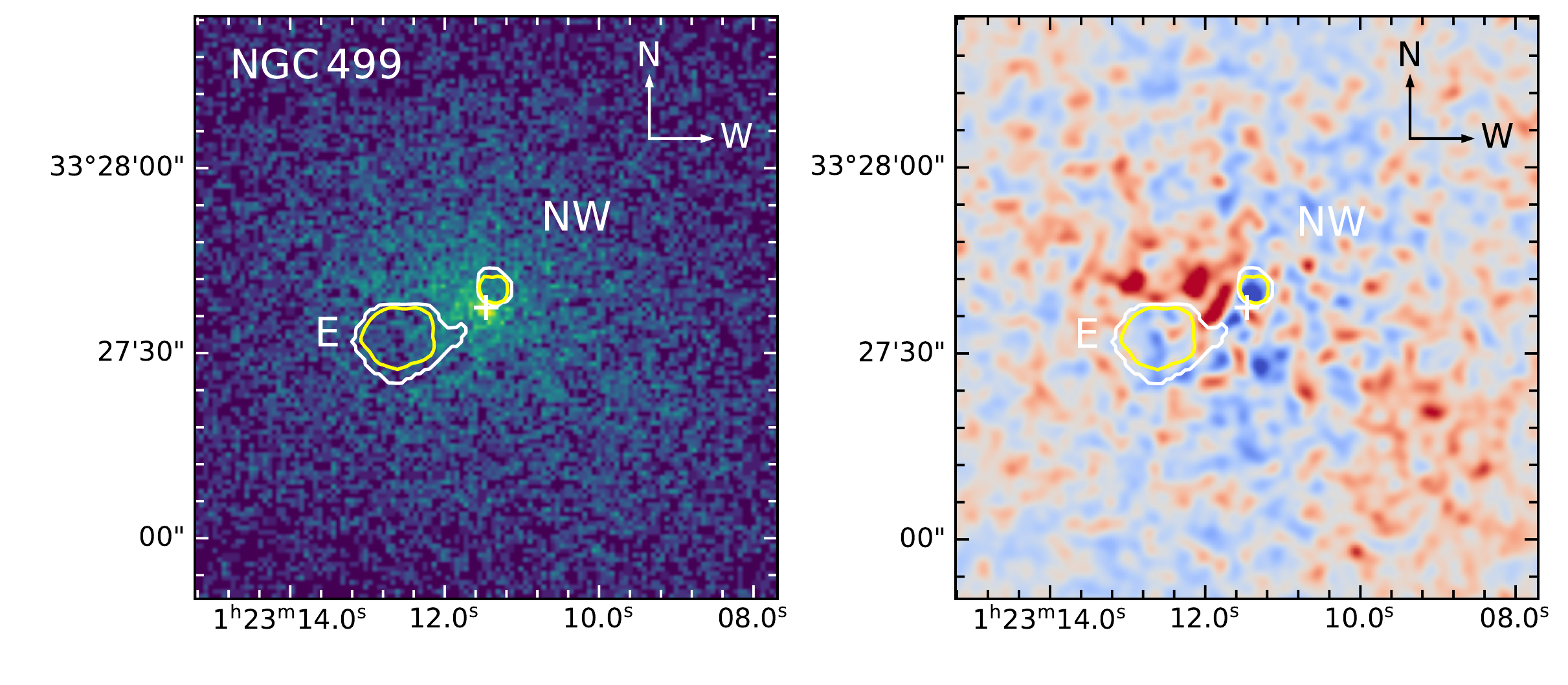}};
    \draw (-6.70, 3.0) node {\includegraphics[height=0.22\textwidth]{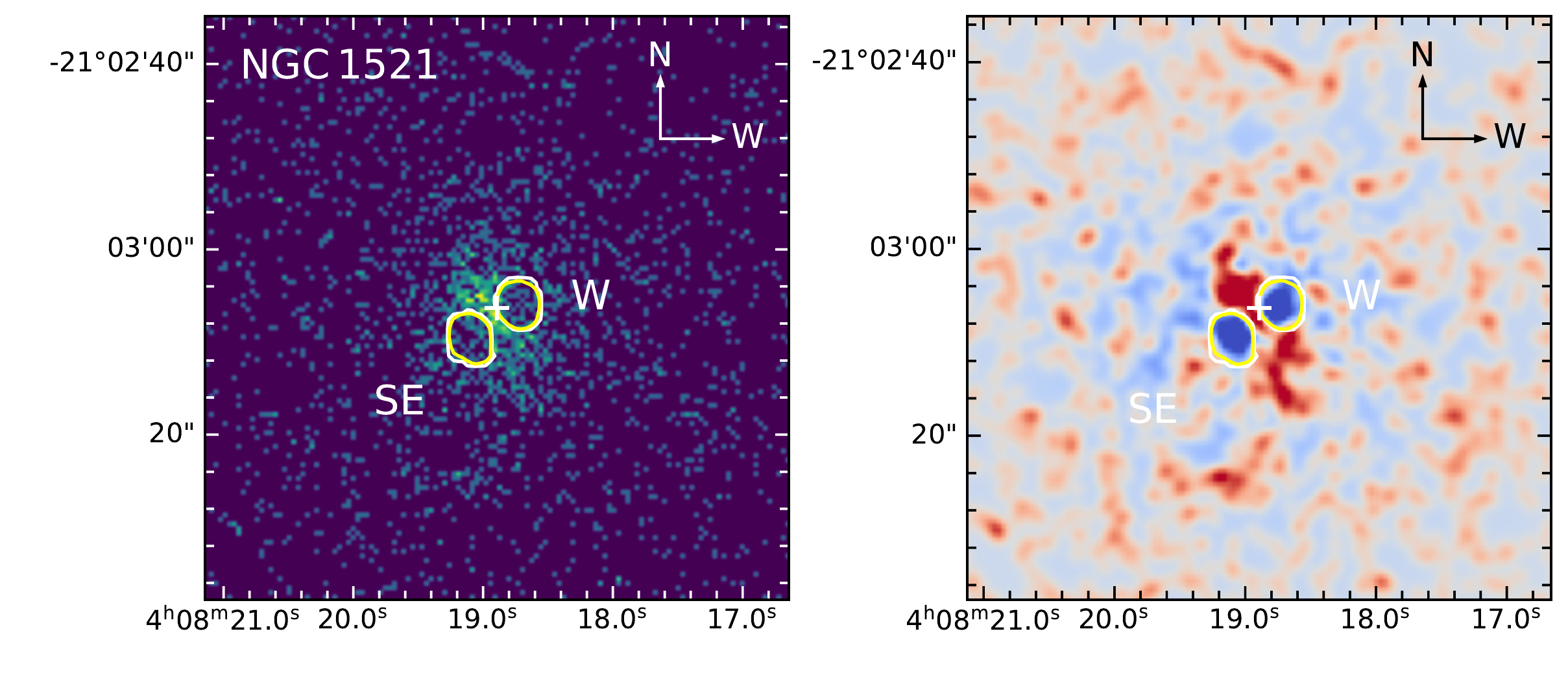}};
    \draw (2.20, 3.0) node {\includegraphics[height=0.22\textwidth]{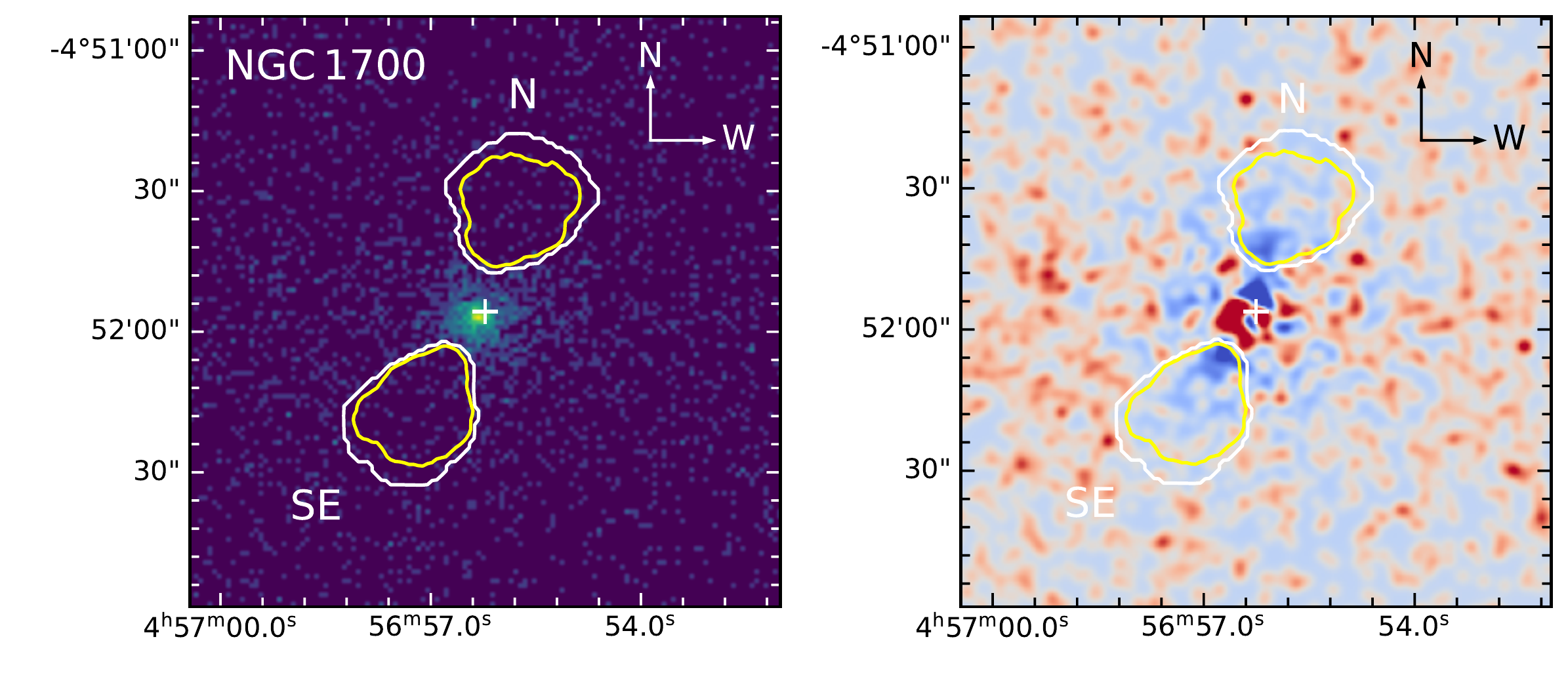}};
    \draw (-6.70, -1.1) node {\includegraphics[height=0.22\textwidth]{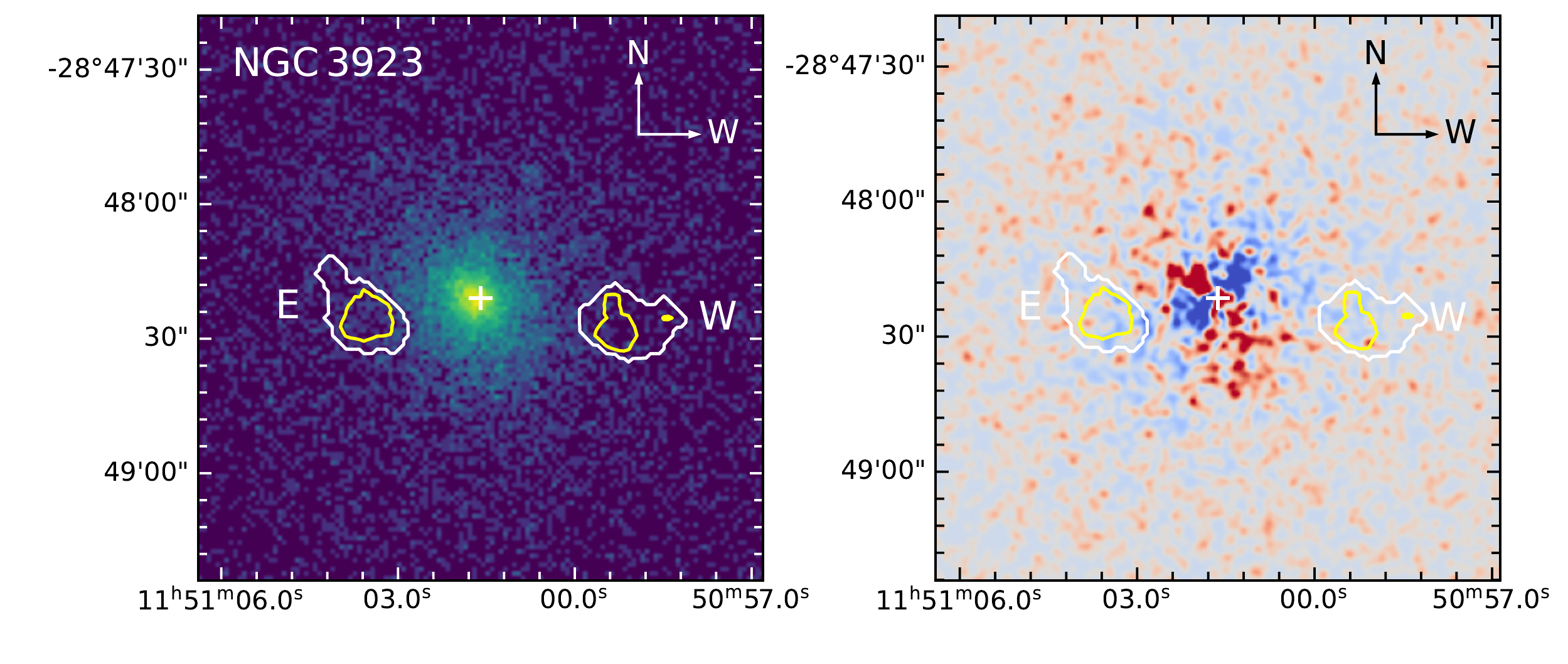}};
    \draw (2.20, -1.1) node {\includegraphics[height=0.22\textwidth]{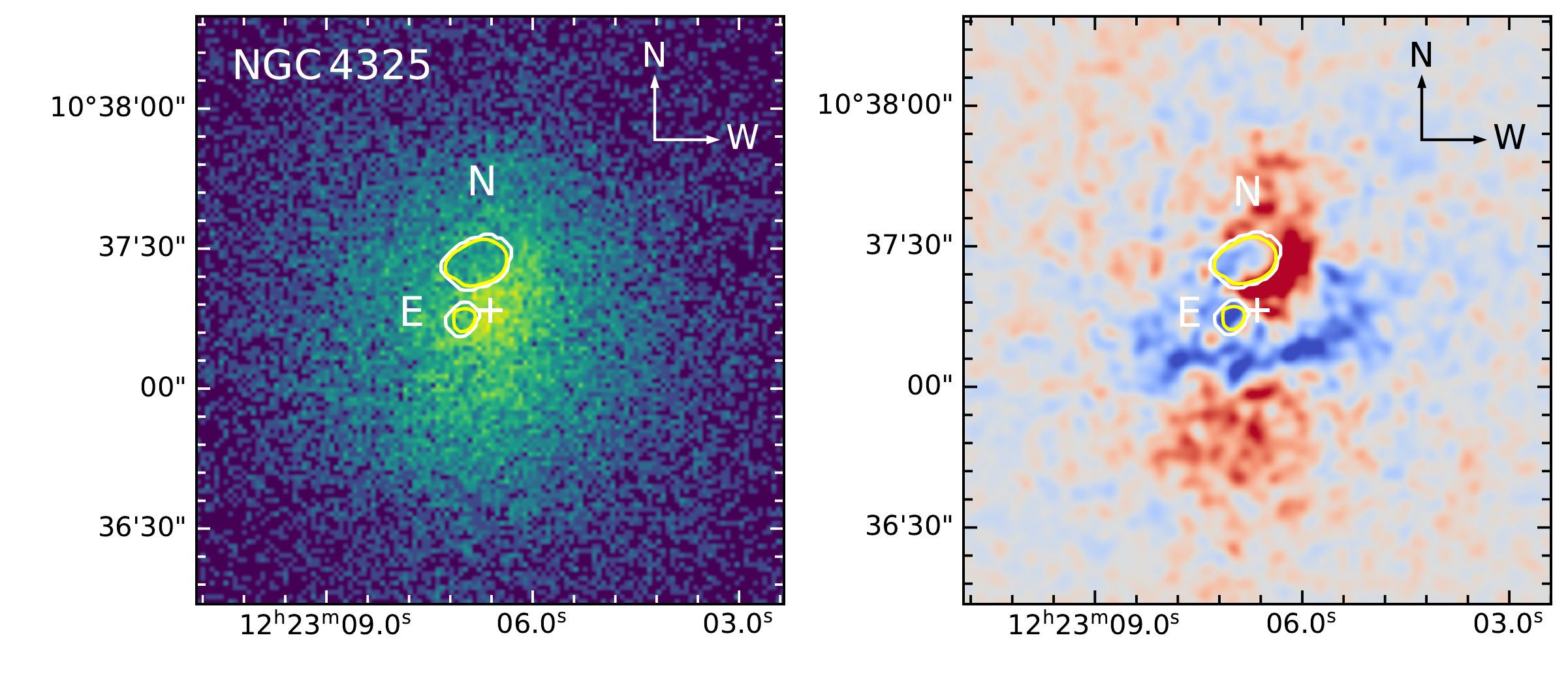}};
    \draw (-6.70, -5.2) node {\includegraphics[height=0.22\textwidth]{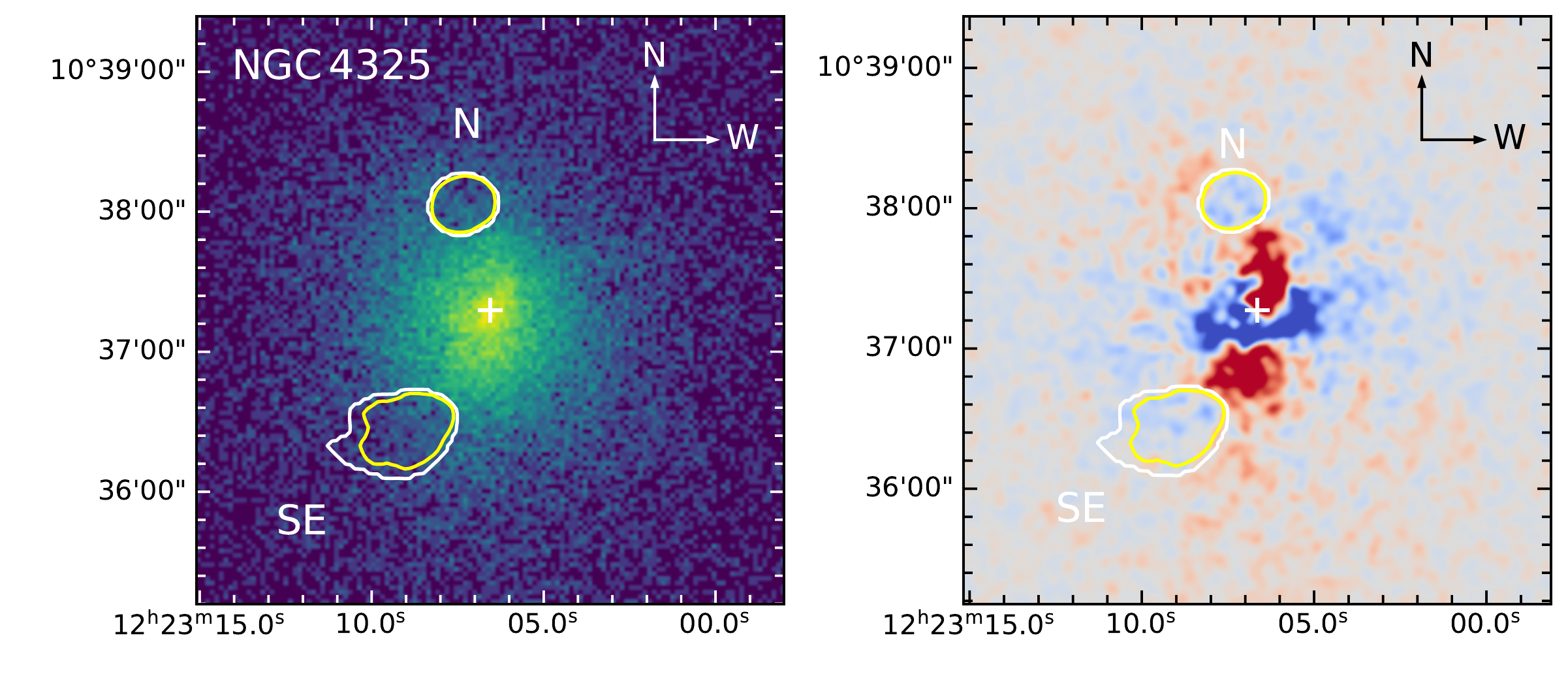}};
    \draw (2.20, -5.2) node {\includegraphics[height=0.22\textwidth]{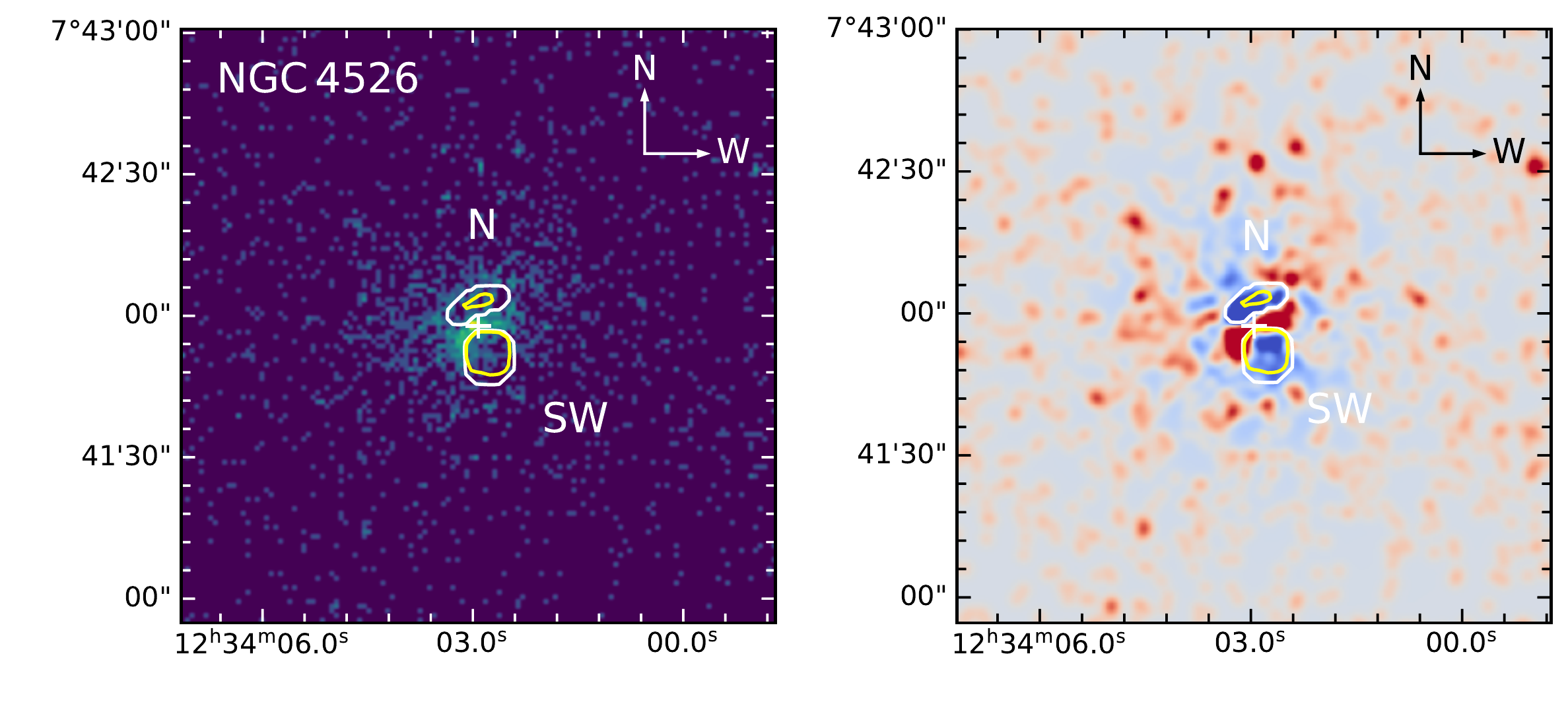}};
    \draw (-6.70, -9.3) node {\includegraphics[height=0.22\textwidth]{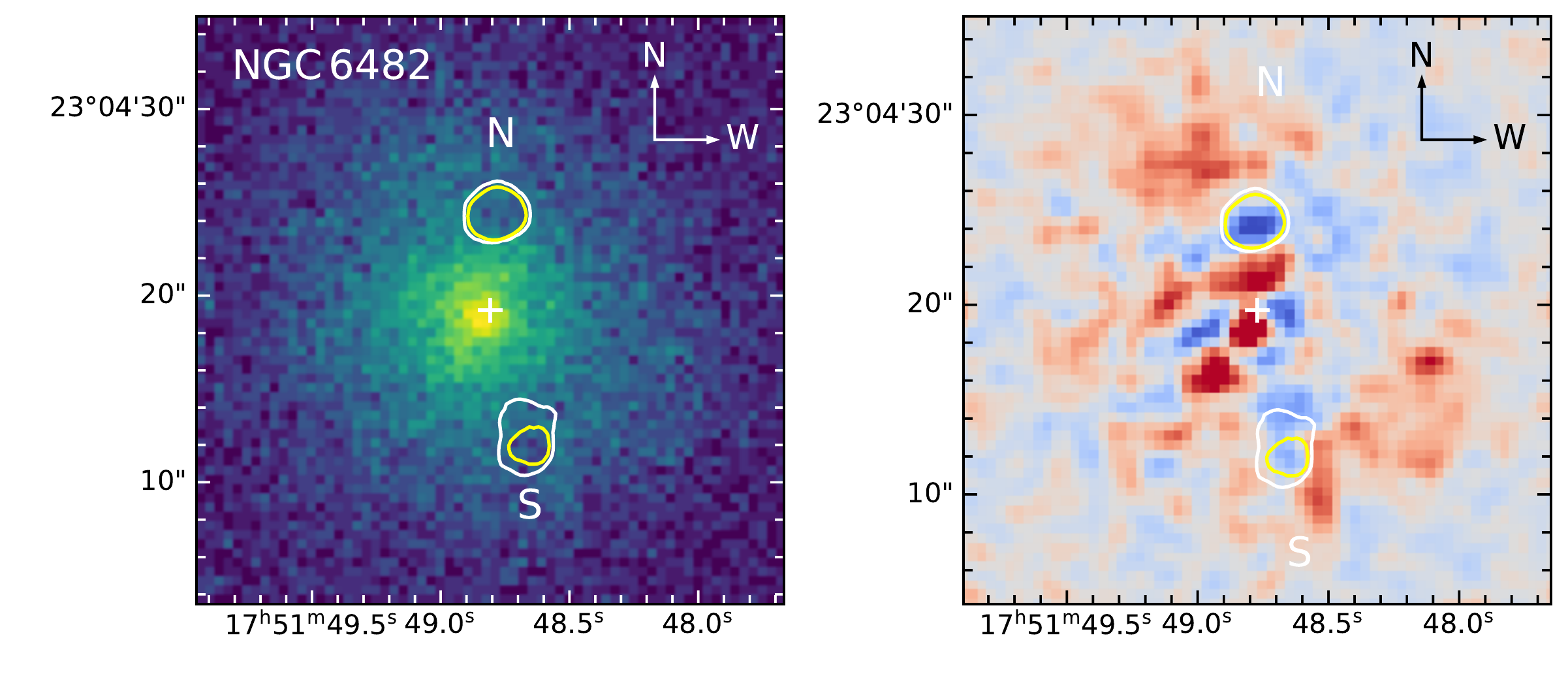}};
    \draw (2.20, -9.3) node {\includegraphics[height=0.22\textwidth]{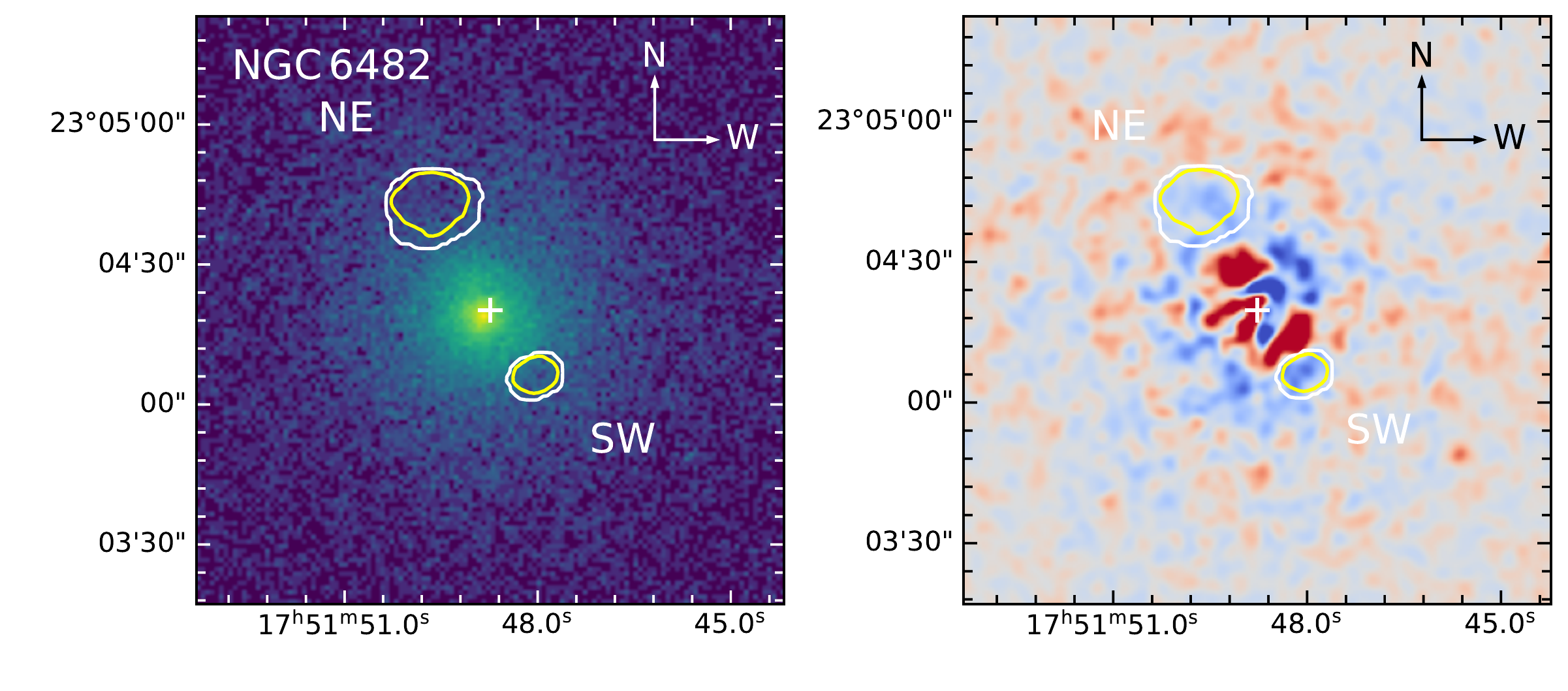}};
    \end{tikzpicture}
    \caption{Newly detected X-ray cavity candidates: IC\,1860, NGC\,499, NGC\,1521, NGC\,1700, NGC\,3923, NGC\,4325, NGC\,4526, NGC\,6482. \textbf{Left subplots}: original X-ray images with filled point sources overlaid by CADET prediction contours, \textbf{right subplots}: smoothed residual images, obtained by dividing the image by its best-fit beta-model, overlaid by the CADET prediction contours.}
    \label{fig:candidates}
\end{figure*}

Based on the combination of CADET predictions, visual inspection of analyzed images, radial and azimuthal photon count statistics (see Appendix \ref{section:significance}), and simulated images with similar properties, we claim the discovery of several new pairs of X-ray cavities (Table \ref{tab:new_detections}) in the following systems: IC\,4765, NGC\,533, NGC\,2300, NGC\,3091, NGC\,4073, NGC\,4125, NGC\,4472, NGC\,5129 (see Figure \ref{fig:new_cavities}). For the following galaxies, we report new cavity candidates for which further confirmation is needed: IC\,1860, NGC\,499, NGC\,1521, NGC\,1700, NGC\,3923, NGC\,4325, NGC\,4526, NGC\,6482 (see Figure \ref{fig:candidates}). For the rest of the galaxies, the CADET predictions are most likely spurious false positive detections -- mainly for the below-discussed reasons (see Section \ref{section:discussion}).

\subsection{Cavities in distant clusters}
 
Besides the sample of 70 nearby early-type galaxies for which the CADET pipeline was optimized, we also aimed to test CADET predictions for sources whose parameters lie outside the distributions used to generate our training images. We, therefore, applied CADET to a sample of well-probed but more distant ($z > 0.05$) galaxy clusters. For the testing, we selected 7 clusters of galaxies known to harbour X-ray cavities in their hot atmospheres: Abell\,2597 \citep{McNamara2001}, Abell\,3017 \citep{Parekh2017,Pandge2021}, Hydra\,A \citep{Kirkpatrick2009}, RBS\,797 \citep{Cavagnolo2011,Ubertosi2021}, MS\,0735.6+7421 \citep{McNamara2005}, SPT-CLJ0509-5342 \citep{Hlavacek2015}, and SPT-CLJ0616-5227 \citep{Hlavacek2015}.

X-ray cavities in distant galaxy clusters are generally harder to detect compared to nearby early-type galaxies and galaxy groups because observations of galaxy clusters are characterised by lower S/N ratio, higher background contribution,
and they often have substantial and complicated sloshing patterns and cold fronts due to more frequent merger activity. However, despite being trained mostly based on parameters of nearby giant elliptical galaxies, for all of the selected galaxy clusters, CADET was able to detect all 14 cavities from the previously known cavity pairs (see Figure 2 in the Supplementary material). For galaxy cluster Abell\,3017, we discovered a possible additional outer pair of X-ray cavities. In the central parts of MS\,0735.6+7421, CADET detected also low-significance inner cavities previously proposed by \cite{Biava2022}.

\section{Discussion}
\label{section:discussion}

We showed that the utilization of machine learning techniques (convolutional neural network) for the detection and size estimation of X-ray cavities on noisy \textit{Chandra} images provides satisfactory results. Although the convolutional neural network was trained purely using elliptical cavity masks (ellipsoidal cavities), it is able to produce arbitrarily shaped cavity predictions. This property can be utilized in a more accurate determination of their total volume and therefore also the corresponding energy required for their inflation ($4pV$).

Even though we only generated images with zero or one pair of cavities for the training dataset, the network is able to find an arbitrary number (also a non-even number) of surface brightness depressions. The advantage of this behaviour is that, for multi-cavity systems, all cavities can be recorded at once and larger-scale cavities are not prioritized over small-scale ones. On the other hand, for many systems, besides real high-significance cavities also low-significance brightness drops and possibly false positive predictions can be obtained.

In the current state, the output of the network needs to be further visually analysed by a human. The scientist should visually assess the reliability of cavity predictions and, for further calculations, choose only scales and cavities of interest. This is necessary since more than two cavities can appear on the image either false-positively or naturally (systems with multiple cavity pairs, e.g. NGC\,5813).

We note that the CADET pipeline has problems with detecting cavities in galaxies with more complex spiral-like structures (IC\,1262, NGC\,1553, NGC\,4636) as well as very bright rims (NGC\,4374). The network is also unable to properly handle X-ray jets or similar elongated structures and tends to predict cavities to be located on either one or both sides of such structures (e.g. NGC\,315, NGC\,383, NGC\,4261). Another type of hardly analyzable images are, as expected, very faint sources with an extremely low number of counts (e.g. NGC\,57, NGC\,410, NGC\,777, NGC\,2305) and due to spaces between the chips also galaxies observed only by the ACIS-I chip (NGC\,1387, NGC\,2563). Proper detection and estimation of cavity extent are nearly impossible also for galaxies that are currently undergoing a noticeable merging event (e.g. NGC\,507, NGC\,741, NGC\,1132, NGC\,4406, NGC\,7618) or ram pressure stripping (NGC\,1404, NGC\,4342, NGC\,4552).

\subsection{Reliability of CADET predictions}
\label{section:cavity_reliability}

The convolutional part of the CADET pipeline produces pixel-wise prediction maps (128x128 pixels) with values in each pixel ranging from 0 to 1. The values are typically highest (close to 1) in cavity centres and decrease towards their edges. By discarding pixels with values lower than a given threshold (discrimination threshold), cavity predictions can be trimmed to obtain only the most `significant' central parts. By setting a pair of discrimination thresholds (see Appendix \ref{section:discrimination_threshold}), the CADET pipeline can be simultaneously calibrated to produce not overestimated nor underestimated cavity predictions while maintaining a reasonable level of true-positive and false-positive rates (Figure~\ref{fig:discrimination_threshold}).

\begin{figure}
    \centering
    \includegraphics[width=0.95\columnwidth]{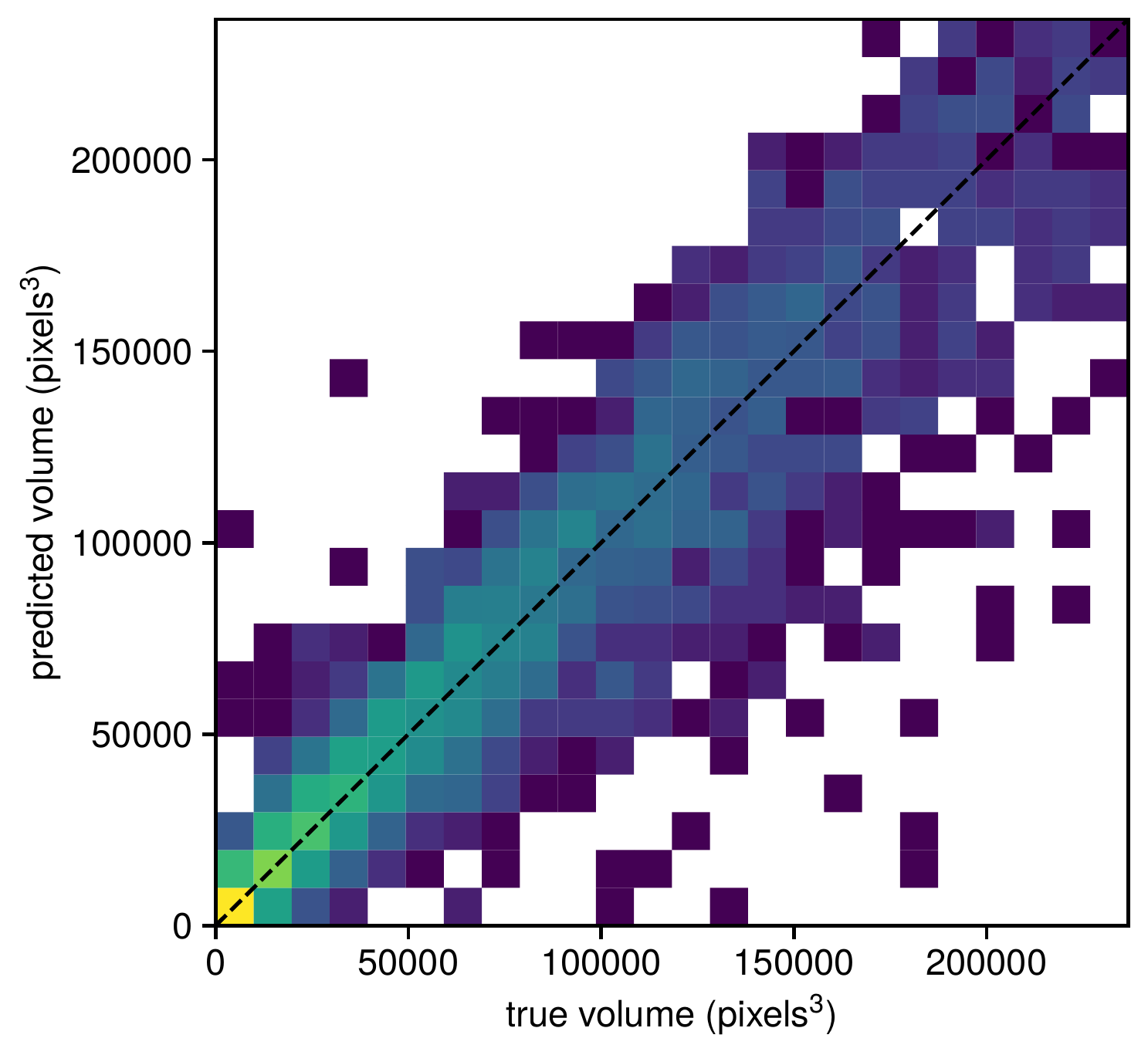}
    \vspace{-2mm}
    \caption{Predicted vs true cavity volumes for a sample of $10^4$ testing images with X-ray cavities obtained for the optimized discrimination thresholds. The black dashed line represents the perfect recovery, the colour indicates the total number of points inside the corresponding bin.}
    \label{fig:true_pred}
\end{figure}

We note, however, that such a pair of discrimination thresholds optimizes the errors, TP and FP rates averaged out over the whole range of possible $\beta$-model and cavity parameters and that using an optimized discrimination threshold might still result in overestimating some of the cavities while underestimating others. For each galaxy, a well-calibrated prediction is, therefore, reached using a slightly different value of the discrimination threshold. In order to investigate how the value of the optimal discrimination threshold depends on the parameters of given images, we binned testing images into groups with a distinct number of counts, parameters of $\beta$-models (core radius $r_0$, slope parameter $\beta$), or parameters of inserted cavities (cavity sizes $r_{\text{cav}}$, see Figure \ref{fig:true_pred}).

\begin{figure}
    \centering
    \includegraphics[width=\columnwidth]{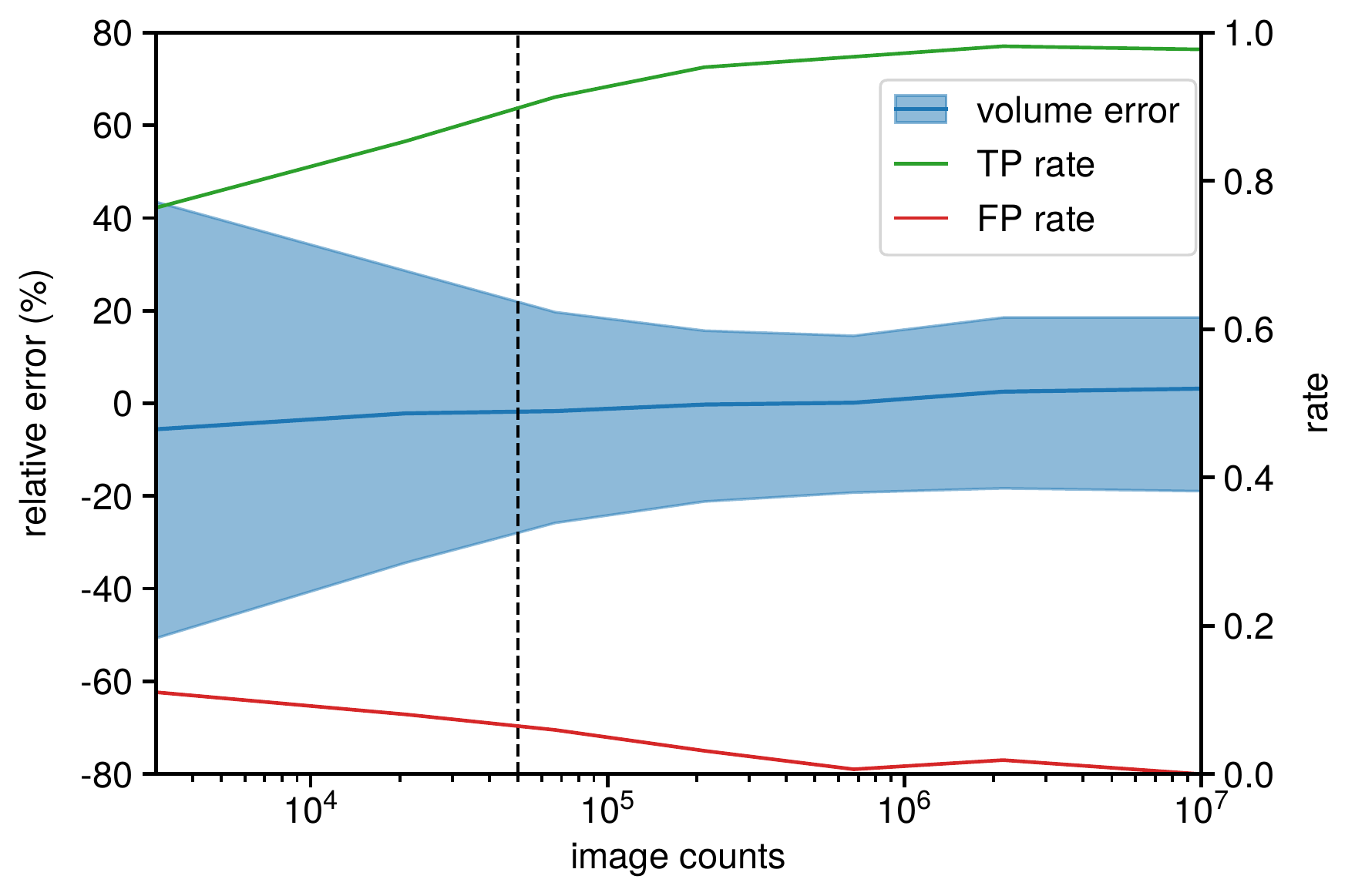}
    \vspace{-5mm}
    \caption{Relative volume error (blue), true-positive (TP) rate (green), and false-positive (FP) rate (red) of CADET predictions as a function of the number of counts in the input images. Volume errors and true-positive rates were estimated for a sample of $10^4$ testing images each containing a single pair of X-ray cavities and false-positive rates were derived from a sample of $2000$ testing images without any X-ray cavities. Parameter distributions used to generate testing images are identical to the parameter distributions of training images. The vertical dashed line corresponds to the division line between low-count images (less than $50\,000$ counts) and high-count images.}
    \label{fig:error_counts}
\end{figure}

Considerably strong dependence was, however, found only for the number of counts of the given images (Figure \ref{fig:error_counts}). We have therefore binned testing images by the number of counts into 2 bins and estimated optimal discrimination thresholds for both bins separately: for images with the number of counts lower than $50\,000$ the optimal thresholds are 0.4 and 0.6, and for images with more than $50\,000$ counts 0.45 and 0.3, respectively. Since for images with more than $50\,000$ counts, the discrimination threshold optimizing TP and FP rates is lower than the threshold optimizing cavity volumes, we only apply the double-threshold approach for images less than $50\,000$ counts.

\subsection{Angular dependence}
\label{section:angular_dependence}

Although jets and X-ray cavities are expected to emerge at random orientations with respect to the observer, when estimating their properties, it is usually assumed that all cavities are located in the plane of the sky. We note, however, that this simplification leads, for many systems, to the underestimation of cavity distances and ages, and when calculating total cavity energy also to the overestimation of surrounding pressures.

Nevertheless, for X-ray cavities launched at higher angles with respect to the plane of the sky, their contrast decreases rapidly with increasing distance from the galactic centre. According to \cite{Enslin2002}, for a spherical cavity launched at the angle of 0 degrees with respect to the plane of the sky, the detectability declines with distance as $d^{-0.3}$, at 45 degrees as $d^{-1}$, and at the angle of 90 degrees as $d^{-3}$. For the detected cavities, there is therefore a much higher chance that they are located closer to the plane of the sky. This idea is supported also by the observed correlation and small scatter between cavity sizes and their distances from the galactic centre (see Figure \hyperlink{figD3}{D3}).

\begin{figure}
    \centering
    \includegraphics[width=0.98\columnwidth]{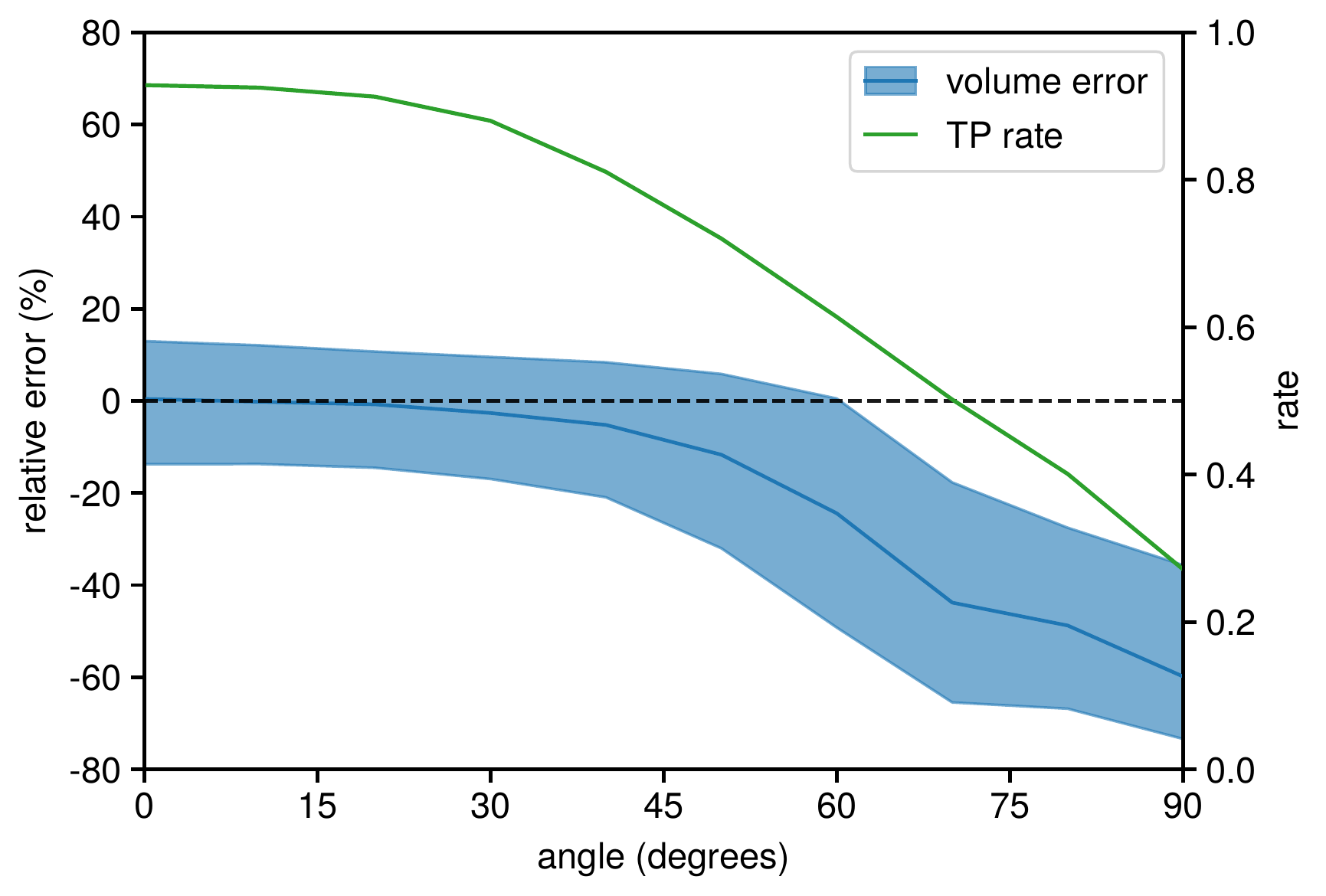}
    \vspace{-2mm}
    \caption{Average reconstructed cavity volume errors and TP rates as a function of cavity launching angle. Parameters of the images were generated from the identical distribution as used to generate the testing images.}
    \label{fig:error_distance_per_angle}
\end{figure}

Using simulated images, we tested the angular dependence of cavity detectability for the CADET pipeline (see Figure \ref{fig:error_distance_per_angle}). We generated images with parameter distributions identical to testing images and we varied the plane-of-the-sky positional angle $\theta$ of the primary cavity in a uniform range from 0 to 90 degrees\footnote{The secondary cavity was launched in an opposite angle $-\theta$ with random gaussian variation ($\mu = 0$, $\sigma = 10^{\circ}$).}. Up to an angle of $\sim 45$ degrees, the detectability (TP rate) and prediction accuracy (cavity volume error) only weakly depend on the plane-of-the-sky positional angle of X-ray cavities. For higher launching angles ($45-90$ degrees), the detectability steeply decreases and reconstructed volumes of true-positive X-ray cavities launched are heavily underestimated. For the angle of 90 degrees, the TP rate only reaches $25\,\%$. Cumulatively, the CADET pipeline was able to detect $66\,\%$ of X-ray cavities compared to $89\,\%$ for testing images with most X-ray cavities close to the plane of the sky. Assuming that the vast majority of early-type galaxies harbour X-ray cavities, with current data quality and methods they can only be detected for approximately 2 out of 3 galaxies. We note that this is consistent with the fraction of previously known X-ray cavities (33 sources) combined with new CADET detections (6 sources) and cavity candidates (8 sources) for our sample of 70 nearby early-type galaxies ($47 / 70 = 67\,\%$).

\section*{Conclusions}
\label{section:conclusion}

We have developed a machine learning pipeline called \textit{Cavity Detection Tool} (CADET), which we trained for finding and size-estimating surface brightness depressions (X-ray cavities) on \textit{Chandra} images of early-type galaxies and galaxy clusters. We have shown that the brute force of modern computer technology combined with the state-of-the-art algorithms represented by convolutional neural networks are capable of automation of more complex astronomical tasks such as detecting elliptical brightness drops in noisy X-ray data. 

The CADET network was trained on a large set ($\approx 10^6$) of artificial images generated to imitate real X-ray data of early-type galaxies as observed by the \textit{Chandra X-ray Observatory}. The density distribution of the simulated galaxies was approximated by an ellipsoidal 3D single or double $\beta$-model from which we cut out either one or zero pairs of ellipsoidal cavities. Parameters of both simulated galaxy models and inserted cavities were generated based on an analysis of 70 nearby early-type galaxies, 33 of which contained one or more pairs of X-ray cavities (50 cavity pairs in total).

By varying the hyper-parameters of the network (fraction of cavities in training images, learning rate, dropout rate) and training all networks for 32 epochs, we found the optimal combination of parameters with the best performance on the testing dataset of mock and real images. For the best-performing network, we further tested the accuracy of its predictions as well as true-positive (TP) and false-positive (FP) rates. Moreover, the best-performing network was calibrated (see Appendix \ref{section:discrimination_threshold}) not to produce overestimated nor underestimated predictions (discrimination threshold $\approx 0.4$) while maintaining a plausible level of FP and TP rates (discrimination threshold $\approx 0.6$). After the calibration, we obtain the resulting average absolute volume reconstruction error of $14^{+13}_{-8}$ per cent and true-positive rate of $89$ per cent for the testing dataset of $10^4$ images with X-ray cavities, and a false positive rate of $5$ per cent for the testing dataset of $2000$ images without any X-ray cavities.

For the sample of 33 galaxies known to harbour one or more pairs of X-ray cavities, the CADET pipeline recovered 91 out of 100 cavities. Furthermore, a comparison between predictions produced by the CADET pipeline and human-made predictions shows a good agreement (see Figure~\ref{fig:CADET_vs_manual}) between estimated cavity volumes with an average difference of $-0.04_{-0.10}^{+0.19}$ dex and $-0.06_{-0.15}^{+0.26}$ dex for this work and \cite{Shin2016}, respectively. 

Furthermore, the CADET network led to the discovery of 8 pairs of new X-ray cavities in nearby early-type galaxies (IC\,4765, NGC\,533, NGC\,2300, NGC\,3091, NGC\,4073, NGC\,4125, NGC\,4472, NGC\,5129), that was further confirmed using azimuthal and radial photon count statistics as well as by simulating mock images with similar properties. The CADET pipeline also discovered another 10 potential cavity pair candidates in 8 sources (IC\,1860, NGC\,499, NGC\,1521, NGC\,1700, NGC\,3923, NGC\,4325, NGC\,4526, NGC\,6482), that need further confirmation in the form of deeper X-ray observations or detection of co-aligned extended radio emission. When applied to images of 7 galaxy clusters harbouring X-ray cavities, the CADET pipeline was able to recover all known X-ray cavities, confirm a possible cavity pair candidate \citep[MS\,0735.6+7421;][]{Biava2022}, and detect a new pair of cavity candidates (Abell\,3017).

\section*{Acknowledgements}

This research was supported by GACR grant 21-13491X. A.S. is supported by the Women In Science Excel (WISE) programme of the Netherlands Organisation for Scientific Research (NWO), and acknowledges the Kavli IPMU for the continued hospitality. SRON Netherlands Institute for Space Research is supported financially by NWO. This research has made use of the NASA/IPAC Extragalactic Database (NED), which is funded by the National Aeronautics and Space Administration and operated by the California Institute of Technology. We thank Laurence Perreault-Levasseur, Yashar Hezavez, Carter Rhea, and Matej Kosiba for valuable advice in the field of convolutional neural networks and machine learning in general. We thank Steve W. Allen for his comments and feedback on the manuscript, and Tony Mroczkowski for discussing the potential application to galaxy clusters.

\section*{Data availability}

The data in this article are available on request to the corresponding author. The source code of training and testing scripts as well as the trained model are available on CADET GitHub page (\url{https://github.com/tomasplsek/CADET}).



\bibliographystyle{mnras}
\bibliography{bibliography}


\appendix

\section{Cavity volume calculation methodology}
\label{section:cavity_volume}

\begin{figure*}
    \centering
    \includegraphics[width=\textwidth]{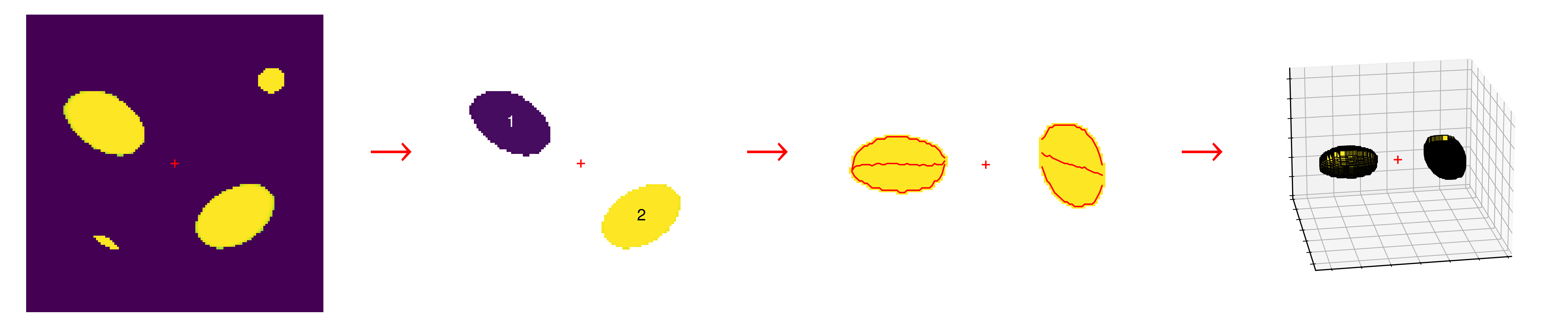}
    \vspace{-2mm}
    \caption{Visualization of the cavity volume estimation process, from left to right: pixel-wise prediction, coordinates of two largest cavities, de-rotated cavities, rotationally symmetric 3D models of both cavities. For visualization purposes, cavity number 1 was de-rotated to the left side of the image while cavity number 2 was to the right side.}
    \label{fig:derotation}
\end{figure*}

When estimating cavity volumes from decomposed CADET predictions for testing images, we select the two largest X-ray cavities if more cavities are present. For predictions on real images, volumes are estimated for cavities of interest specified visually by the author. To calculate volumes of thus selected cavities, we use the exact predicted cavity shape obtained after applying both discrimination thresholds and we do not further approximate cavities with an elliptical shape. To estimate the depth of the cavity, we only assume rotational symmetry around the direction towards the centre of the image (coordinate $x=63.5$, $y=63.5$) and therefore we assume the depth of the cavity at every point along the galactocentric direction to be equal to its width at that direction. 

This is technically realized by splitting the cavity prediction into two $128\times128$ matrices, each containing a single cavity. Each cavity matrix is then rotated so its centre of mass is at $y=63.5$. For this `de-rotated' image, the central pixel and width are derived in each pixel-row along the x-axis and the depth of the cavity in each pixel-row is assumed to be equal to its width (see Figure \ref{fig:derotation}). We note that the code published on the CADET GitHub page is robust also for more complex cavity shapes with a gap in between two cavity parts in the same pixel row (e.g. banana shape).

The resulting 3D representations of cavities in the form of $128\times128\times128$ cubes are saved into a binary NumPy file format and can be further combined with measured pressure profiles to calculate the energy contained in X-ray cavities. As already tested in \cite{Plsek2022}, such an approach provides more accurate total energy estimates compared to calculating the energy solely from central pressure and total cavity volumes.

\section{Optimization of the discrimination threshold}
\label{section:discrimination_threshold}

Pixel-wise predictions obtained from the CNN part of the CADET pipeline are being further decomposed into individual cavities using the DBSCAN algorithm. Nevertheless, before the decomposition, the pixel-wise predictions need to be further trimmed so that only pixels above a given value (discrimination threshold) are accounted for. The choice of this discrimination threshold will, however, strongly affect predicted areas of detected cavities and also the amount of true positive and false positive predictions.

\begin{figure}
    \centering
    \includegraphics[width=\columnwidth]{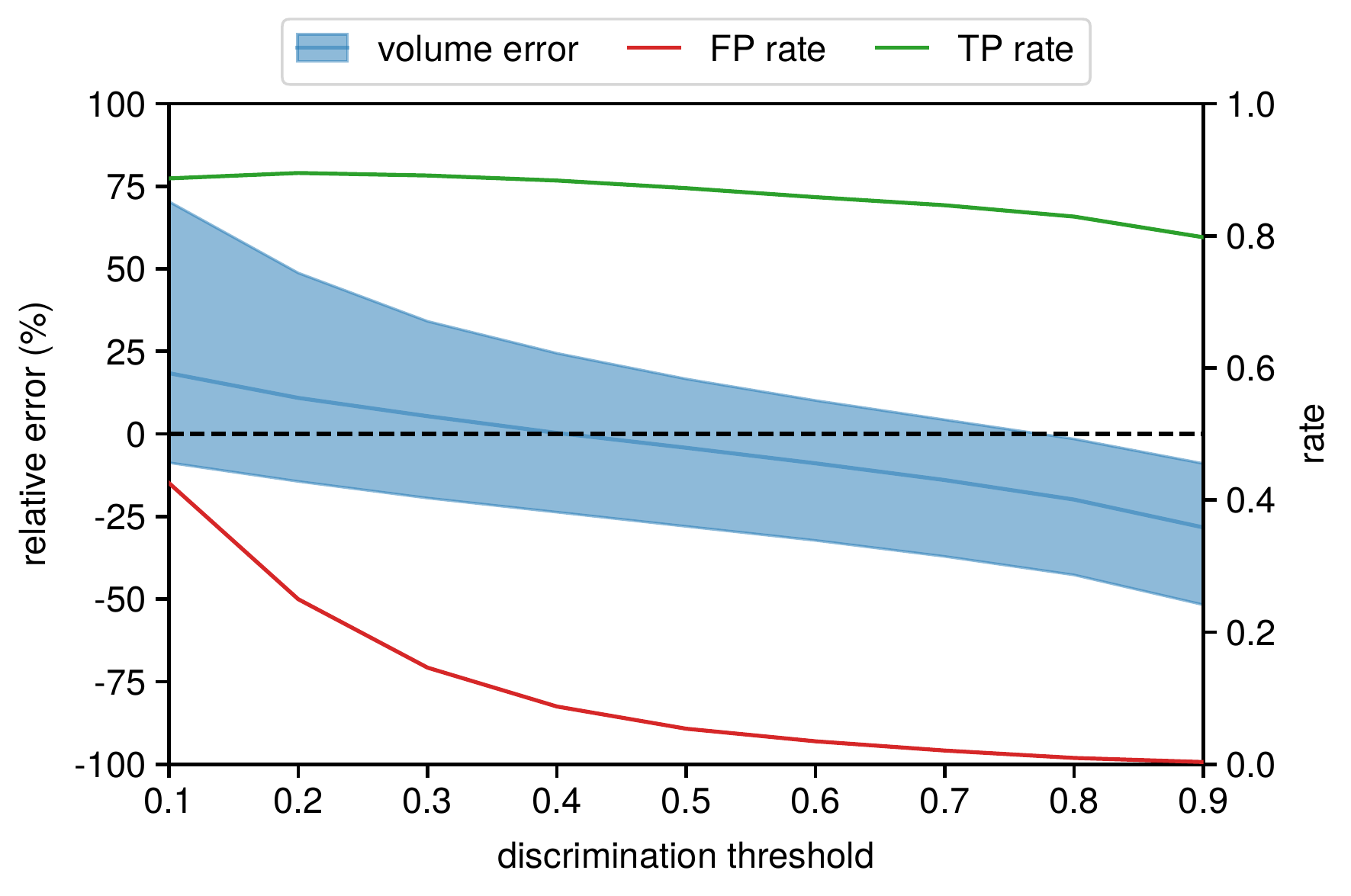}
    \vspace{-3mm}
    \caption{Relative cavity volume error (blue), false positive rate (red) and true positive rate (green) as a function of discrimination threshold. Relative errors and true positive rates were calculated for a sample of $10^4$ testing images with single pair of X-ray cavities and false positive rates were estimated from a sample of $10^3$ images without any X-ray cavities. For cavity errors, the blue line is the median value, while the blue area represents the 1$\sigma$ confidence interval. Meanwhile, area and volume error distributions centre around zero for a discrimination threshold of $\approx 0.4$, to obtain optimal FP and TP rates a higher discrimination threshold of $\approx 0.6$ is required.}
    \label{fig:discrimination_threshold}
\end{figure}

For the testing set of $10^4$ images with X-ray cavities, we, therefore, estimated the amount of true positive\footnote{Predictions were accounted as true positive if at least $20\%$ of the original cavity area has been recovered.} predictions (TP rate) for 9 different values of the discrimination threshold ($0.1-0.9$). For true-positive cavities, we further expressed their relative area and volume reconstruction errors and using interpolation, we found an optimal value of the discrimination threshold for which the distribution of relative volume errors is clustered around zero ($\approx 0.4$; see Figure~\ref{fig:discrimination_threshold}).

Nevertheless, setting a relatively low discrimination threshold that properly calibrates reconstructed volumes of true positive cavity detections ($\approx 0.4$) would, for many networks, result in accounting for plenty of low-significance predictions and thus in producing a lot of false positives (FP rate). For this reason, all trained networks were also applied on a set of $2000$ images without any X-ray cavities and the rate of false positive predictions was estimated for the same set of discrimination thresholds as used when estimating the true positive rate. An ideal discrimination threshold that optimizes both FP and TP rates was found ($\approx 0.6$) by simultaneously minimizing the false positive rate (maximally $5\%$) and maximizing the true positive rate (minimally $80\%$). 

By applying both thresholds consequently, volumes of detected X-ray cavities can be calibrated by the lower discrimination threshold, and low-significance and hopefully false positive predictions can subsequently be discarded by the higher discrimination threshold\footnote{In the case of the higher discrimination threshold, the cavity is taken as valid if the mean value is above this threshold.}.

\section{Cavity significance estimation}
\label{section:significance}

\begin{figure*}
\begin{tikzpicture}[remember picture]
    \draw (0, 4.2) node {\includegraphics[width=\textwidth]{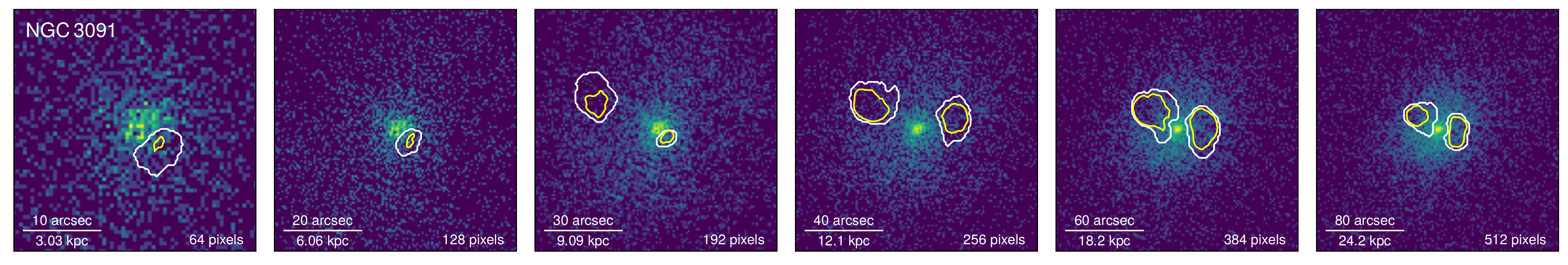}};
    \draw (0, -3.0) node {\includegraphics[width=0.95\textwidth]{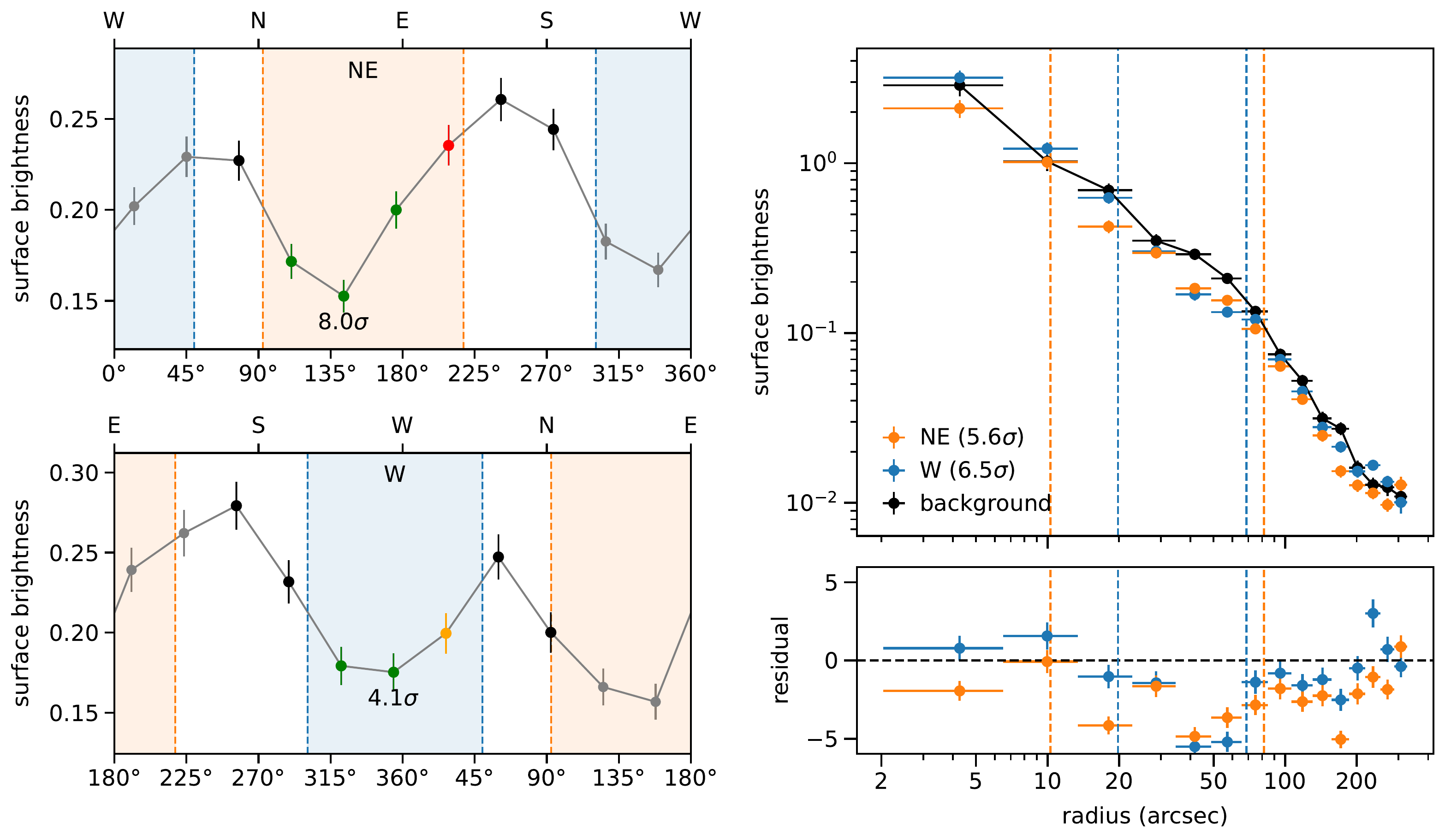}};
    \draw (6.24,2.87) -- (-7.15,1.37);
    \draw (8.825,2.85) -- (8.350,1.33);
\end{tikzpicture}
\caption{Cavity detection and significance estimation for NGC\,3091 in which we detected a new pair of X-ray cavities. \textbf{Upper}: Pixel-wise CADET predictions of newly detected X-ray cavities overlaid over X-ray images of NGC\,3091. Contours correspond to 0.4 (white) and 0.65 (yellow) levels. Individual subplots show images probed on different size scales, from left to right: 0.5 binning ($64 \times 64$ pixels), 1 binning ($128 \times 128$ pixels), 1.5 binning ($196 \times 196$ pixels), 2 binning ($256 \times 256$ pixels), 3 binning ($384 \times 384$ pixels) and 4 binning ($512 \times 512$ pixels). \textbf{Lower left}: azimuthal profiles of detected X-ray cavities. The number of azimuthal bins was chosen so that every cavity is divided into at least 3 bins and another 2 surrounding bins on both sides (black points) were used to estimate the level of background. The significance of individual cavity bins was calculated as in \protect\cite{Ubertosi2021} and is visualized by green (>$3\sigma$), yellow (>$1\sigma$), and red (<$1\sigma$) data points. The significance of the bin with the highest decrement is stated below the corresponding data point. Borders of X-ray cavities are marked by dashed lines and distinguished by colour to the W cavity (blue) and NE cavity (orange). \textbf{Lower right}: comparison of an azimuthally averaged radial profile excluding detected cavities, which served as a background for radial significance calculations, with radial profiles across detected X-ray cavities. Significances of individual radial bins were calculated similarly as in azimuthal profiles. We state the significance of the most significant radial bins for individual cavities in the legend. Borders of individual X-ray cavities are marked by blue (N cavity) and orange (S cavity) dashed lines.}
\label{fig:cavity_significance}
\end{figure*}

The reliability of detected X-ray cavities was tested using azimuthal and radial count statistics. To obtain the significance of cavity predictions, we derived azimuthal and radial surface brightness profiles from source-filled images and we expressed the significance of azimuthal and radial bins inside X-ray cavities with respect to the surrounding background \citep[see][]{Hlavacek2015,Ubertosi2021}. As valid cavity predictions, we only chose cavities with a depth lower than or equal to $3\sigma$ below the surrounding background (see Figure \ref{fig:cavity_significance}). We note, however, that such an approach is not robust for highly elliptical $\beta$-models or galaxies with strong sloshing patterns and shock fronts, and conversely, due to poor count statistics it can also result in overestimating the significance of random brightness drops. 

For these reasons, we have performed additional reliability testing based on simulating images of galaxies with similar properties as the given input images. For each galaxy, we generated $10000$ similar images by sampling the parameters of simulated $\beta$-models from the results of the beta-modelling analysis (parameter values were varied within estimated uncertainties). Into thus obtained galaxy models, we inserted cavities with the same sizes and properties as derived from CADET predictions. For each cavity, we then expressed the false positive rate (chance of detecting cavity at a similar distance with an area bigger than half of the area of the detected cavity), true positive rate (chance of recovering at least 20\% of the original cavity area), and average volume reconstruction error. We note, however, that the estimated false-positive rate represents rather a lower limit and for realistic images, the FP rate can be higher due to the much more complex structure of real galaxies, groups and clusters.

The cavity was marked as successfully confirmed when passing at least 3 out of 4 conditions: radial and azimuthal count statistics is for at least a single bin in each cavity 3$\sigma$ under the surrounding background, the false positive rate is lower than 5\% and the true positive rate is higher than 80\%.

\clearpage


\setlength{\tabcolsep}{2.2pt}
\renewcommand{\arraystretch}{1.6}
\begin{table*}
    \vspace{1mm}
    \begin{flushleft}
    {\large \bf APPENDIX D: PARAMETER DISTRIBUTIONS}
    \end{flushleft}
    \vspace{1mm}

    \centering
    \begin{flushleft}
    \justifying \textbf{Table D1:} 
    \hypertarget{tableD1}{Parameters}
    Parameters obtained from $\beta$-modelling the whole sample of galaxies: galaxy name, original size of the image region in \textit{Chandra} pixels (1 pixel ~ 0.492 arcsec), the total number of counts; first component parameters: core radius $r_{\text{c}}$, amplitude $A$, beta parameter $\beta$, ellipticity $e$; second component parameters: core radius $r_{\text{c},2}$, amplitude $A_2$, beta parameter $\beta_2$; and the background level $c_0$.
    \end{flushleft}
    \begin{tabular}{c c c c c c c c c c c}
\hline
Galaxy & Original size & Counts & \HP$r_{\text{c}}$\HP & \HP$A$\HP & \HP$\alpha$\HP & \HP$e$\HP & \HP$r_{\text{c},2}$\HP & \HP$A_2$\HP & \HP$\alpha_2$\HP & \HP$c0$\HP\\
\hline
3C449 & 768 & 46202 & $0.49_{-0.05}^{+0.04}$ & $453_{-53}^{+60}$ & $1\pm0.02$ & $0.18\pm0.01$ & $200\pm10$ & $0.208\pm0.005$ & $1\pm0.1$ & $0.049\pm0.006$\\
IC1262 & 512 & 118849 & $12.3_{-0.5}^{+0.6}$ & $4.8\pm0.2$ & $0.98\pm0.01$ & $0.109\pm0.005$ & $85.6\pm2.1$ & $1.77\pm0.03$ & $0.99\pm0.1$ & $0.1\pm0.01$\\
IC1459 & 512 & 18660 & $0.49\pm0.01$ & $4410_{-200}^{+250}$ & $1.36_{-0.006}^{+0.014}$ & $0.11\pm0.01$ & $58.6_{-3.7}^{+4.3}$ & $0.19\pm0.02$ & $1.4\pm0.1$ & $0.022\pm0.001$\\
IC1860 & 512 & 27841 & $5.9\pm0.4$ & $5.5\pm0.3$ & $0.65\pm0.01$ & $0.15\pm0.01$ & 0 & 0 & 0 & $0.013\pm0.002$\\
IC4296 & 512 & 9632 & $0.98\pm0.04$ & $294_{-17}^{+18}$ & $1.13\pm0.01$ & $0.18\pm0.02$ & 0 & 0 & 0 & $0.017\pm0.002$\\
IC4765 & 512 & 6090 & $4.1\pm0.6$ & $3.4_{-0.4}^{+0.5}$ & $0.94\pm0.05$ & 0 & $127.1_{-4.8}^{+4.7}$ & $0.24\pm0.02$ & $10\pm1$ & $0.01\pm0.001$\\
NGC57 & 128 & 823 & $4.2\pm0.5$ & $3.1\pm0.4$ & $1.06\pm0.04$ & $0.15\pm0.05$ & 0 & 0 & 0 & $0.12\pm0.01$\\
NGC193 & 512 & 29406 & $0.72_{-0.05}^{+0.04}$ & $280_{-21}^{+25}$ & $1.43\pm0.04$ & 0 & $274.6_{-5.7}^{+6}$ & $0.19\pm0.002$ & $1.4\pm0.1$ & $0.068\pm0.007$\\
NGC383 & 512 & 13471 & $0.35_{-0.06}^{+0.04}$ & $670_{-140}^{+180}$ & $1.16\pm0.04$ & 0 & $3.9\pm0.4$ & $11.1_{-1.4}^{+1.9}$ & $1.1\pm0.1$ & $0.037\pm0.001$\\
NGC315 & 512 & 16673 & $0.208_{-0.006}^{+0.009}$ & $6990_{-550}^{+220}$ & $0.99_{-0.004}^{+0.003}$ & $0.05\pm0.01$ & 0 & 0 & 0 & $0.01\pm0.001$\\
NGC499 & 640 & 34211 & $2.6_{-0.3}^{+0.4}$ & $6\pm0.9$ & $0.84\pm0.01$ & $0.087_{-0.01}^{+0.011}$ & $51_{-2}^{+2.3}$ & $0.7\pm0.03$ & $0.84\pm0.08$ & $0.13\pm0.01$\\
NGC507 & 512 & 56224 & $2.2\pm0.2$ & $31.4_{-2.5}^{+2.6}$ & $1.06\pm0.03$ & $0.02_{-0.01}^{+0.011}$ & $176.6\pm3.8$ & $0.432\pm0.005$ & $1.1\pm0.1$ & $0.088\pm0.009$\\
NGC533 & 512 & 21546 & $3.7\pm0.4$ & $13.8_{-0.9}^{+1.1}$ & $0.82\pm0.02$ & 0 & $44.1_{-6.4}^{+7}$ & $1.2\pm0.2$ & $2.6_{-0.4}^{+0.5}$ & $0.017_{-0.002}^{+0.001}$\\
NGC708 & 640 & 425100 & $18.5_{-0.3}^{+0.4}$ & $16.6\pm0.2$ & $0.67\pm0.006$ & $0.233\pm0.003$ & 0 & 0 & 0 & $0.234\pm0.007$\\
NGC720 & 640 & 35400 & $6.4\pm0.4$ & $6.0_{-0.3}^{+0.4}$ & $0.72\pm0.01$ & $0.12_{-0.01}^{+0.02}$ & 0 & 0 & 0 & $0.028\pm0.001$\\
NGC741 & 512 & 40443 & $2.2\pm0.1$ & $54\pm3$ & $0.81\pm0.01$ & $0.07\pm0.01$ & 0 & 0 & 0 & $0.063\pm0.001$\\
NGC777 & 512 & 3474 & $4.3_{-0.4}^{+0.3}$ & $3.4_{-0.3}^{+0.4}$ & $0.84\pm0.01$ & $0.11\pm0.03$ & 0 & 0 & 0 & $0.13\pm0.01$\\
NGC1132 & 512 & 21967 & $4.4\pm0.4$ & $9.6\pm0.4$ & $1.06_{-0.02}^{+0.03}$ & $0.11\pm0.02$ & $51.7_{-5.2}^{+5.6}$ & $0.17\pm0.01$ & $0.35_{-0.03}^{+0.04}$ & $0.13\pm0.01$\\
NGC1275 & 768 & 15323417 & $1.5\pm0.03$ & $3251_{-61}^{+63}$ & $1.65\pm0.02$ & $0.11\pm0.01$ & $193.8\pm0.4$ & $91.88\pm0.07$ & $1.268\pm0.002$ & $0.093\pm0.009$\\
NGC1316 & 512 & 58438 & $0.94_{-0.06}^{+0.07}$ & $164_{-10}^{+11}$ & $0.764\pm0.007$ & 0 & $52.1\pm0.8$ & $6.5\pm0.3$ & $10\pm1$ & $0.096\pm0.002$\\
NGC1380 & 512 & 7017 & $7\pm0.6$ & $2.6\pm0.2$ & $1.07\pm0.04$ & $0.1\pm0.04$ & 0 & 0 & 0 & $0.018\pm0.002$\\
NGC1387 & 512 & 5354 & $1.3_{-0.1}^{+0.2}$ & $7_{-0.9}^{+1}$ & $0.67\pm0.02$ & $0.22\pm0.03$ & 0 & 0 & 0 & $0.006\pm0.001$\\
NGC1399 & 640 & 328646 & $5.44\pm0.06$ & $190.5_{-2.3}^{+2.4}$ & $0.942\pm0.003$ & 0 & 0 & 0 & 0 & $0.336\pm0.002$\\
NGC1404 & 512 & 305846 & $4.88\pm0.05$ & $296\pm3.3$ & $0.902\pm0.002$ & 0 & 0 & 0 & 0 & $0.113\pm0.003$\\
NGC1407 & 512 & 23152 & $0.22_{-0.02}^{+0.03}$ & $246_{-34}^{+38}$ & $0.67\pm0.07$ & $0.1\pm0.01$ & $59.9\pm1.2$ & $0.81\pm0.04$ & $2.3\pm0.2$ & $0.013\pm0.001$\\
NGC1521 & 512 & 9164 & $5.3_{-0.5}^{+0.6}$ & $2.7\pm0.3$ & $0.82\pm0.02$ & $0.21\pm0.03$ & 0 & 0 & 0 & $0.018\pm0.001$\\
NGC1550 & 640 & 230601 & $3.9\pm0.3$ & $31.4_{-2}^{+1.9}$ & $1.07\pm0.07$ & $0.235\pm0.004$ & $16.4\pm0.5$ & $10.3\pm0.5$ & $0.616\pm0.003$ & $0.059\pm0.006$\\
NGC1553 & 640 & 5878 & $0.46\pm0.06$ & $408_{-68}^{+89}$ & $1.37_{-0.05}^{+0.06}$ & $0.16\pm0.03$ & $8.7\pm0.9$ & $0.151\pm0.002$ & $0.4\pm0.04$ & $0.037\pm0.004$\\
NGC1600 & 512 & 50980 & $18.5_{-1.4}^{+1.6}$ & $9.9_{-0.3}^{+0.4}$ & $2.3\pm0.2$ & $0.13_{-0.01}^{+0.02}$ & $52.2\pm5.2$ & $0.27\pm0.02$ & $0.5\pm0.1$ & $0.08_{-0.03}^{+0.01}$\\
NGC1700 & 512 & 6361 & $1.6\pm0.2$ & $13.4\pm1.6$ & $0.87\pm0.02$ & $0.27\pm0.03$ & 0 & 0 & 0 & $0.015_{-0.001}^{+0.002}$\\
NGC2300 & 512 & 29865 & $8.8\pm0.4$ & $6.6\pm0.3$ & $0.92\pm0.02$ & $0.06\pm0.02$ & 0 & 0 & 0 & $0.057\pm0.001$\\
NGC2305 & 512 & 3756 & $2.1\pm0.2$ & $6\pm0.8$ & $0.78\pm0.02$ & $0.08_{-0.04}^{+0.03}$ & 0 & 0 & 0 & $0.003\pm0.0003$\\
NGC2563 & 512 & 6910 & $2.7\pm0.3$ & $6.8\pm0.8$ & $0.87\pm0.02$ & $0.1\pm0.04$ & 0 & 0 & 0 & $0.016\pm0.002$\\
NGC3091 & 512 & 10206 & $3.7_{-0.4}^{+0.5}$ & $5.7\pm0.6$ & $0.87_{-0.04}^{+0.05}$ & 0 & $160.6_{-3.7}^{+3.8}$ & $0.2\pm0.01$ & $4.8\pm0.5$ & $0.008\pm0.001$\\
NGC3402 & 768 & 49730 & $5.7\pm0.2$ & $16.2\pm0.6$ & $0.755\pm0.003$ & $0.061\pm0.009$ & 0 & 0 & 0 & $0.06\pm0.006$\\
NGC3923 & 512 & 27664 & $6.1\pm0.2$ & $25.1\pm0.8$ & $1.09\pm0.01$ & $0.16\pm0.01$ & 0 & 0 & 0 & $0.044\pm0.001$\\
\hline
    \end{tabular}
    \label{tab:beta_params}
\end{table*}

\setlength{\tabcolsep}{2.6pt}
\renewcommand{\arraystretch}{1.6}
\begin{table*}
    \begin{flushleft}
    \textbf{Table D1.} Continued.
    \end{flushleft}

    \centering
    \begin{tabular}{c c c c c c c c c c c}
\hline
Galaxy & Original size & Counts & \HP$r_{\text{c}}$\HP & \HP$A$\HP & \HP$\alpha$\HP & \HP$e$\HP & \HP$r_{\text{c},2}$\HP & \HP$A_2$\HP & \HP$\alpha_2$\HP & \HP$c0$\HP\\
\hline
NGC4073 & 768 & 59275 & $11.4_{-0.3}^{+0.4}$ & $7.3\pm0.2$ & $0.796_{-0.007}^{+0.008}$ & $0.173_{-0.008}^{+0.007}$ & 0 & 0 & 0 & $0.016\pm0.001$\\
NGC4104 & 512 & 13991 & $4.7\pm0.2$ & $14.1\pm0.7$ & $1.11\pm0.02$ & $0.25\pm0.02$ & 0 & 0 & 0 & $0.033\pm0.001$\\
NGC4125 & 512 & 14521 & $1.1\pm0.1$ & $16.8_{-1.6}^{+1.9}$ & $0.595_{-0.008}^{+0.01}$ & $0.21\pm0.02$ & 0 & 0 & 0 & $0.002\pm0.001$\\
NGC4261 & 512 & 50099 & $1.19\pm0.03$ & $756_{-24}^{+23}$ & $1.038\pm0.004$ & 0 & 0 & 0 & 0 & $0.076\pm0.001$\\
NGC4291 & 512 & 7468 & $0.54_{-0.07}^{+0.08}$ & $8.2\pm1.1$ & $0.505\pm0.006$ & $0.21_{-0.03}^{+0.02}$ & 0 & 0 & 0 & $0.12\pm0.01$\\
NGC4325 & 768 & 42117 & $6.6_{-1.1}^{+1.3}$ & $1.7_{-0.3}^{+0.4}$ & $1.41\pm0.02$ & $0.155_{-0.008}^{+0.007}$ & $48.2\pm1.3$ & $2.3\pm0.06$ & $1.4\pm0.1$ & $0.013\pm0.001$\\
NGC4342 & 512 & 13071 & $0.2\pm0.01$ & $677\pm68$ & $1.01\pm0.03$ & $0.18\pm0.04$ & $59.5_{-4.3}^{+4.8}$ & $0.11\pm0.01$ & $1\pm0.1$ & $0.032\pm0.001$\\
NGC4374 & 512 & 212811 & $0.26\pm0.02$ & $3790_{-350}^{+420}$ & $0.774\pm0.004$ & 0 & $92.1\pm1$ & $3.3\pm0.1$ & $3.4\pm0.3$ & $0.354_{-0.005}^{+0.004}$\\
NGC4382 & 512 & 7780 & $0.14\pm0.02$ & $26.5_{-3.8}^{+3}$ & $0.5\pm0.01$ & $0.07\pm0.04$ & 0 & 0 & 0 & $0.006\pm0.001$\\
NGC4406 & 512 & 514275 & $4.1_{-0.2}^{+0.3}$ & $35.9_{-1.7}^{+1.9}$ & $0.76\pm0.04$ & $0.315\pm0.003$ & $135\pm7.1$ & $3.8_{-0.1}^{+0.2}$ & $1.8\pm0.1$ & $1\pm0.02$\\
NGC4472 & 640 & 734419 & $8.6\pm0.1$ & $175.8_{-1.8}^{+1.9}$ & $0.944\pm0.007$ & 0 & $331_{-20}^{+21}$ & $1.42_{-0.03}^{+0.04}$ & $1.17_{-0.08}^{+0.09}$ & $0.11\pm0.01$\\
NGC4477 & 512 & 19800 & $1\pm0.1$ & $30.6_{-4.6}^{+5.1}$ & $1.14_{-0.06}^{+0.07}$ & $0.14\pm0.03$ & $33.3_{-2.2}^{+2.3}$ & $0.51\pm0.03$ & $1.1\pm0.1$ & $0.052\pm0.001$\\
NGC4486 & 640 & 3605386 & $0.4_{-0.03}^{+0.02}$ & $1247_{-83}^{+96}$ & $0.731\pm0.004$ & $0.168\pm0.001$ & $40.1\pm0.3$ & $62.4\pm0.3$ & $0.73\pm0.07$ & $2.14\pm0.03$\\
NGC4526 & 512 & 4955 & $1.1\pm0.1$ & $39_{-4.9}^{+6.2}$ & $1.7\pm0.1$ & $0.22\pm0.03$ & $20.1\pm1.7$ & $0.78\pm0.06$ & $1.7\pm0.2$ & $0.013\pm0.001$\\
NGC4552 & 512 & 61952 & $3.4\pm0.1$ & $66.2\pm3.2$ & $1.62\pm0.02$ & 0 & $22.6\pm0.5$ & $14\pm0.3$ & $1.6\pm0.2$ & $0.089\pm0.001$\\
NGC4555 & 512 & 4958 & $1.8\pm0.2$ & $16.6_{-1.7}^{+2}$ & $0.98\pm0.02$ & $0.05_{-0.03}^{+0.04}$ & 0 & 0 & 0 & $0.012\pm0.001$\\
NGC4636 & 640 & 299549 & $71.2\pm0.4$ & $21\pm0.2$ & $10\pm1$ & 0 & $393.2\pm1.2$ & $3.19\pm0.02$ & $10\pm1$ & $0.222\pm0.001$\\
NGC4649 & 512 & 281760 & $9.21\pm0.08$ & $152.5\pm1.4$ & $1.033\pm0.003$ & $0.056\pm0.003$ & 0 & 0 & 0 & $0.123\pm0.003$\\
NGC4696 & 768 & 6853500 & $13.1_{-0.08}^{+0.07}$ & $278.8\pm1.3$ & $0.58\pm0.06$ & $0.132\pm0.001$ & $153.3\pm0.6$ & $42.3\pm0.2$ & $10\pm1$ & $0.12\pm0.01$\\
NGC4778 & 512 & 148213 & $11.5\pm0.4$ & $18.6\pm0.4$ & $0.83_{-0.01}^{+0.02}$ & $0.043\pm0.005$ & $242.9_{-2.9}^{+3.2}$ & $1.75\pm0.05$ & $10\pm1$ & $0.088_{-0.006}^{+0.007}$\\
NGC4936 & 512 & 4330 & $0.16_{-0.03}^{+0.04}$ & $34_{-7.3}^{+11.4}$ & $0.61\pm0.02$ & $0.15\pm0.05$ & 0 & 0 & 0 & $0.008\pm0.001$\\
NGC5044 & 512 & 1314389 & $14.4\pm0.5$ & $31.4\pm0.7$ & $0.81\pm0.01$ & 0 & $146.1\pm0.5$ & $13.17\pm0.06$ & $1.3\pm0.1$ & $0.096\pm0.01$\\
NGC5129 & 512 & 33361 & $1.06\pm0.08$ & $80_{-6.5}^{+7.5}$ & $0.87\pm0.02$ & 0 & $240_{-29}^{+33}$ & $0.136_{-0.008}^{+0.011}$ & $0.86\pm0.09$ & $0.01_{-0.02}^{+0.01}$\\
NGC5171 & 512 & 6678 & $3.1_{-0.7}^{+0.8}$ & $1_{-0.2}^{+0.3}$ & $0.8_{-0.06}^{+0.07}$ & $0.22_{-0.09}^{+0.08}$ & 0 & 0 & 0 & $0.023\pm0.001$\\
NGC5813 & 512 & 953760 & $35.5_{-1.5}^{+1.6}$ & $47.3\pm0.6$ & $2.9\pm0.2$ & $0.131\pm0.002$ & $415.9_{-1}^{+1.1}$ & $10.88_{-0.06}^{+0.05}$ & $10\pm1$ & $0.852\pm0.006$\\
NGC5846 & 512 & 88165 & $11.6\pm0.4$ & $8.9\pm0.2$ & $0.665\pm0.008$ & $0.209_{-0.006}^{+0.007}$ & 0 & 0 & 0 & $0.01\pm0.005$\\
NGC6107 & 512 & 5168 & $1.6_{-0.3}^{+0.2}$ & $3.3\pm0.5$ & $0.62\pm0.02$ & $0.06_{-0.04}^{+0.05}$ & 0 & 0 & 0 & $0.007\pm0.001$\\
NGC6166 & 512 & 954398 & $18.8_{-0.8}^{+1}$ & $26.9_{-0.6}^{+0.5}$ & $1.12_{-0.05}^{+0.06}$ & 0 & $334_{-15}^{+13}$ & $6.5\pm0.1$ & $4.4_{-0.4}^{+0.3}$ & $1.03\pm0.02$\\
NGC6338 & 512 & 33798 & $3.7\pm0.3$ & $15.6_{-1.4}^{+1.5}$ & $1.58_{-0.05}^{+0.06}$ & $0.18\pm0.01$ & $29.1_{-1.2}^{+1.3}$ & $3.7\pm0.1$ & $1.6\pm0.2$ & $0.064\pm0.001$\\
NGC6482 & 512 & 9424 & $4.4\pm0.3$ & $9.7_{-0.6}^{+0.7}$ & $0.87\pm0.01$ & $0.02_{-0.01}^{+0.02}$ & 0 & 0 & 0 & $0.002\pm0.001$\\
NGC6868 & 640 & 20924 & $0.066_{-0.007}^{+0.008}$ & $1720_{-250}^{+300}$ & $0.748_{-0.007}^{+0.008}$ & $0.16\pm0.03$ & 0 & 0 & 0 & $0.034\pm0.001$\\
NGC7618 & 512 & 34814 & $2.8_{-0.2}^{+0.3}$ & $33.4_{-3.1}^{+3.3}$ & $1.35_{-0.05}^{+0.06}$ & $0.06\pm0.01$ & $48.7_{-2.2}^{+2.4}$ & $1.32\pm0.04$ & $1.4\pm0.1$ & $0.047\pm0.002$\\
NGC7619 & 640 & 22579 & $2.6\pm0.1$ & $21.5\pm1.4$ & $0.793\pm0.008$ & $0.11\pm0.02$ & 0 & 0 & 0 & $0.02\pm0.001$\\
NGC7796 & 512 & 12166 & $1.12\pm0.1$ & $24.1\pm2.6$ & $0.76\pm0.01$ & 0 & 0 & 0 & 0 & $0.026\pm0.001$\\
\hline
    \end{tabular}
    \label{tab:beta_params_2}
\end{table*}

\clearpage

\begin{figure*}
    \centering
    \vspace{40mm}
    \includegraphics[scale=0.43]{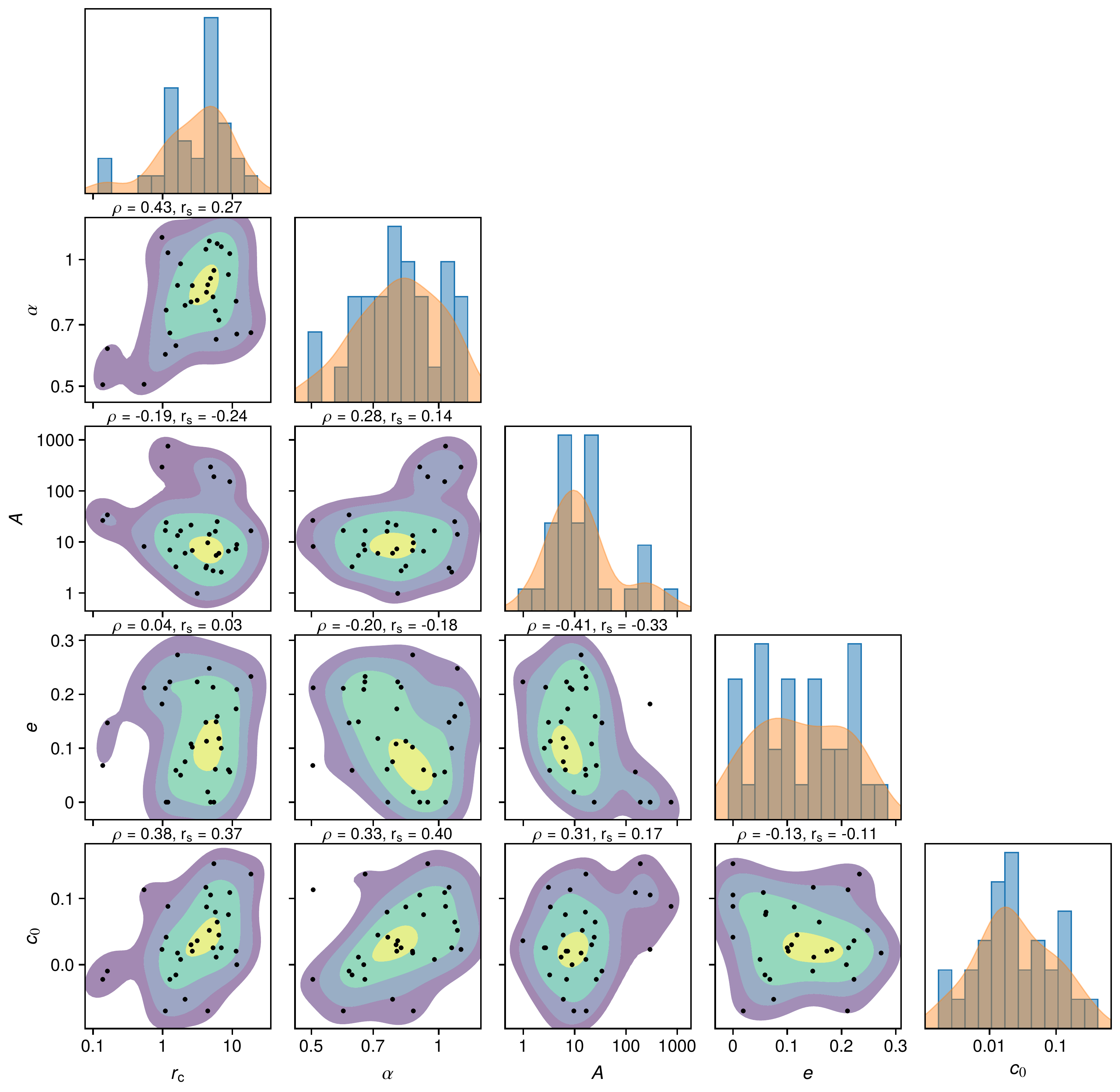}
    \begin{flushleft}
    \justifying \textbf{Figure D1:} \hypertarget{figD1}{Corner} plot showing distributions of $\beta$-model parameters of galaxies fitted by single $\beta$-model: core radius $r_{\text{c}}$, alpha parameter $\alpha$, amplitude $A$, ellipticity $e$ and level of background $c_0$. Contours in parameter pair plots correspond to $90\,\%$, $75\,\%$, $50\,\%$ and $10\,\%$ intervals in terms of cumulative probability of Gaussian KDE. Pearson and Spearman correlation coefficients are stated above individual plots of parameter pairs.
    \end{flushleft}
\end{figure*}

\begin{figure*}
    \centering
    \includegraphics[scale=0.43]{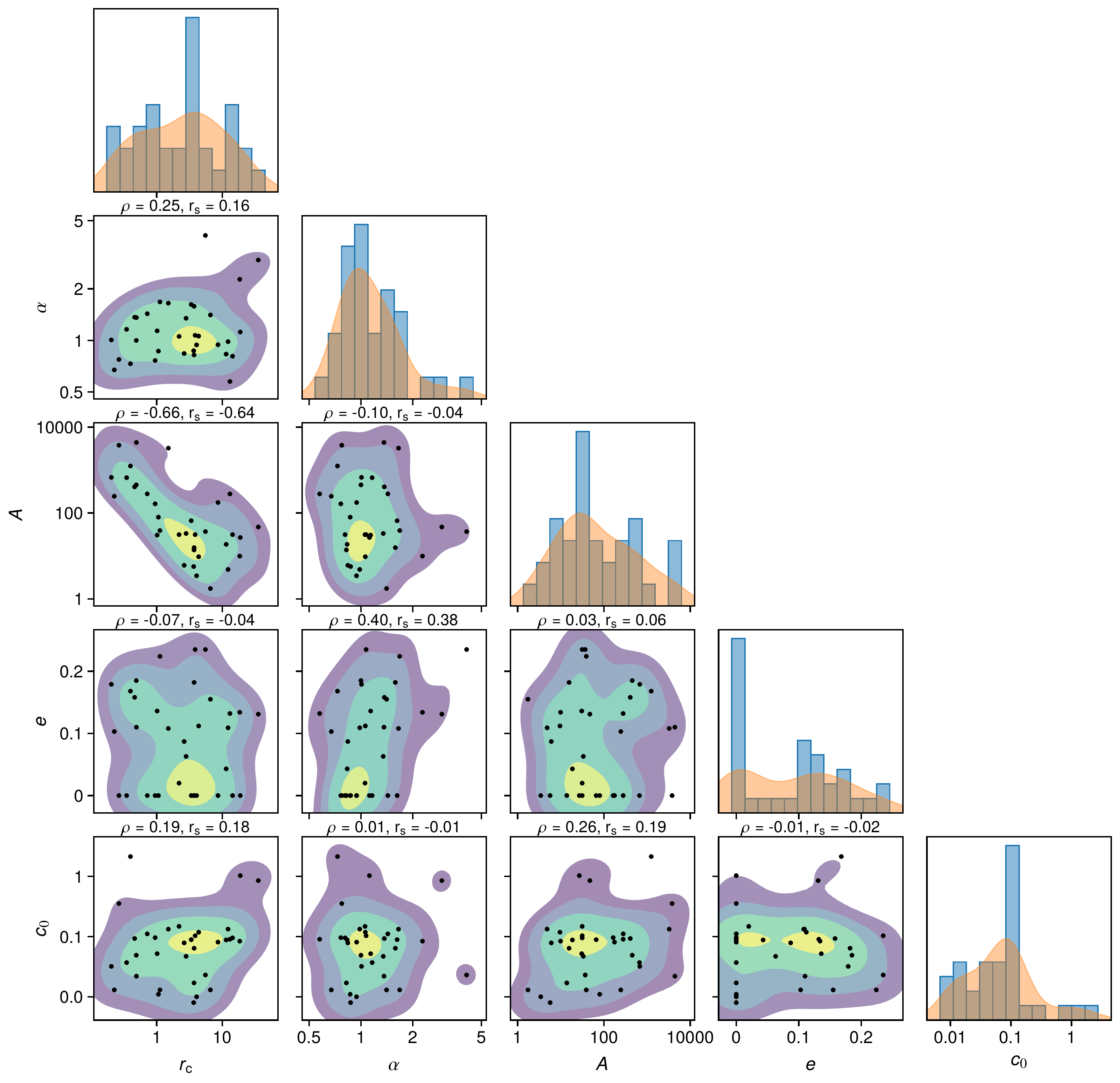}
    \vspace{4mm}
    \includegraphics[scale=0.43]{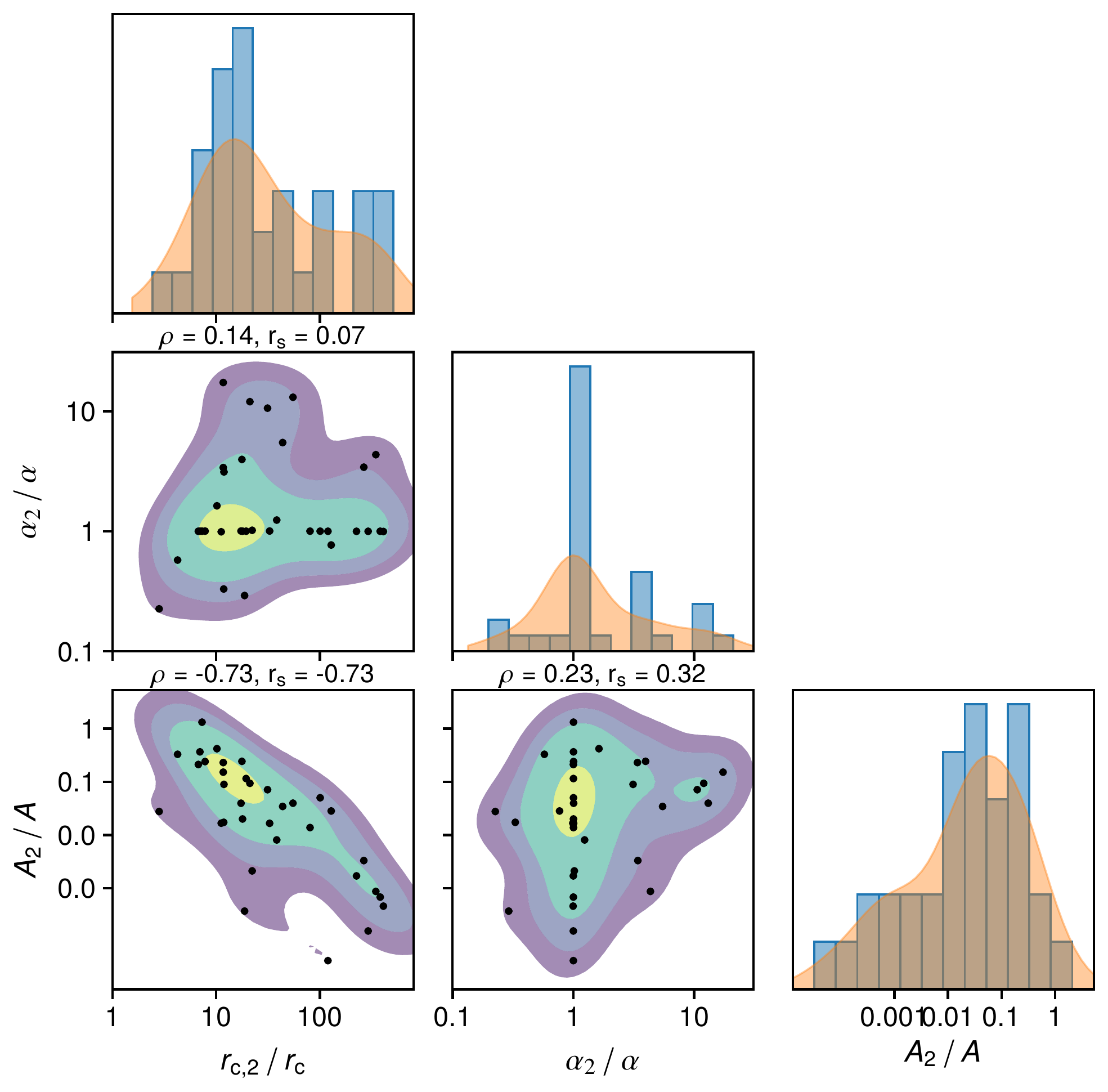}
    \vspace{-4mm}
    \begin{flushleft}
    \justifying \textbf{Figure D2:} Upper: \hypertarget{figD2}{corner} plot showing distributions of the first $\beta$-component parameters for galaxies fitted by double $\beta$-model: core radius $r_{\text{c}}$, alpha parameter $\alpha$, amplitude $A$, ellipticity $e$ and level of background $c_0$. Lower: corner plot showing distributions of ratios of core radii $r_{\text{c},2} / r_{\text{c}}$, alpha parameters $\alpha_2 / \alpha$ and amplitudes $A_2 / A$ between the second and the first $\beta$-component. Contours in parameter pair plots correspond to $90\,\%$, $75\,\%$, $50\,\%$ and $10\,\%$ intervals in terms of cumulative probability of Gaussian KDE. Pearson and Spearman correlation coefficients are stated above individual plots of parameter pairs.
    \end{flushleft}
\end{figure*}

\clearpage

\setlength{\tabcolsep}{5.0pt}
\renewcommand{\arraystretch}{1.046}
\tablefirsthead{\multicolumn{5}{l}{\small \hspace{-4mm} \textbf{Table D2.} \hypertarget{tableD2}{Parameters} of visually estimated X-ray cavities.\vspace{2mm}}\\
\toprule Galaxy & Cavity & $R_{\text{W}}\,(")$ & $R_{\text{L}}\,(")$ & $R\,(")$ & $\phi\,(^{\circ})$  & $\varphi\,(^{\circ})$\\ \toprule}
\tablehead{\multicolumn{2}{l}{{\small \hspace{-4mm} \textbf{Table D2.} Continued. \vspace{2mm}}} \\ \toprule
\tablelasttail{\bottomrule}
Galaxy & Cavity & $R_{\text{W}}\,(")$ & $R_{\text{L}}\,(")$ & $d\,(")$ & $\phi\,(^{\circ})$  & $\varphi\,(^{\circ})$\\ \toprule}
\label{tab:cavities}
\begin{supertabular}{c c c c c c c}
3C\,449 & E1 & 24.1 & 32.0 & 96.5 & 282 & 16\\
  & N1 & 24.8 & 32.5 & 114.7 & 89 & 14\\
IC\,1262 & NE1 & 5.45 & 10.7 & 15.3 & 126 & 27.8\\
  & E1 & 4.9 & 9.24 & 19.8 & 199 & 40.5\\
IC\,1459 & SE1 & 1.49 & 2.26 & 5.13 & 230 & 33.5\\
  & N1 & 1.66 & 2.61 & 3.67 & 80.9 & 4.36\\
IC\,4296 & NW1 & 2.74 & 2.12 & 4.52 & 43.2 & 1.06\\
  & SE1 & 2.96 & 1.83 & 4.47 & 234 & 70.9\\
NGC\,193 & NW1 & 24.8 & 35.7 & 36.2 & 41.9 & 83.8\\
  & S1 & 24.4 & 46.2 & 31.1 & 282 & 16.4\\
NGC\,499 & S1 & 20.1 & 27.5 & 48.3 & 251 & 39.2\\
  & N1 & 16.6 & 28.7 & 56.6 & 98.4 & 50\\
NGC\,507 & N1 & 11.7 & 18.2 & 21.4 & 108 & 24.1\\
  & SW1 & 14.6 & 19.7 & 24.5 & 315 & 13.9\\
  & NW2 & 17.4 & 26.9 & 56.9 & 37.2 & 35.4\\
  & SE2 & 35.9 & 26.3 & 56 & 234 & 35.6\\
NGC\,533 & E1 & 1.28 & 1.44 & 2.08 & 158 & 60.5\\
  & SW1 & 1.85 & 2.27 & 3.42 & 333 & 50.6\\
NGC\,708 & E1 & 10.9 & 7.81 & 13.3 & 174 & 30.7\\
  & W1 & 10.6 & 8.83 & 17.8 & 7.76 & 35.8\\
  & E2 & 25.4 & 29.8 & 72.8 & 172 & 84\\
  & W2 & 27.6 & 35 & 92.4 & 339 & 87.4\\
NGC\,1132 & N1 & 1.71 & 1.35 & 2.94 & 89.9 & 89.9\\
  & SW1 & 1.44 & 1.49 & 4.06 & 333 & 63.3\\
NGC\,1275 & SW1 & 19.3 & 19.9 & 26.4 & 299 & 18.6\\
  & N1 & 16.8 & 15.8 & 17.1 & 102 & 69.9\\
  & NW2 & 20.5 & 38.8 & 74.7 & 33.2 & 81.4\\
  & S2 & 29.3 & 37.5 & 95.7 & 262 & 6.28\\
NGC\,1316 & SE1 & 5.86 & 4.99 & 6.66 & 224 & 16.3\\
  & NW1 & 5.61 & 4.24 & 6.78 & 44.6 & 40.7\\
NGC\,1399 & N1 & 4.5 & 2.25 & 6.06 & 81 & 3.82\\
  & S1 & 4.58 & 2.42 & 6.3 & 265 & 1.89\\
  & N2 & 33.5 & 11.8 & 71 & 86.2 & 5\\
  & S2 & 37.8 & 12.7 & 77.2 & 262 & 0.599\\
NGC\,1407 & NE1 & 7.55 & 17.4 & 13.8 & 126 & 80\\
  & W1 & 7.44 & 17.9 & 22 & 349 & 82.9\\
NGC\,1553 & N1 & 26.5 & 49.7 & 43.6 & 70.7 & 10.9\\
  & SE1 & 19.2 & 44.8 & 37.9 & 236 & 11.6\\
NGC\,1600 & N1 & 2.23 & 3.67 & 2.72 & 68.2 & 85\\
  & S1 & 3.01 & 2.23 & 3.99 & 286 & 35.4\\
NGC\,4261 & E1 & 1.62 & 1.16 & 1.92 & 159 & 6.64\\
  & W1 & 2.19 & 1.5 & 2.74 & 348 & 4.21\\
NGC\,4325 & W1 & 8.78 & 13 & 15.7 & 355 & 55.9\\
  & SE1 & 15.7 & 12.7 & 16.9 & 206 & 25.8\\
NGC\,4374 & NE1 & 21.3 & 17.4 & 24.6 & 125 & 2.26\\
  & S1 & 30.4 & 22 & 59 & 280 & 31.3\\
NGC\,4472 & N1 & 2.58 & 5.47 & 7.72 & 111 & 15.6\\
  & W1 & 4.43 & 2.51 & 8.11 & 342 & 3.22\\
  & NW2 & 14.3 & 29.5 & 43.7 & 25.8 & 80.4\\
  & E2 & 28.3 & 15 & 47.3 & 159 & 31.1\\
\hline \\
NGC\,4477 & SE1 & 7.61 & 10.9 & 13.4 & 210 & 78.8\\
  & NW1 & 6.03 & 9.12 & 10.6 & 25.3 & 74.5\\
NGC\,4486 & E1 & 4.64 & 7.08 & 5.49 & 193 & 86.6\\
  & W1 & 6.84 & 3.86 & 14.5 & 357 & 31.4\\
  & E2 & 14.6 & 13.8 & 15.6 & 196 & 2.63\\
  & W2 & 16.3 & 8.39 & 25.3 & 7.27 & 20\\
NGC\,4552 & N1 & 3.67 & 7.42 & 8.38 & 95.9 & 82.1\\
  & SW1 & 4.44 & 9.37 & 6.67 & 298 & 73\\
  & E2 & 22.4 & 28.1 & 33.8 & 186 & 53.8\\
  & W2 & 36.5 & 19.5 & 38.8 & 350 & 57\\
NGC\,4636 & E1 & 1.69 & 2.78 & 2.11 & 190 & 89.9\\
  & NW1 & 1.79 & 3.56 & 5.49 & 43.3 & 87.9\\
  & NE2 & 4.68 & 5.38 & 9.22 & 117 & 78.2\\
  & SW2 & 4.53 & 6.84 & 14.6 & 329 & 81.8\\
  & E3 & 27.3 & 18.5 & 40.3 & 198 & 9.22\\
  & W3 & 20.1 & 27.9 & 56.5 & 10.8 & 41.6\\
  & NE4 & 38.1 & 23 & 90.8 & 156 & 20.4\\
  & SW4 & 39.8 & 23.2 & 87.2 & 332 & 9.38\\
NGC\,4649 & N1 & 4.67 & 6.01 & 12.5 & 98.9 & 2.27\\
  & S1 & 6.45 & 5.03 & 12.7 & 287 & 78.5\\
NGC\,4696 & SE1 & 11.3 & 7.62 & 10 & 230 & 47.1\\
  & W1 & 7.36 & 12.3 & 18.1 & 351 & 66.1\\
  & NE2 & 8.68 & 15.8 & 24 & 132 & 13.6\\
  & SW2 & 13.7 & 11.8 & 48 & 296 & 82\\
NGC\,4778 & N1 & 1.39 & 2.33 & 7.19 & 79.2 & 21.5\\
  & SE1 & 2.16 & 1.1 & 4.39 & 230 & 89.2\\
  & S2 & 9.63 & 13 & 19.6 & 288 & 86.5\\
  & NE2 & 18.6 & 11.9 & 35.2 & 127 & 39.3\\
NGC\,5044 & SE1 & 1.5 & 1.9 & 2.22 & 243 & 20.3\\
  & NW1 & 1.27 & 2.71 & 3.73 & 26.9 & 62.5\\
  & NE2 & 9.73 & 3.58 & 20 & 124 & 28.9\\
  & SW2 & 9.74 & 4.95 & 19.1 & 321 & 18.7\\
NGC\,5044 & S3 & 20.8 & 15.9 & 27.2 & 248 & 42\\
  & NW3 & 15.7 & 18.6 & 25.9 & 53.7 & 88.7\\
  & E4 & 19.6 & 35.9 & 46.5 & 179 & 77.4\\
  & W4 & 18.4 & 22.9 & 55.3 & 347 & 88.9\\
NGC\,5813 & N1 & 5.79 & 3.99 & 7.38 & 98.8 & 19.9\\
  & SW1 & 4.42 & 5.78 & 7.1 & 307 & 86.2\\
  & NE2 & 15.8 & 10.9 & 29.7 & 127 & 20.6\\
  & SW2 & 25.9 & 24.9 & 44.8 & 322 & 6.23\\
NGC\,5813 & NE3 & 21.4 & 50.7 & 121 & 133 & 72.7\\
  & SW3 & 22.1 & 47.4 & 154 & 310 & 72.2\\
NGC\,5846 & NE1 & 3.73 & 6.35 & 5.38 & 138 & 82.3\\
  & SW1 & 3.74 & 5.29 & 7.53 & 304 & 88.8\\
  & SE2 & 20 & 14.8 & 47.1 & 214 & 49.9\\
  & W2 & 16.1 & 17.8 & 39.4 & 353 & 43.2\\
NGC\,6166 & W1 & 3.55 & 2 & 4.6 & 10.2 & 3.5\\
  & NE1 & 2.19 & 3.13 & 4.81 & 126 & 89.9\\
  & E2 & 20.5 & 13.2 & 33.9 & 182 & 16.2\\
  & W2 & 17.2 & 14 & 38.3 & 358 & 17.7\\
NGC\,6338 & SW1 & 7.15 & 6.1 & 9.6 & 318 & 77\\
  & NE1 & 11.7 & 6.89 & 12.8 & 157 & 9.21\\
\end{supertabular}

\begin{figure*}
    \centering
    \vspace{40mm}
    \includegraphics[scale=0.45]{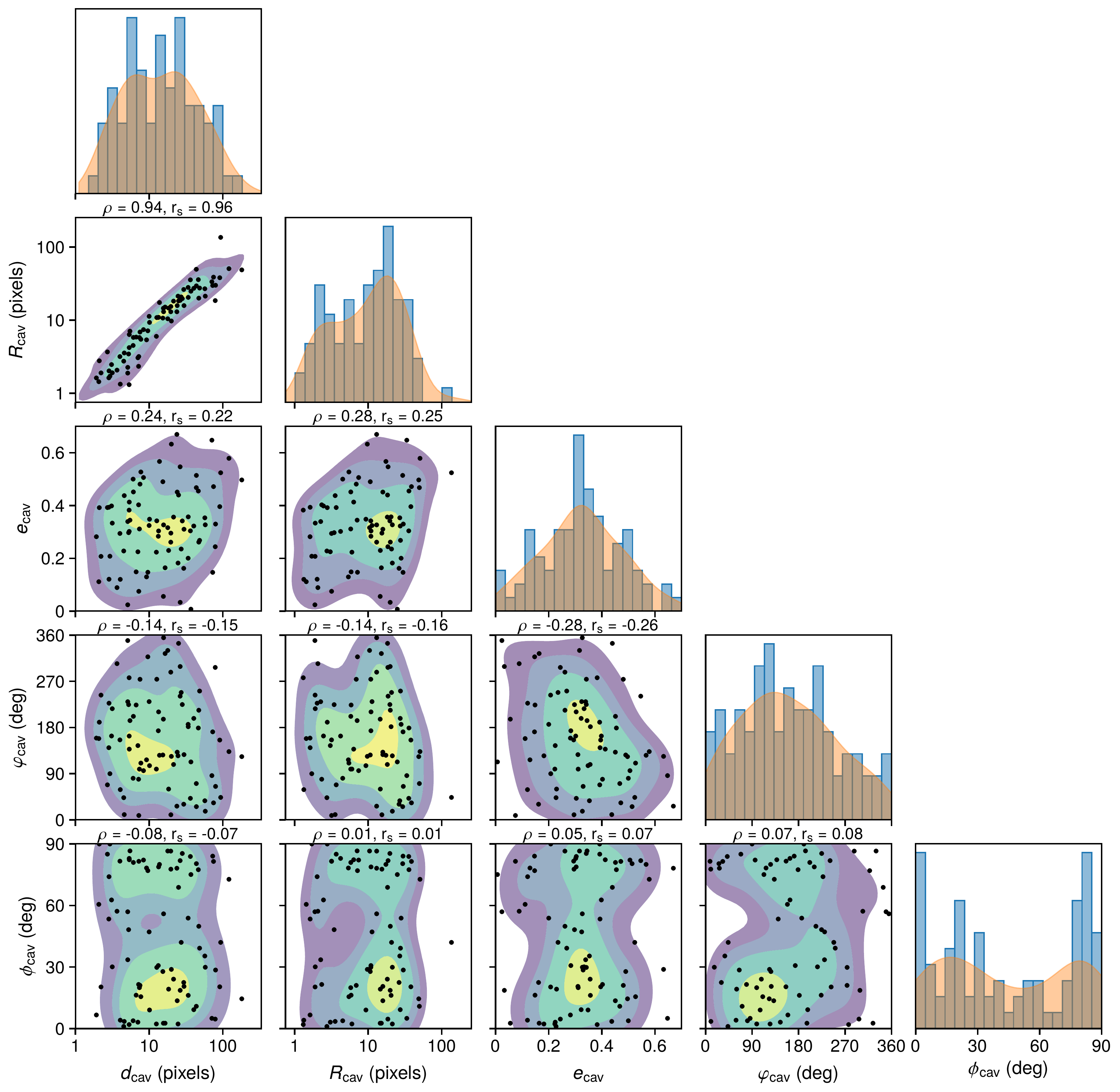}
    \begin{flushleft}
    \justifying \textbf{Figure D3:} \hypertarget{figD3}{Corner} plot showing distributions and correlations between primary cavity parameters: distance from the centre of the galaxy $d_{\text{cav}}$, semi-major axis $R_{\text{cav}}$, ellipticity~$e_{\text{cav}}$ and rotational angle $\phi_{\text{cav}}$ (angle between semi-major axis and the direction towards the centre of galaxy). Contours in parameter pair plots correspond to $90\,\%$, $75\,\%$, $50\,\%$ and $10\,\%$ intervals in terms of cumulative probability of Gaussian KDE. Pearson and Spearman correlation coefficients are stated above individual plots of parameter pairs.
    \end{flushleft}
\end{figure*}

\begin{figure*}
    \vspace{190mm}
    \begin{tikzpicture}[overlay]
    \draw (0.05, 7.8) node {\includegraphics[scale=0.45]{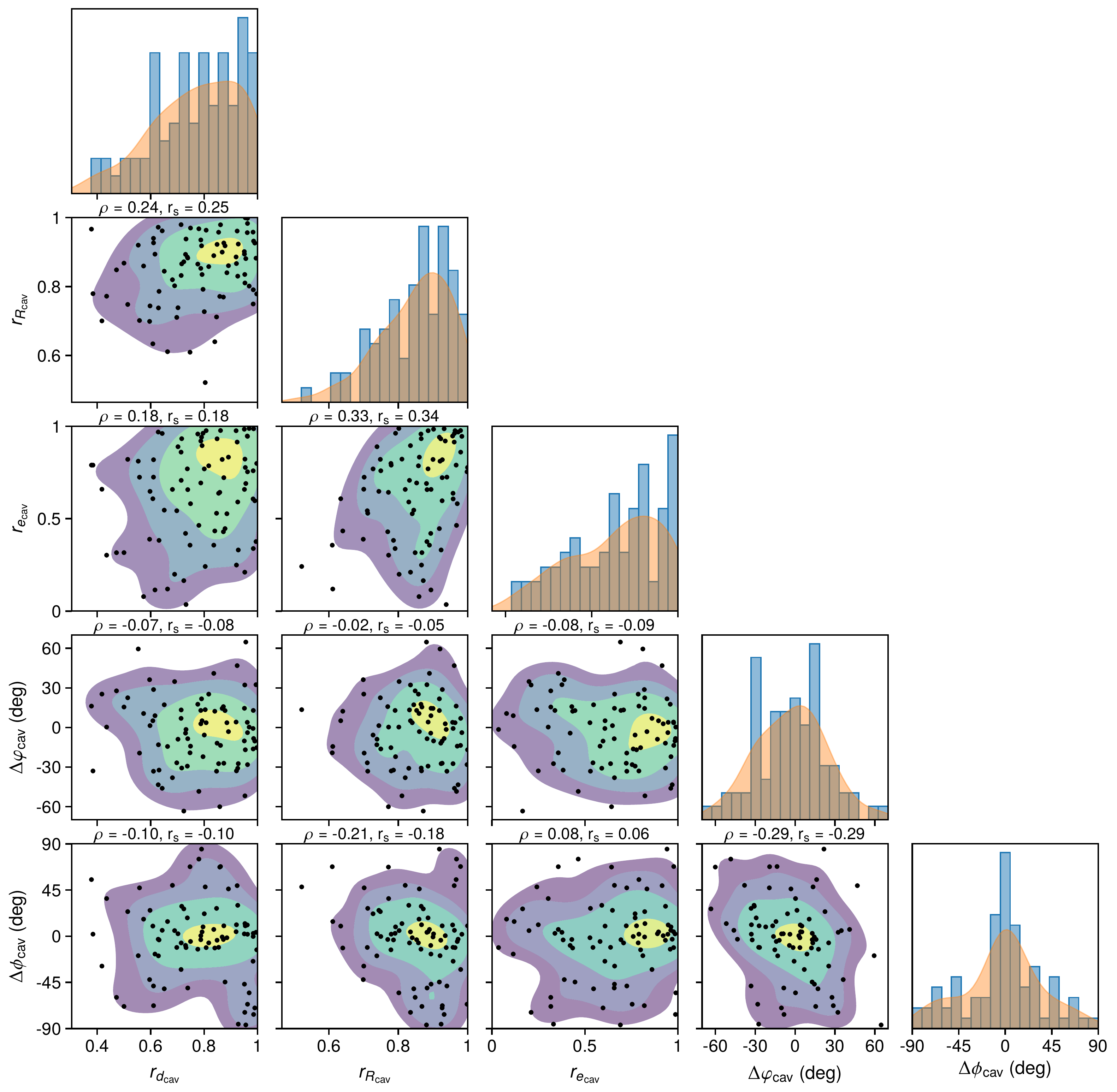}};
    \end{tikzpicture}
    \vspace{-5mm}
    \begin{flushleft}
    \justifying \textbf{Figure D4:} \hypertarget{figD4}{Corner} plot showing distributions and correlations of secondary cavity parameters (ratios and differences between individual cavity parameters within a cavity pair): relative distance $r_d$, relative semi-major axis $r_R$, difference between ellipticity $r_{e_{\text{cav}}}$, difference between positional angles $d\phi$ and difference between rotational angles $d\varphi$. Contours in parameter pair plots correspond to $90\,\%$, $75\,\%$, $50\,\%$ and $10\,\%$ intervals in terms of cumulative probability of Gaussian KDE. Pearson and Spearman correlation coefficients are stated above individual plots of parameter pairs.
    \end{flushleft}
\end{figure*}

\begin{figure*}
    \vspace{3mm}
    \begin{flushleft}
    {\large \bf APPENDIX E: EXEMPLARY MOCK IMAGES}
    \end{flushleft}
    \vspace{6mm}    
    
    \begin{tikzpicture}
    \draw (0.5, 8.1) node {\includegraphics[width=0.92\textwidth]{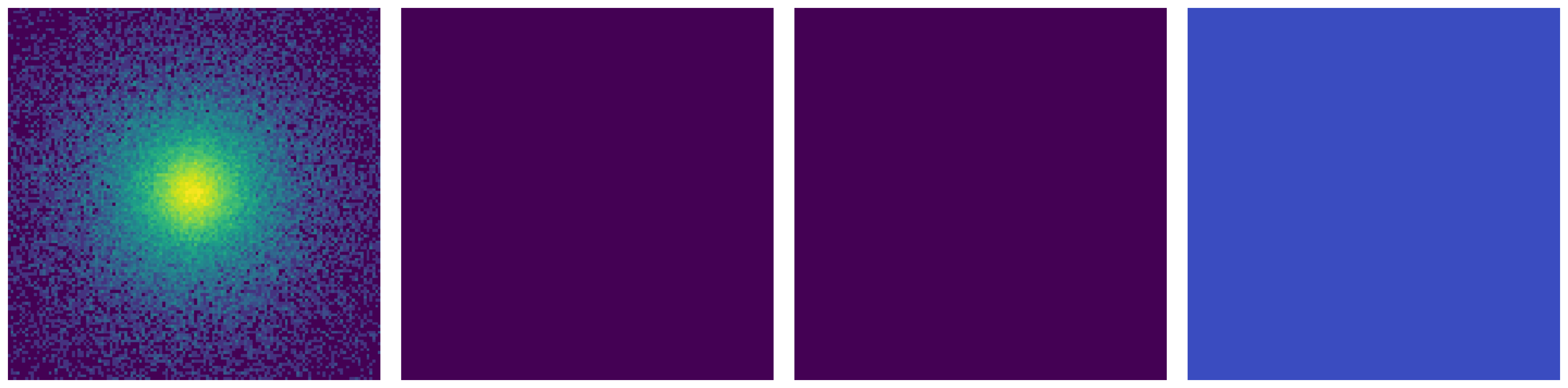}};
    \draw (0.5, 4.10) node {\includegraphics[width=0.92\textwidth]{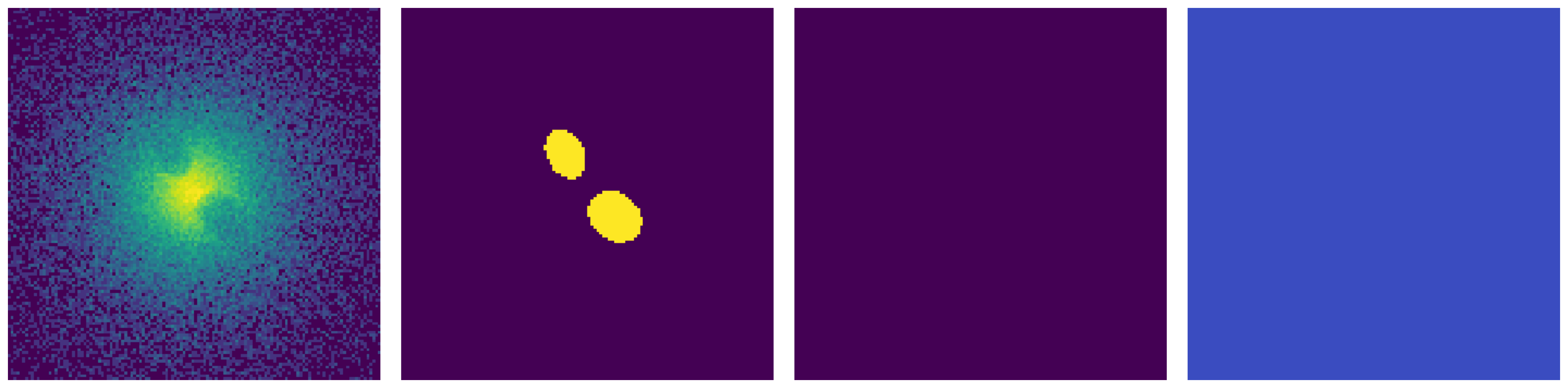}};
    \draw (0.5, 0.1) node {\includegraphics[width=0.92\textwidth]{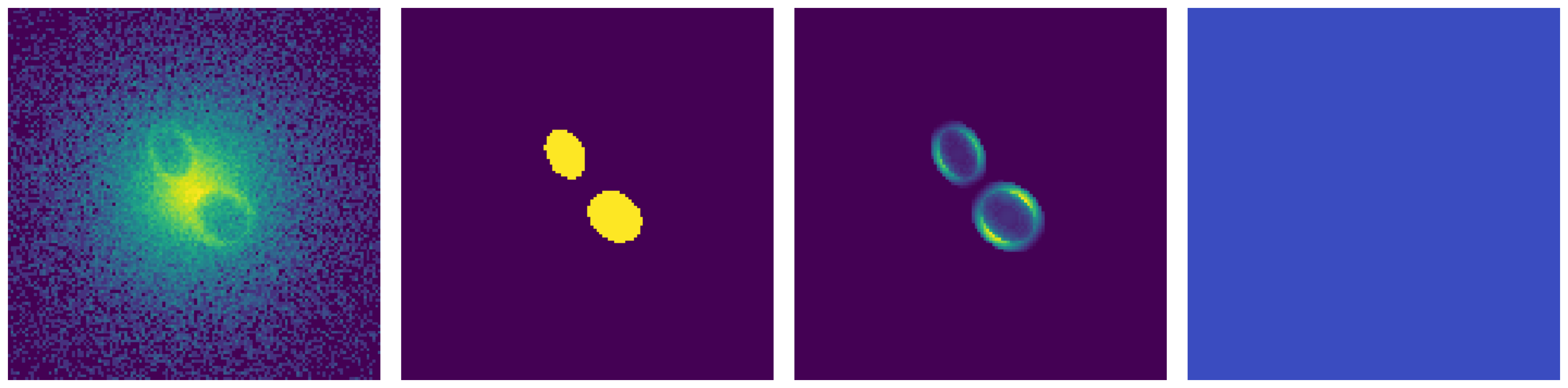}};
    \draw (0.5, -3.90) node {\includegraphics[width=0.92\textwidth]{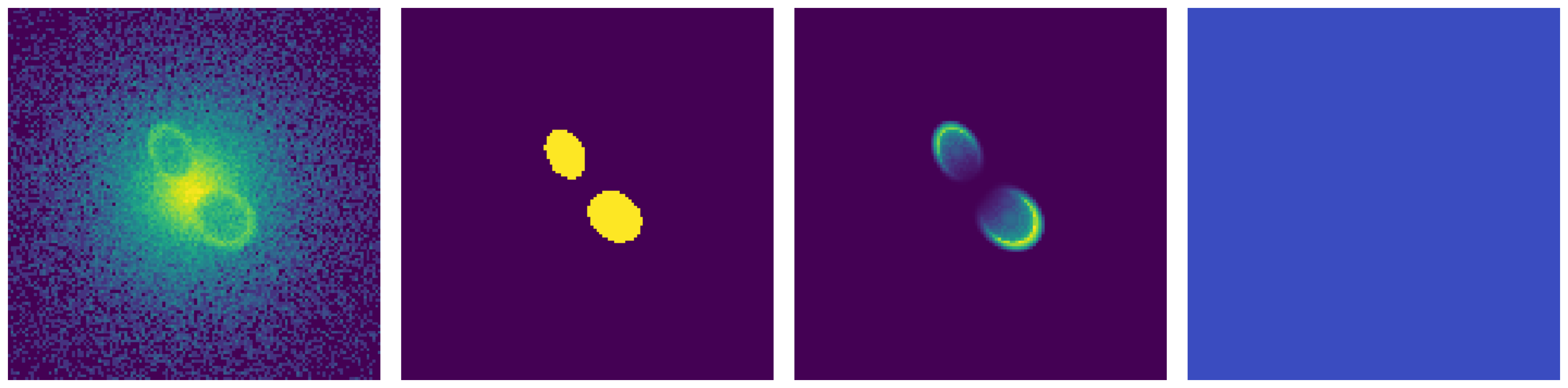}};
    \draw (0.5, -7.9) node {\includegraphics[width=0.92\textwidth]{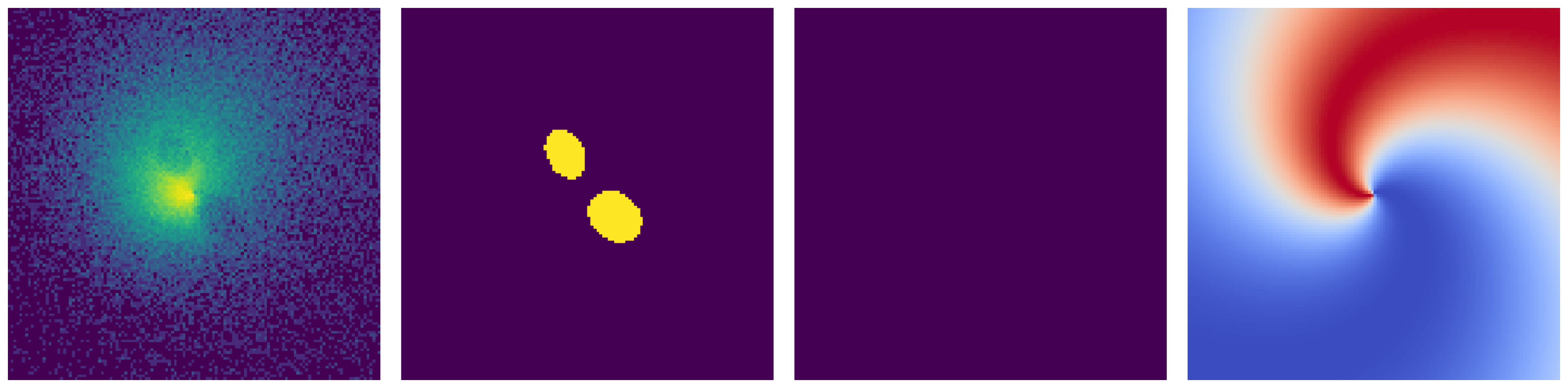}};
    \draw (-7.1, 9.65) node {\large \color{white}a)};
    \draw (-7.1, 5.65) node {\large \color{white}b)};
    \draw (-7.1, 1.65) node {\large \color{white}c)};
    \draw (-7.1, -2.35) node {\large \color{white}d)};
    \draw (-7.1, -6.35) node {\large \color{white}e)};
    \end{tikzpicture}
    \vspace{2mm}
    \begin{flushleft}
    \justifying \textbf{Figure E1:} \hypertarget{figE1}{Examples} of artificially generated images with various features: \textbf{a)} single beta model without cavities, \textbf{b)} beta model with cavities, \textbf{c)} beta model with cavities and type I rims, \textbf{d)} beta model with cavities and type II rims, \textbf{e)} beta model with cavities and gas sloshing. Individual columns show: the resulting noisy image, binary cavity mask, cavity rim pattern, and  sloshing pattern.
    \end{flushleft}
\end{figure*}

\begin{figure*}
    \vspace{2mm}
    \begin{flushleft}
    {\large \bf APPENDIX F: CADET ARCHITECTURE}
    \end{flushleft}
    \vspace{2mm}
    
    \centering
    \includegraphics[width=0.88\textwidth]{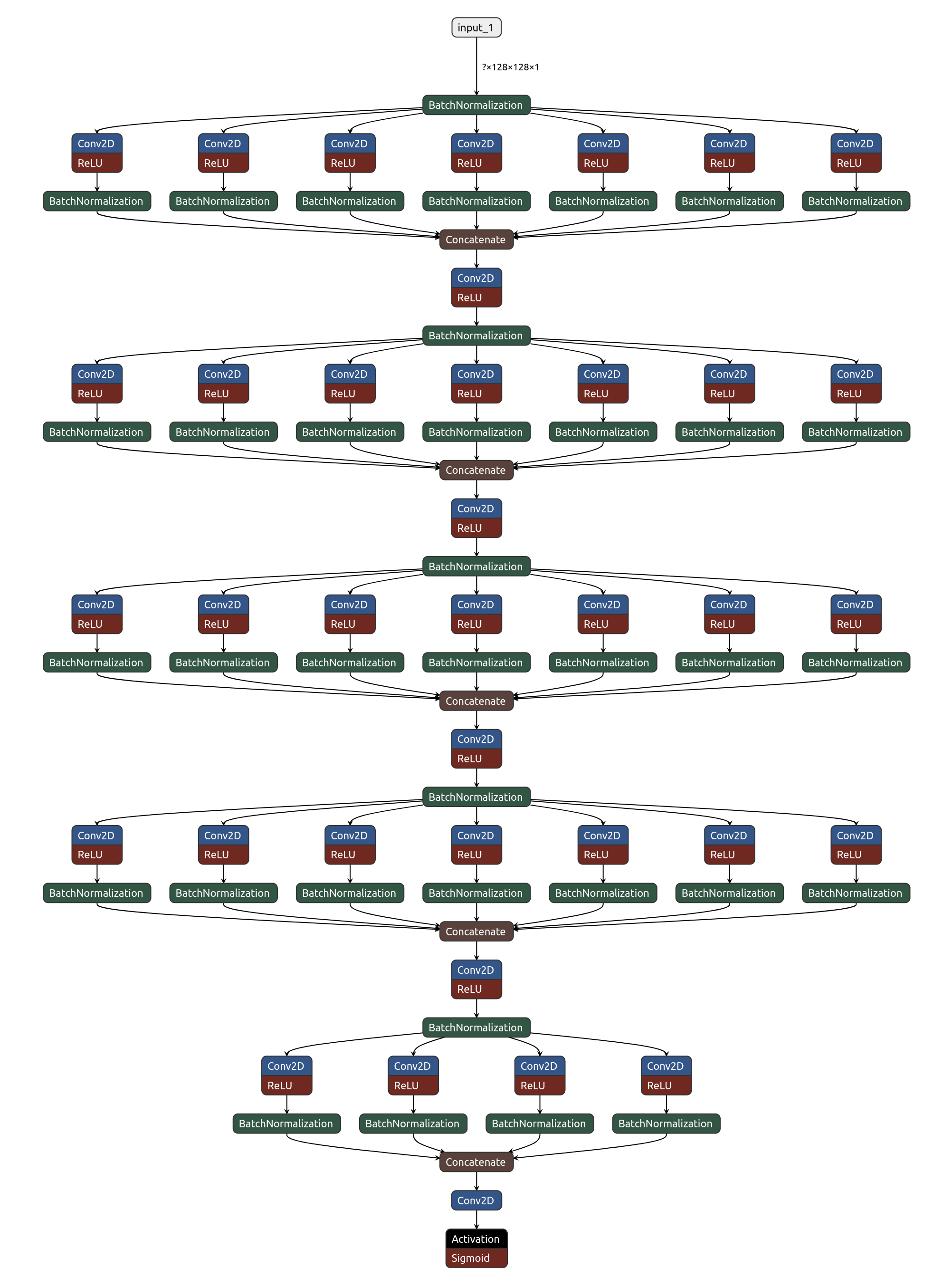}
    \vspace{2mm}
    \begin{flushleft}
    \justifying \textbf{Figure F1:} \hypertarget{figF1}{The} schematic picture of the convolutional neural network composed of 5 Inception-like blocks. Each block consists of a series of parallel convolutional layers each composed of various numbers of convolutional filters with various sizes. The output of all parallel convolutional layers is then concatenated into a single output, followed by a convolutional layer with 32 of $1 \times 1$ filters and a dropout layer, which was for simplicity omitted. The scheme was created using the \href{https://netron.app/}{Netron} visualisation tool.
    \end{flushleft}
\end{figure*}


\bsp	
\label{lastpage}
\end{document}



\begin{figure*}
\begin{tikzpicture}[remember picture,overlay,shift={(current page.center)}]
    \draw (0, 12.1) node {\large \bf Application of CADET on real \textit{Chandra} images};
    \draw (0, 10.2) node {\includegraphics[height=85pt]{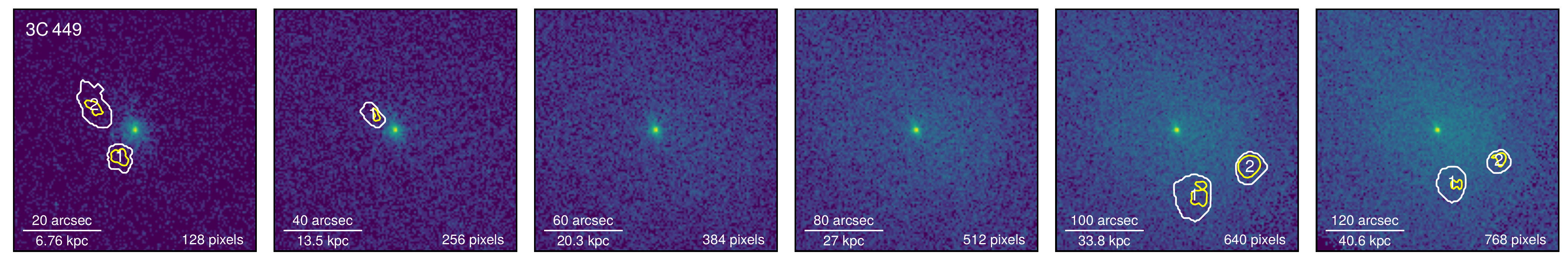}};
    \draw (0, 7.2) node {\includegraphics[height=85pt]{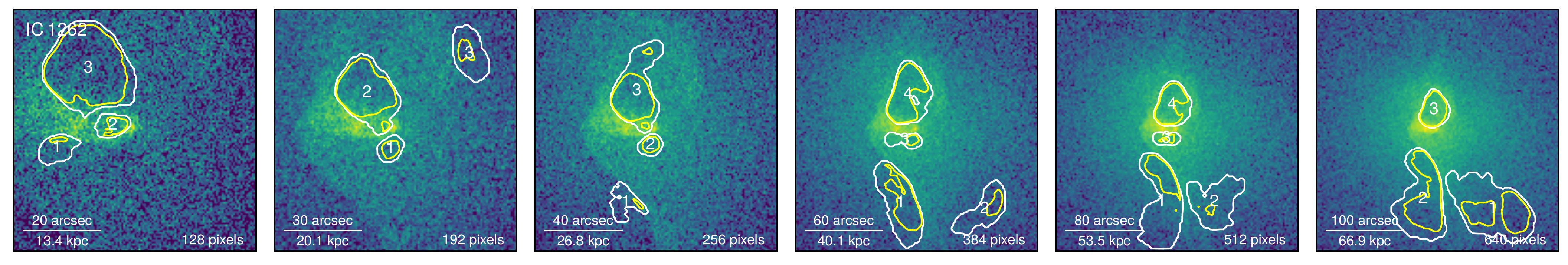}};
    \draw (0, 4.2) node {\includegraphics[height=85pt]{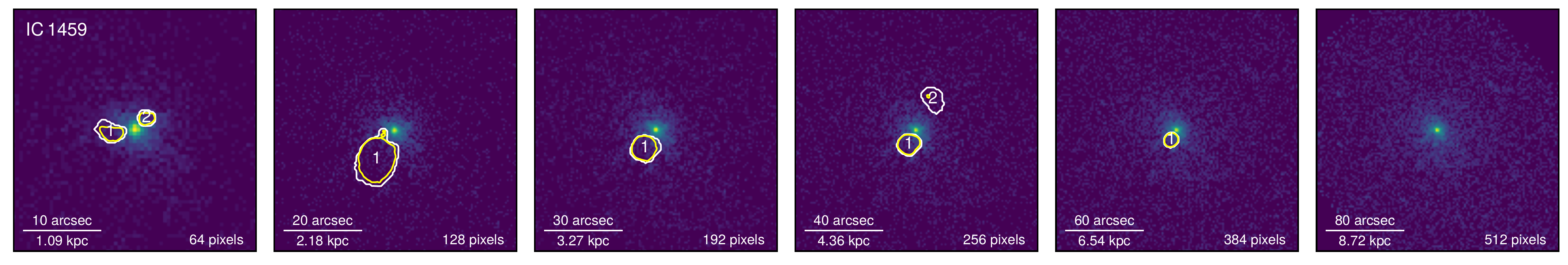}};
    \draw (0, 1.2) node {\includegraphics[height=85pt]{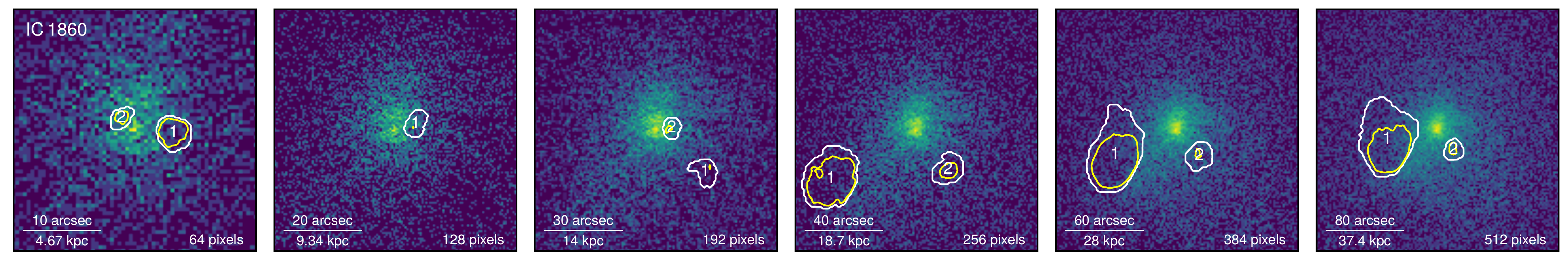}};
    \draw (0, -1.8) node {\includegraphics[height=85pt]{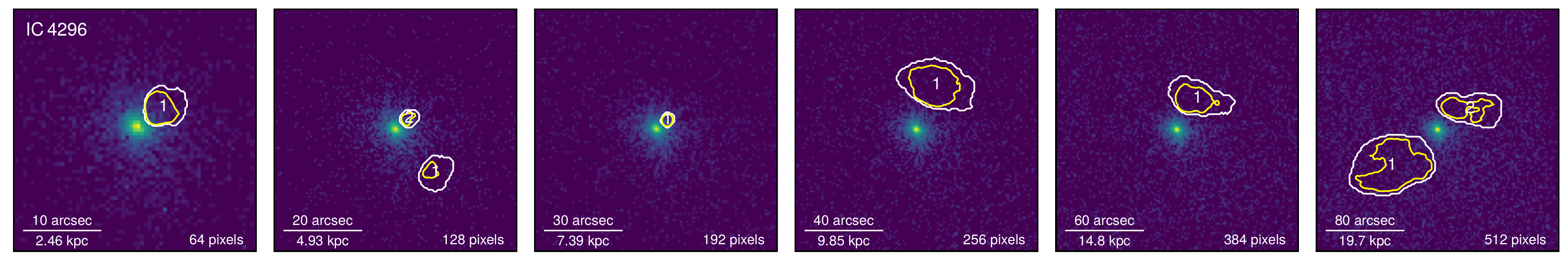}};
    \draw (0, -4.8) node {\includegraphics[height=85pt]{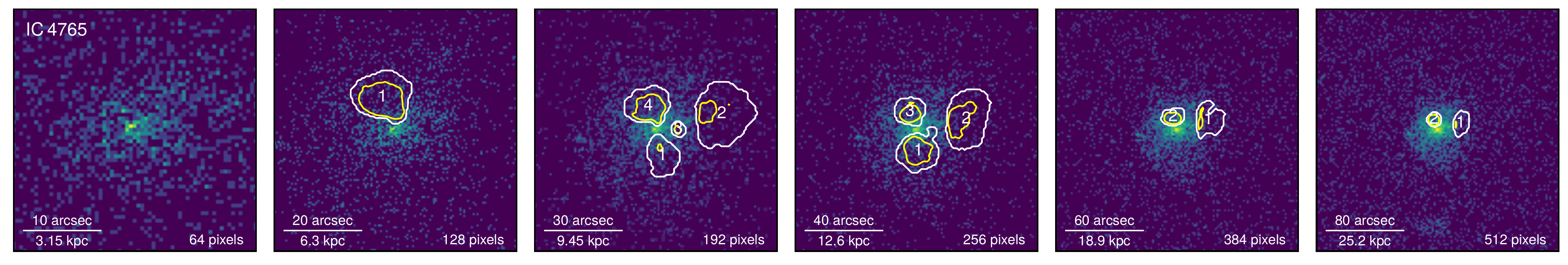}};
    \draw (0, -7.8) node {\includegraphics[height=85pt]{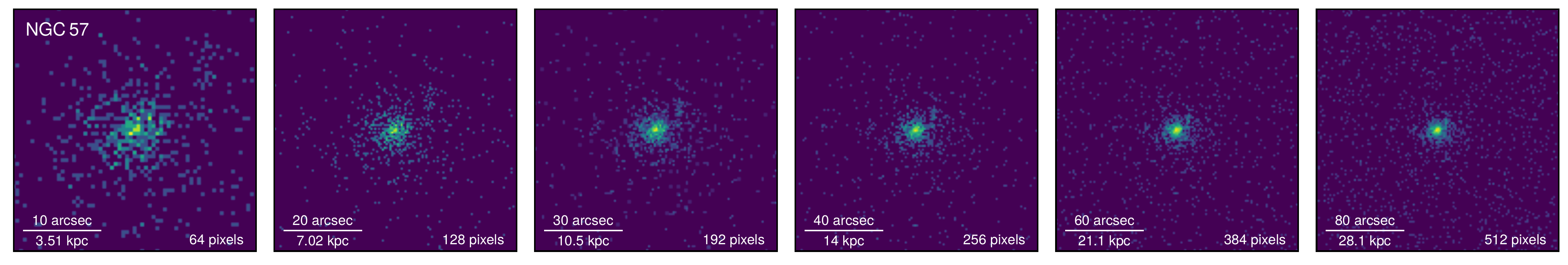}};
    \draw (0, -10.8) node {\includegraphics[height=85pt]{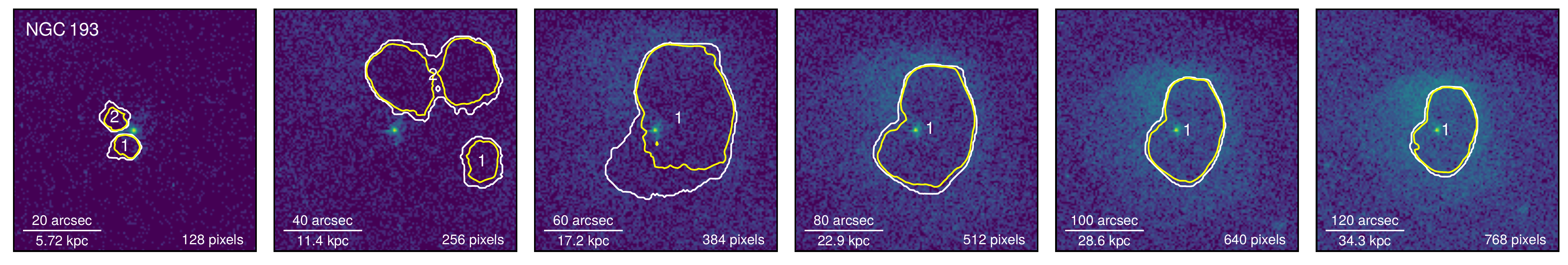}};
\end{tikzpicture}
\end{figure*}

\clearpage

\begin{figure*}
\begin{tikzpicture}[remember picture,overlay,shift={(current page.center)}]
    \draw (0, 10.2) node {\includegraphics[height=85pt]{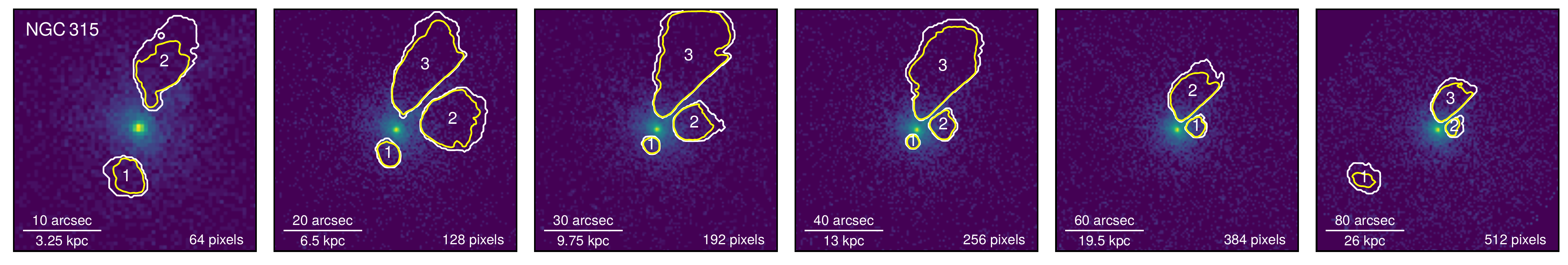}};
    \draw (0, 7.2) node {\includegraphics[height=85pt]{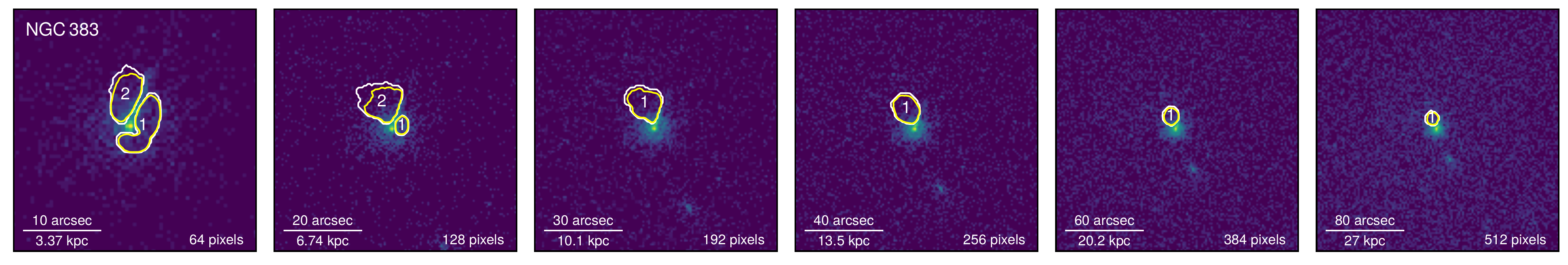}};
    \draw (0, 4.2) node {\includegraphics[height=85pt]{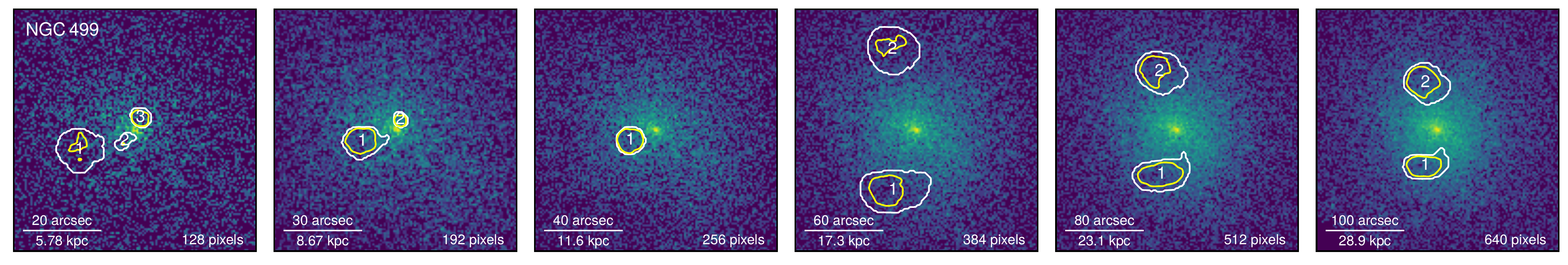}};
    \draw (0, 1.2) node {\includegraphics[height=85pt]{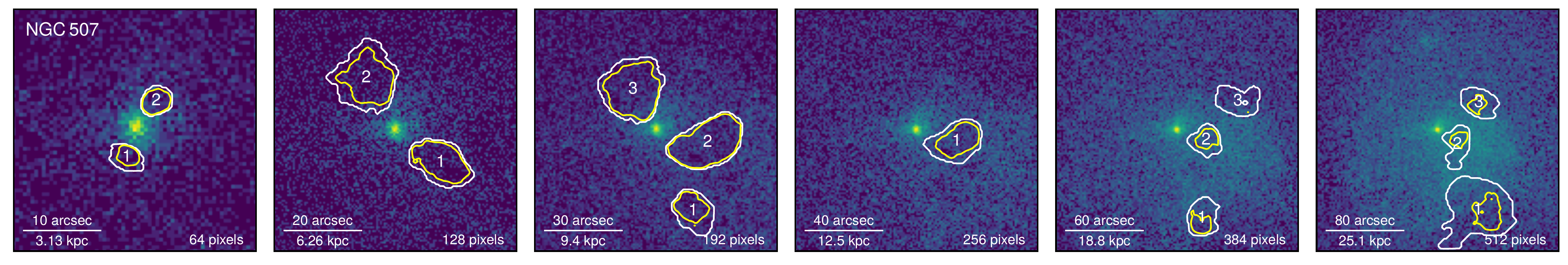}};
    \draw (0, -1.8) node {\includegraphics[height=85pt]{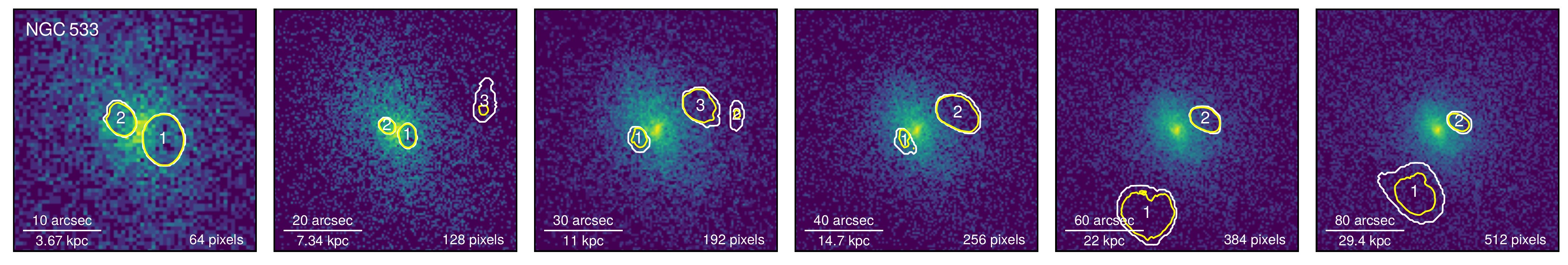}};
    \draw (0, -4.8) node {\includegraphics[height=85pt]{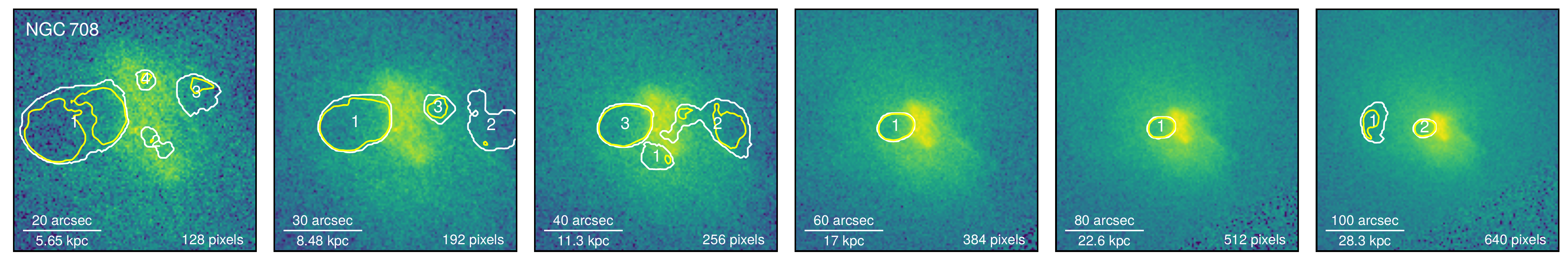}};
    \draw (0, -7.8) node {\includegraphics[height=85pt]{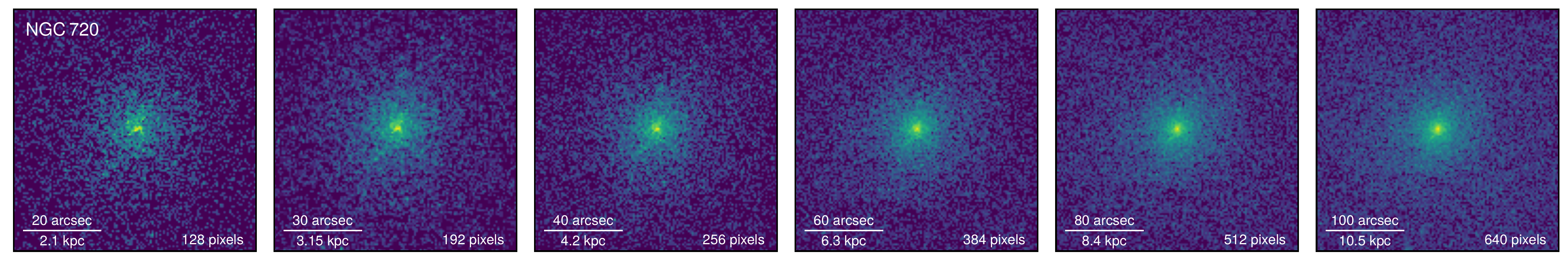}};
    \draw (0, -10.8) node {\includegraphics[height=85pt]{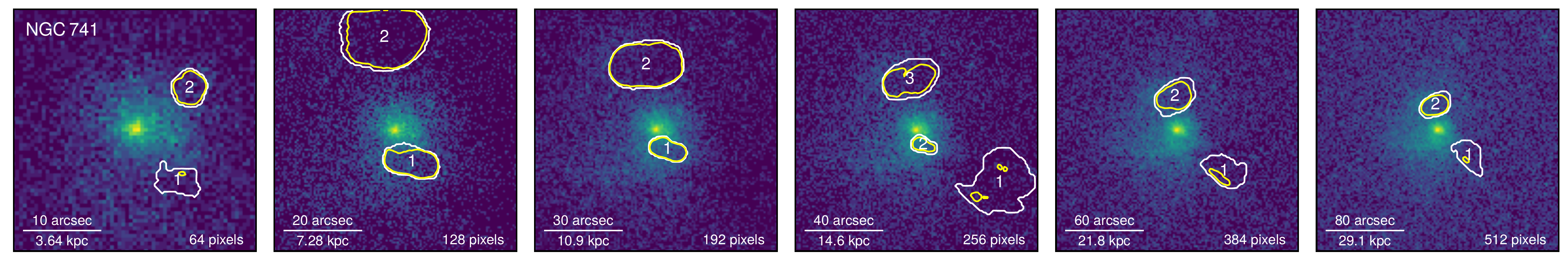}};
\end{tikzpicture}
\end{figure*}

\clearpage

\begin{figure*}
\begin{tikzpicture}[remember picture,overlay,shift={(current page.center)}]
    \draw (0, 10.2) node {\includegraphics[height=85pt]{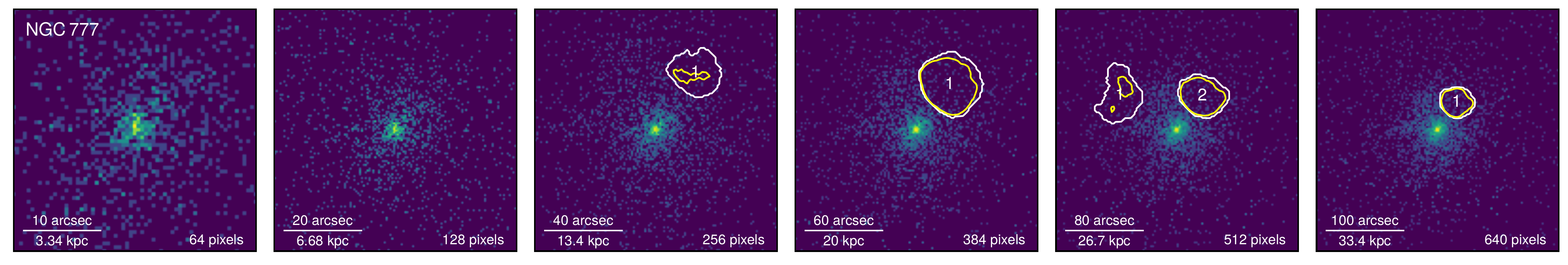}};
    \draw (0, 7.2) node {\includegraphics[height=85pt]{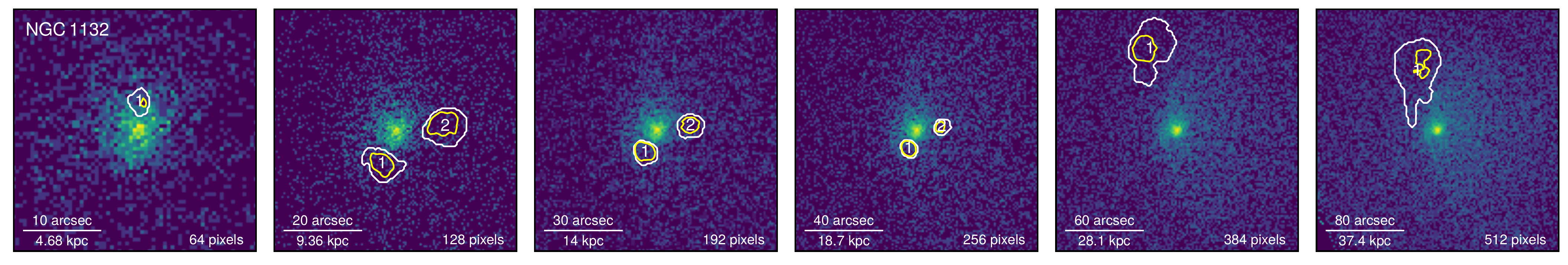}};
    \draw (0, 4.2) node {\includegraphics[height=85pt]{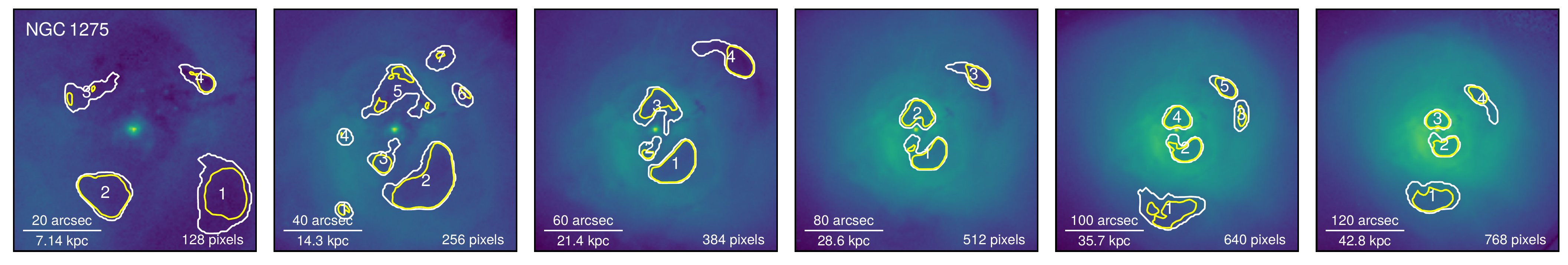}};
    \draw (0, 1.2) node {\includegraphics[height=85pt]{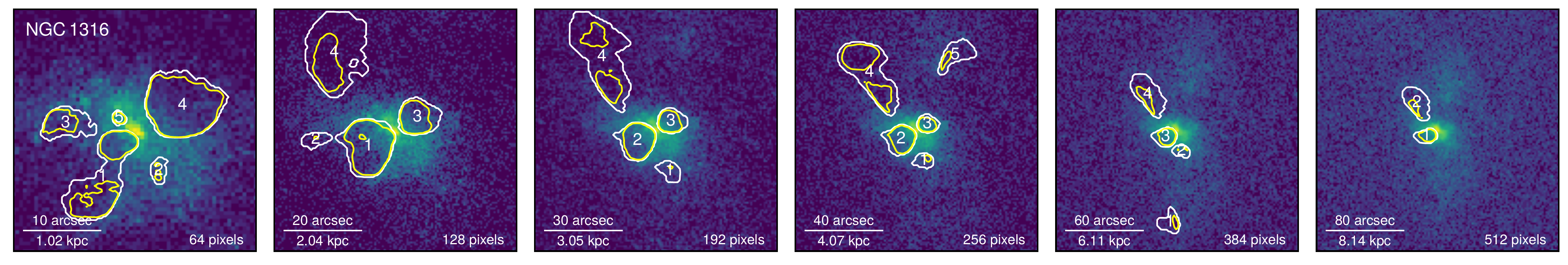}};
    \draw (0, -1.8) node {\includegraphics[height=85pt]{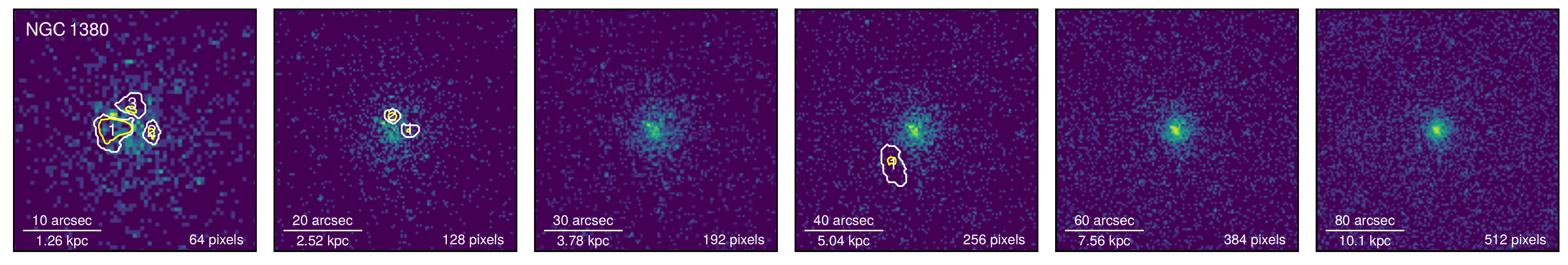}};
    \draw (0, -4.8) node {\includegraphics[height=85pt]{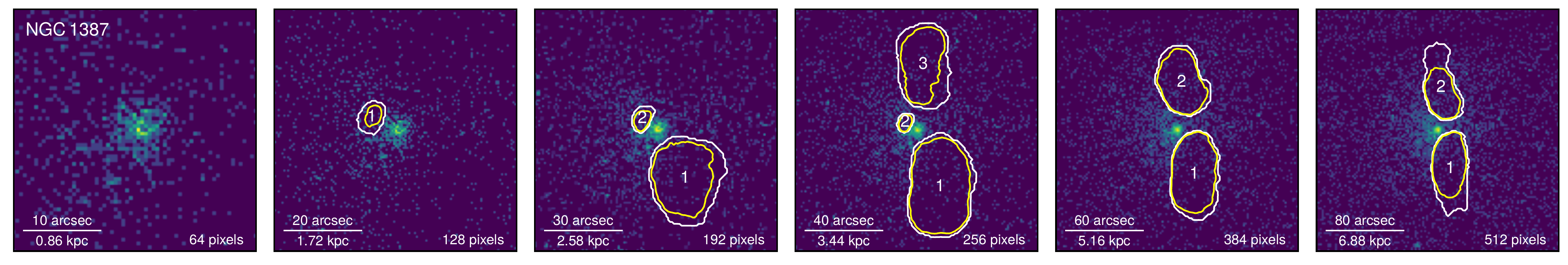}};
    \draw (0, -7.8) node {\includegraphics[height=85pt]{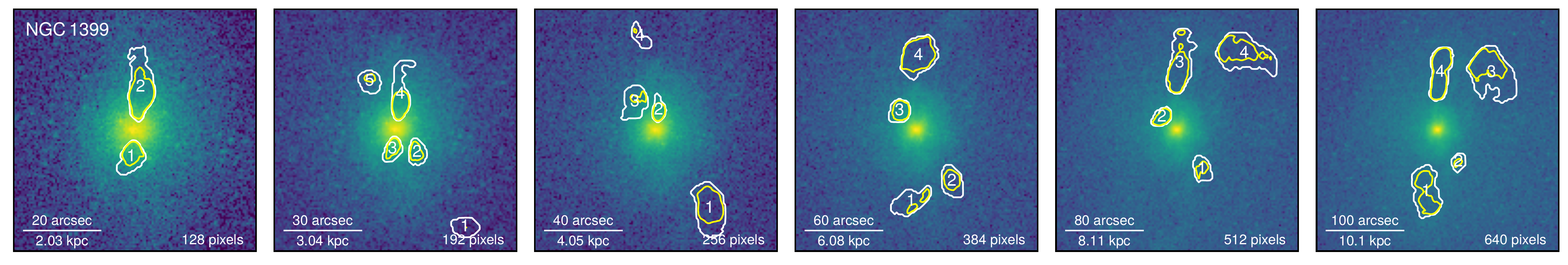}};
    \draw (0, -10.8) node {\includegraphics[height=85pt]{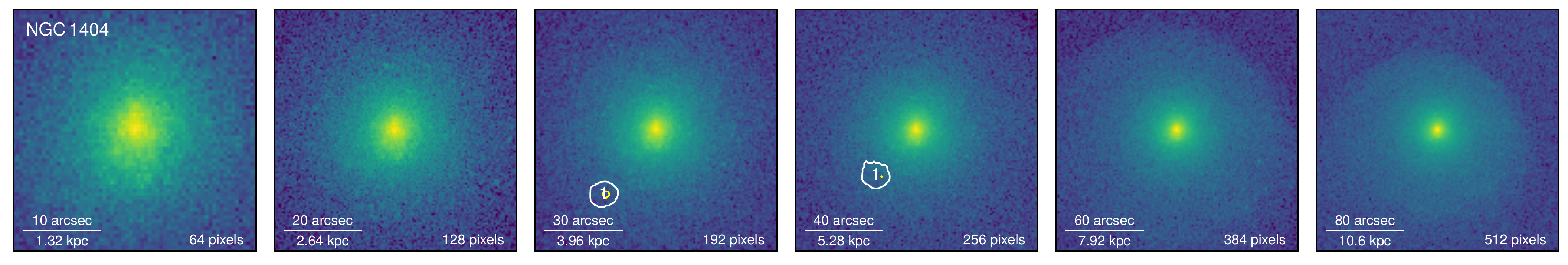}};
\end{tikzpicture}
\end{figure*}

\clearpage

\begin{figure*}
\begin{tikzpicture}[remember picture,overlay,shift={(current page.center)}]
    \draw (0, 10.2) node {\includegraphics[height=85pt]{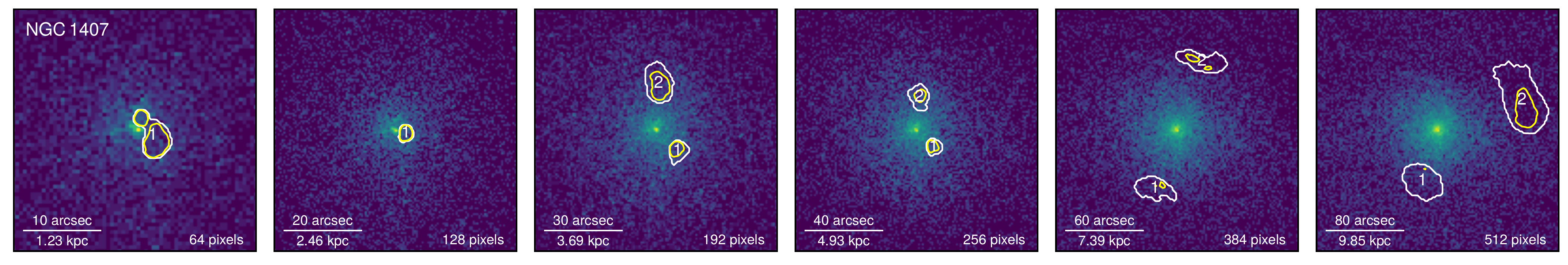}};
    \draw (0, 7.2) node {\includegraphics[height=85pt]{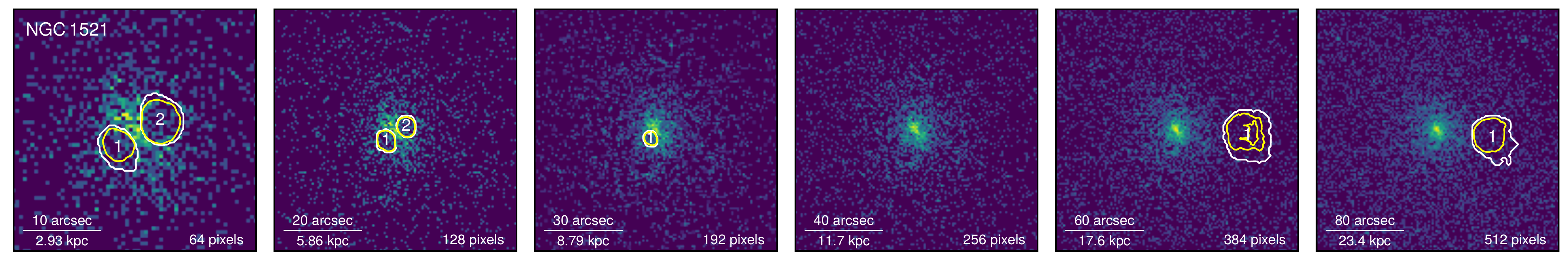}};
    \draw (0, 4.2) node {\includegraphics[height=85pt]{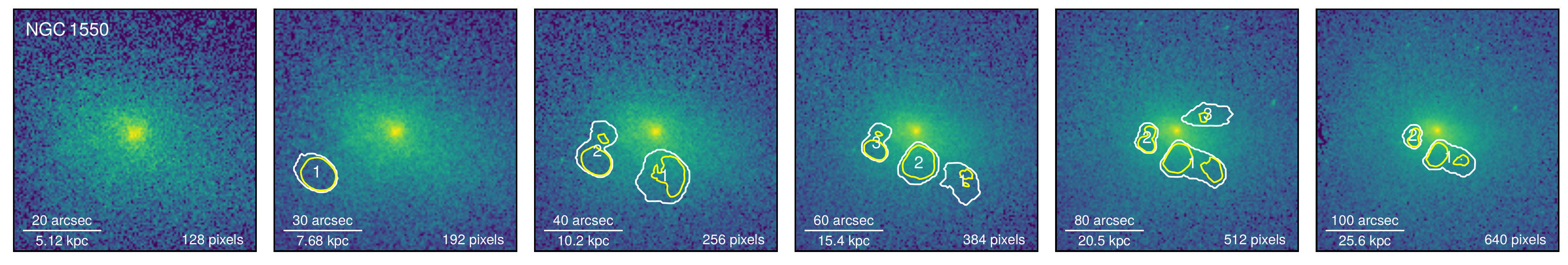}};
    \draw (0, 1.2) node {\includegraphics[height=85pt]{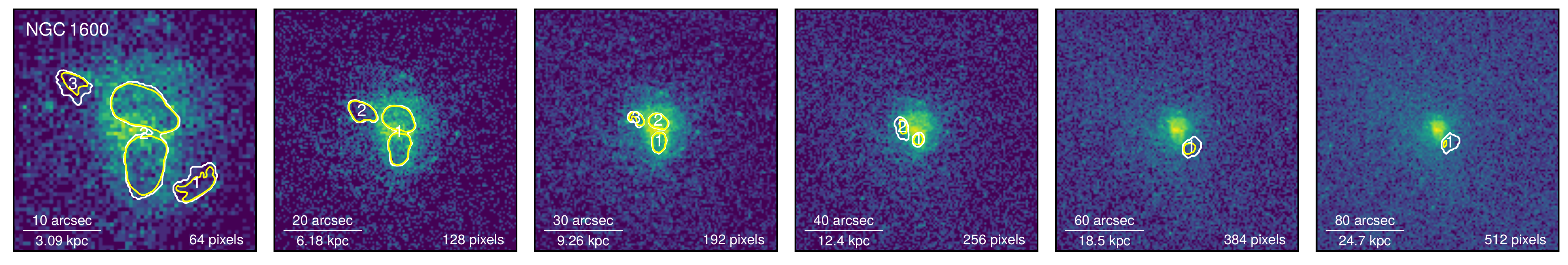}};
    \draw (0, -1.8) node {\includegraphics[height=85pt]{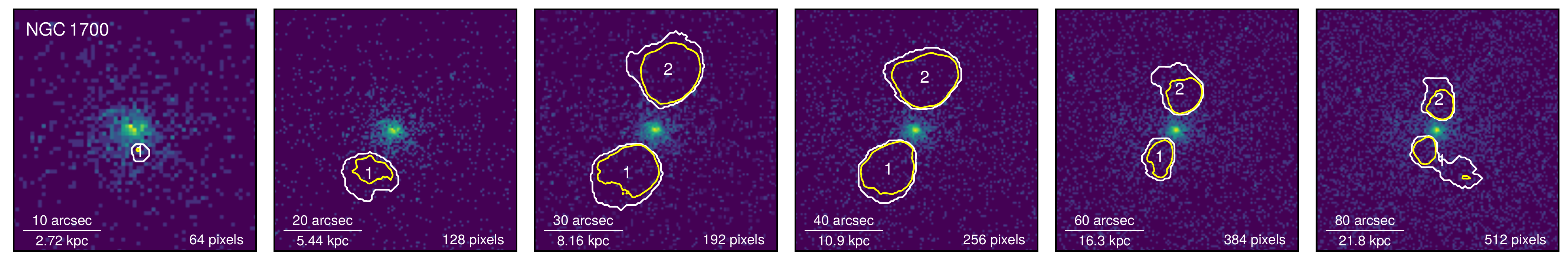}};
    \draw (0, -4.8) node {\includegraphics[height=85pt]{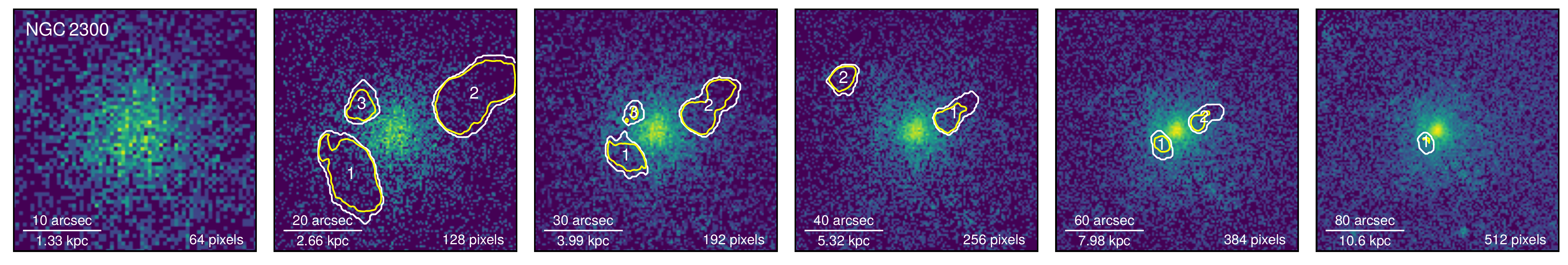}};
    \draw (0, -7.8) node {\includegraphics[height=85pt]{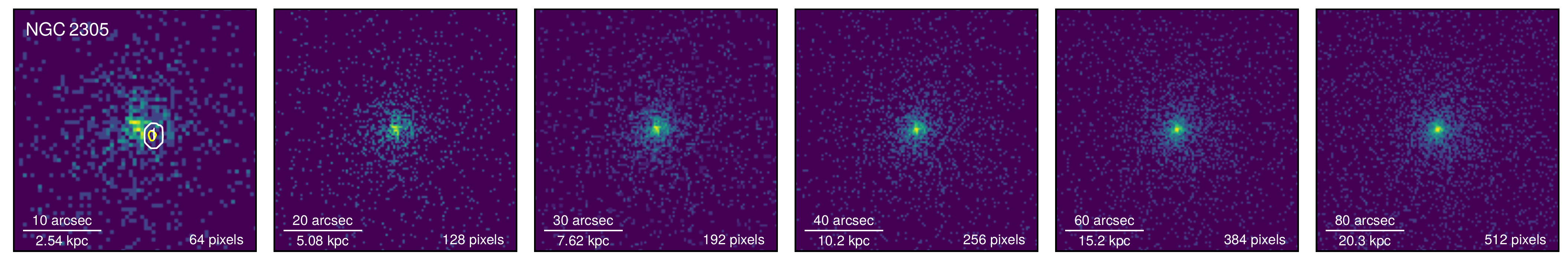}};
    \draw (0, -10.8) node {\includegraphics[height=85pt]{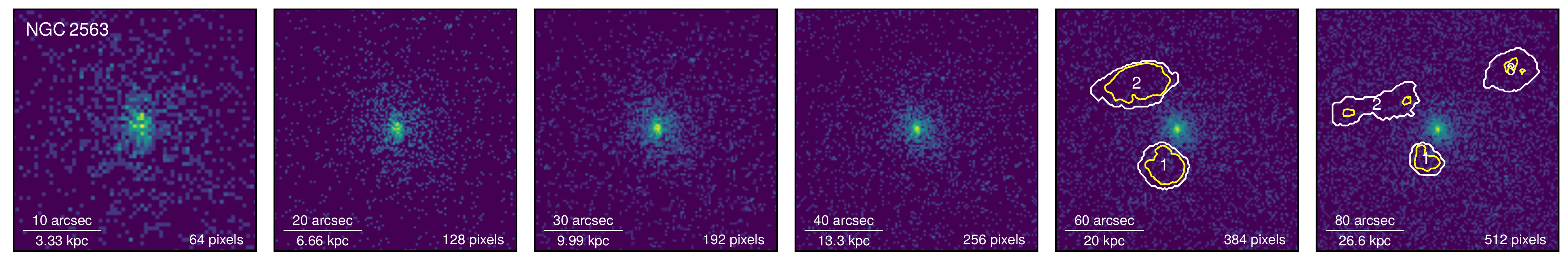}};
\end{tikzpicture}
\end{figure*}

\clearpage

\begin{figure*}
\begin{tikzpicture}[remember picture,overlay,shift={(current page.center)}]
    \draw (0, 10.2) node {\includegraphics[height=85pt]{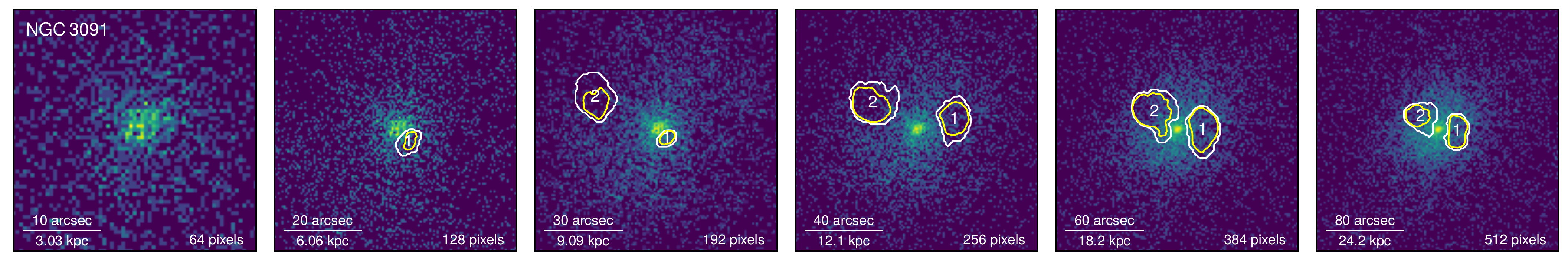}};
    \draw (0, 7.2) node {\includegraphics[height=85pt]{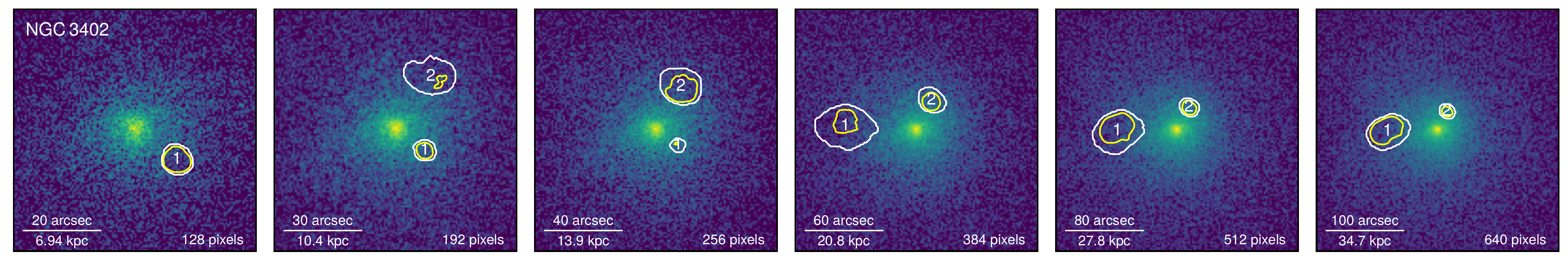}};
    \draw (0, 4.2) node {\includegraphics[height=85pt]{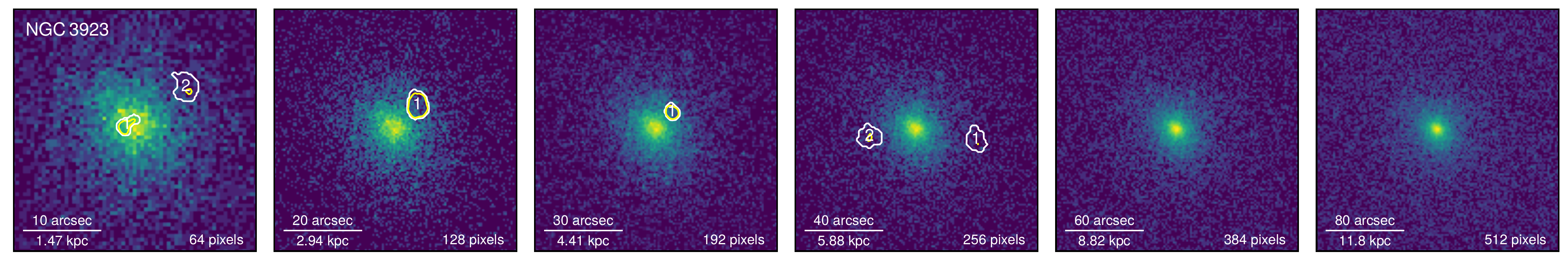}};
    \draw (0, 1.2) node {\includegraphics[height=85pt]{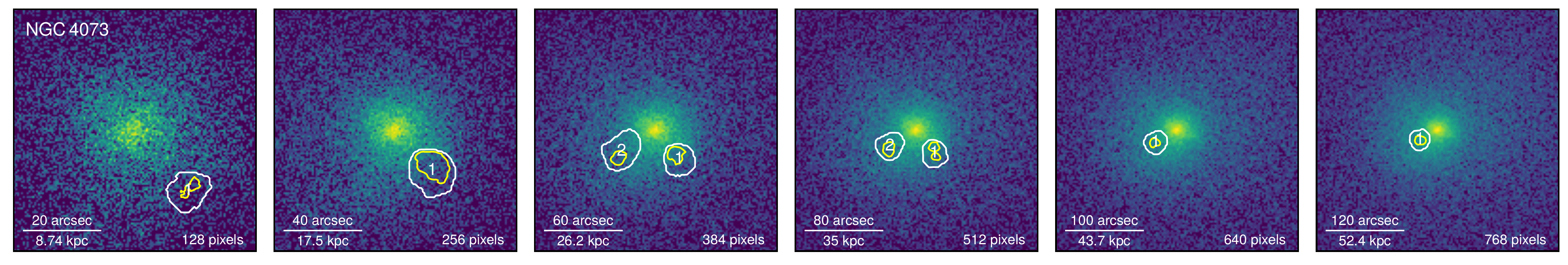}};
    \draw (0, -1.8) node {\includegraphics[height=85pt]{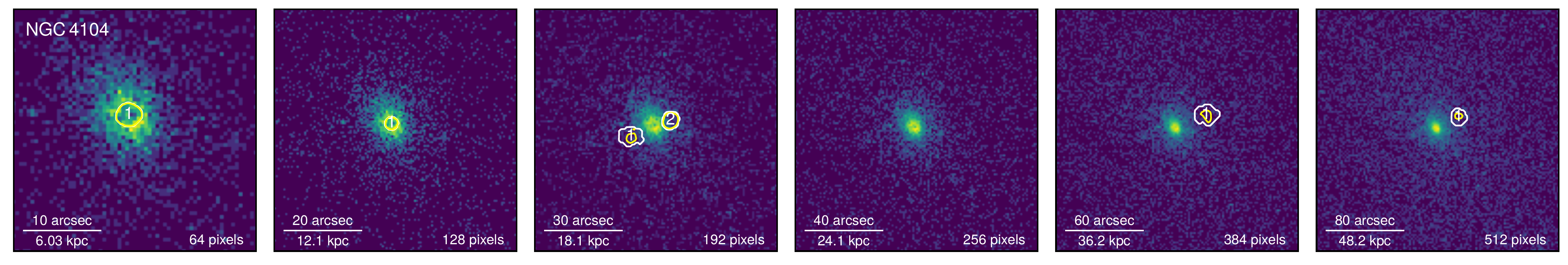}};
    \draw (0, -4.8) node {\includegraphics[height=85pt]{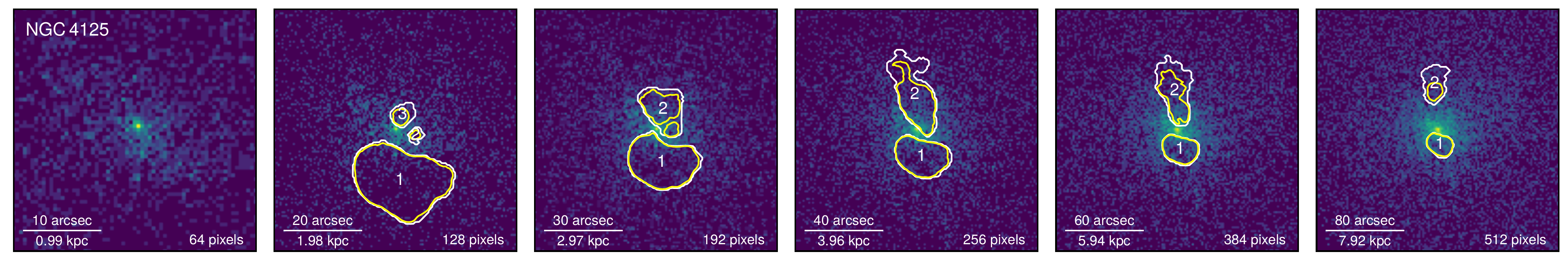}};
    \draw (0, -7.8) node {\includegraphics[height=85pt]{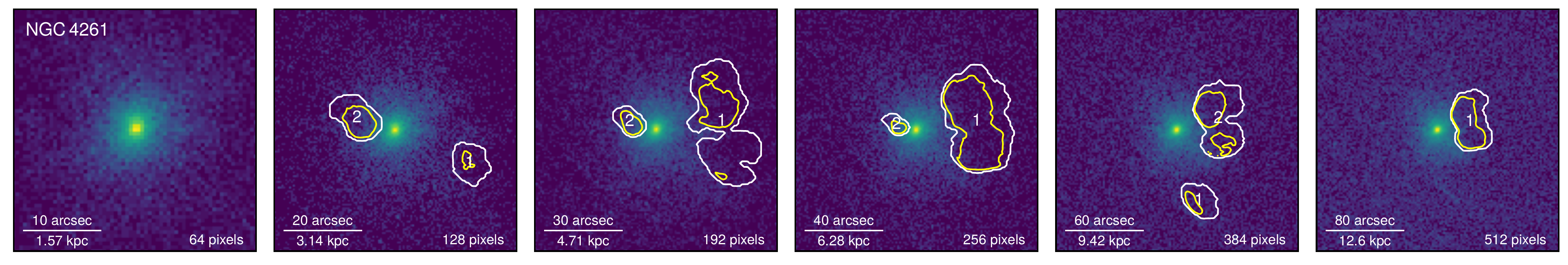}};
    \draw (0, -10.8) node {\includegraphics[height=85pt]{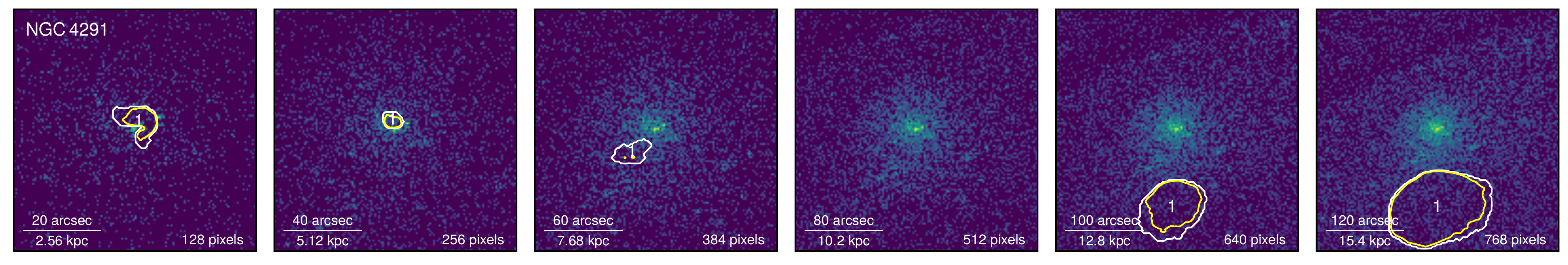}};
\end{tikzpicture}
\end{figure*}

\clearpage

\begin{figure*}
\begin{tikzpicture}[remember picture,overlay,shift={(current page.center)}]
    \draw (0, 10.2) node {\includegraphics[height=85pt]{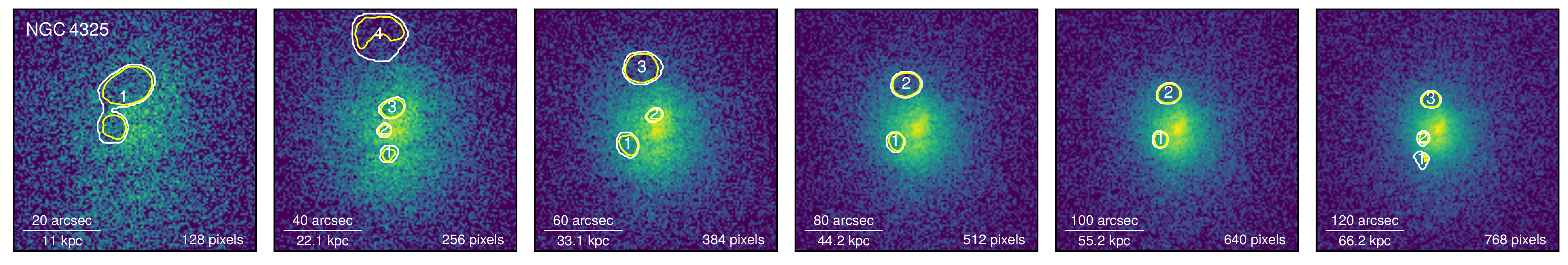}};
    \draw (0, 7.2) node {\includegraphics[height=85pt]{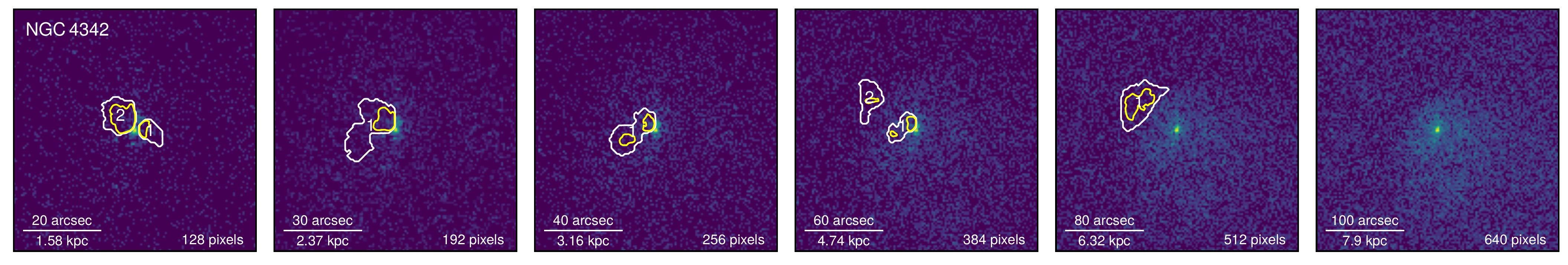}};
    \draw (0, 4.2) node {\includegraphics[height=85pt]{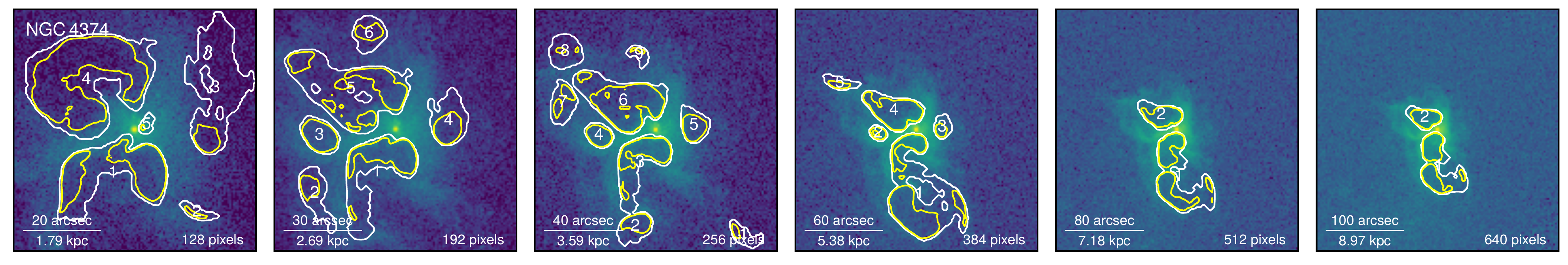}};
    \draw (0, 1.2) node {\includegraphics[height=85pt]{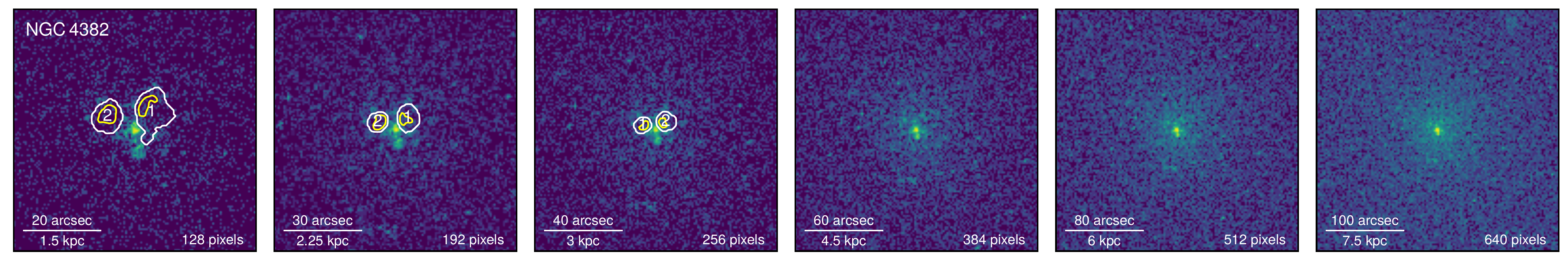}};
    \draw (0, -1.8) node {\includegraphics[height=85pt]{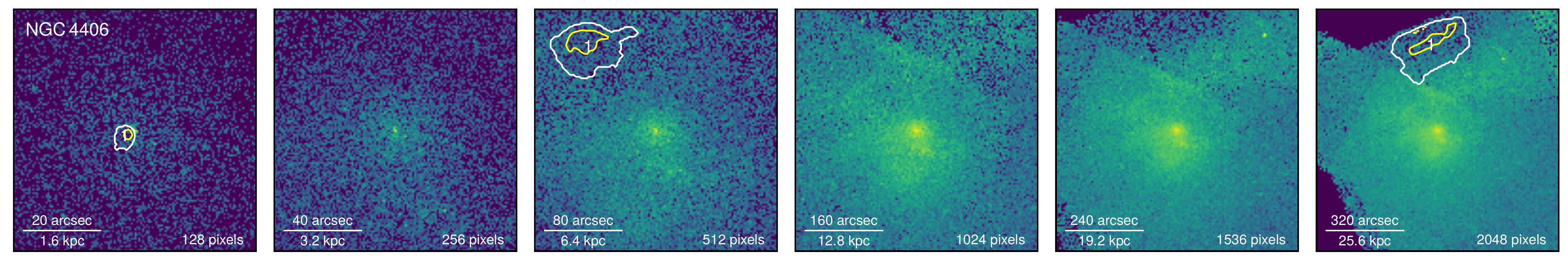}};
    \draw (0, -4.8) node {\includegraphics[height=85pt]{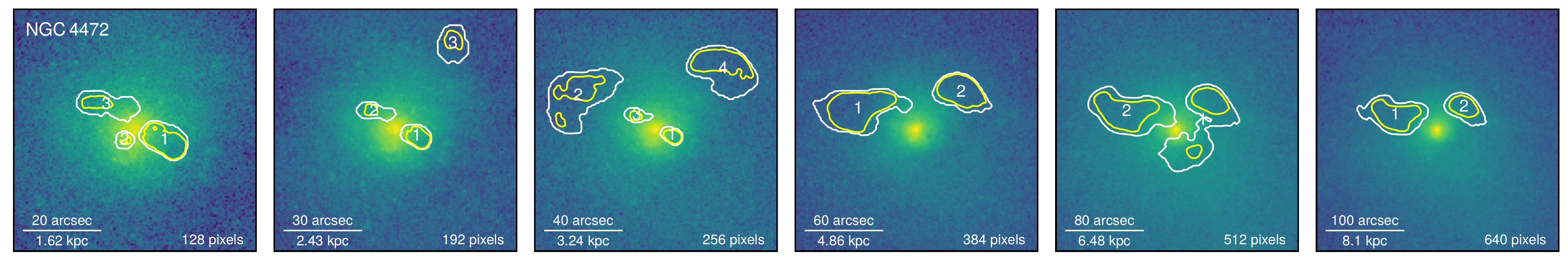}};
    \draw (0, -7.8) node {\includegraphics[height=85pt]{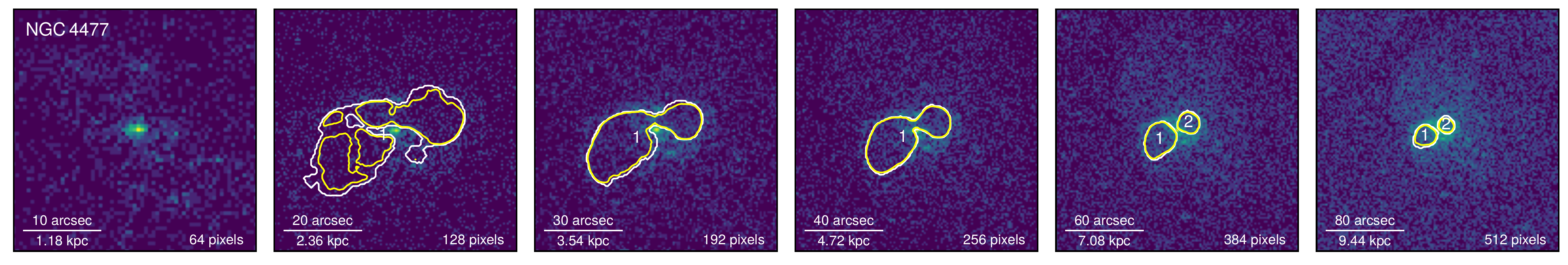}};
    \draw (0, -10.8) node {\includegraphics[height=85pt]{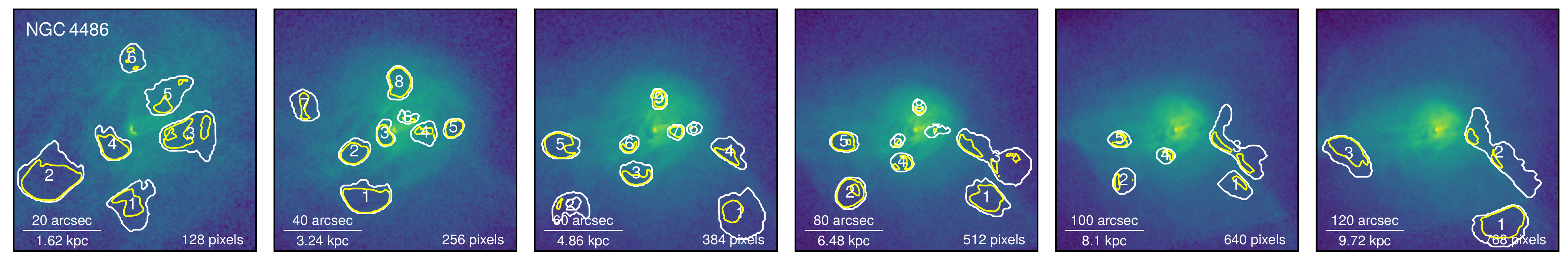}};
\end{tikzpicture}
\end{figure*}

\clearpage

\begin{figure*}
\begin{tikzpicture}[remember picture,overlay,shift={(current page.center)}]
    \draw (0, 10.2) node {\includegraphics[height=85pt]{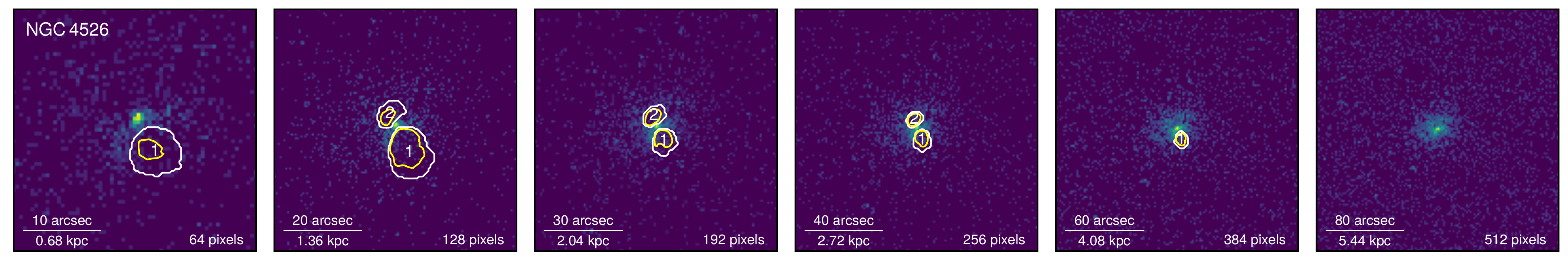}};
    \draw (0, 7.2) node {\includegraphics[height=85pt]{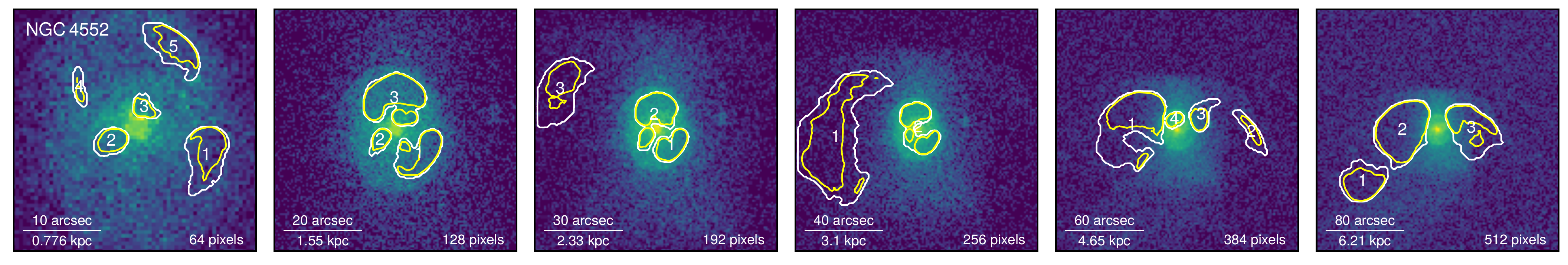}};
    \draw (0, 4.2) node {\includegraphics[height=85pt]{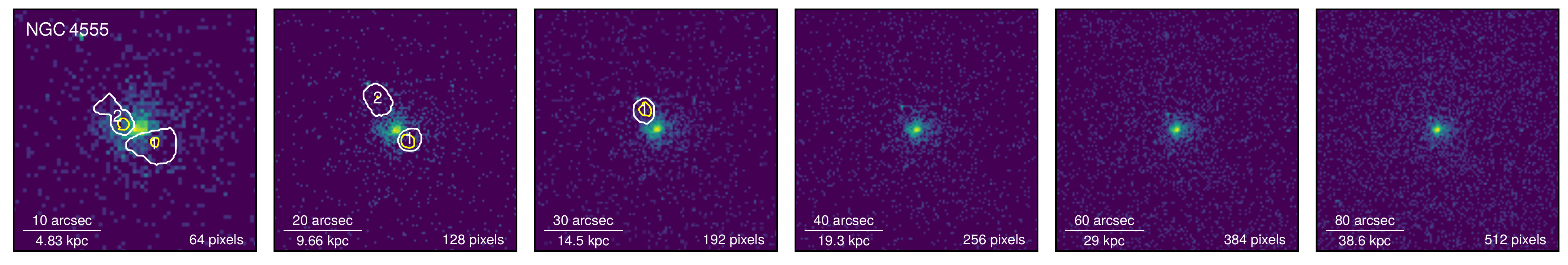}};
    \draw (0, 1.2) node {\includegraphics[height=85pt]{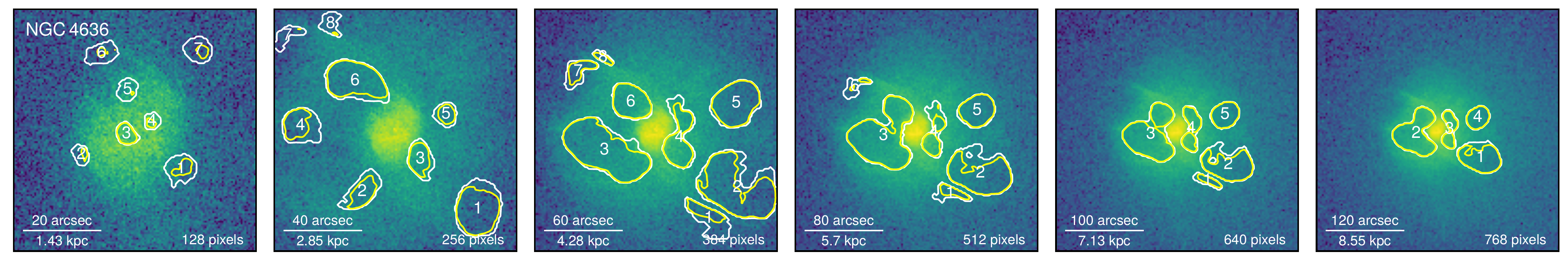}};
    \draw (0, -1.8) node {\includegraphics[height=85pt]{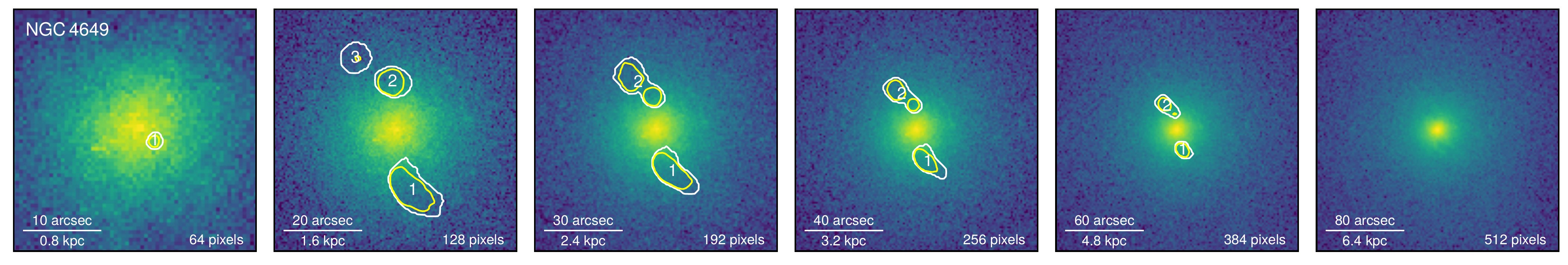}};
    \draw (0, -4.8) node {\includegraphics[height=85pt]{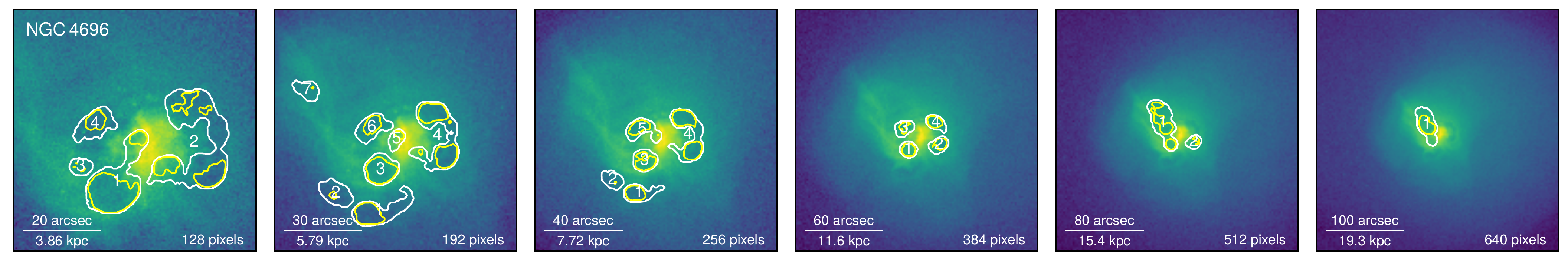}};
    \draw (0, -7.8) node {\includegraphics[height=85pt]{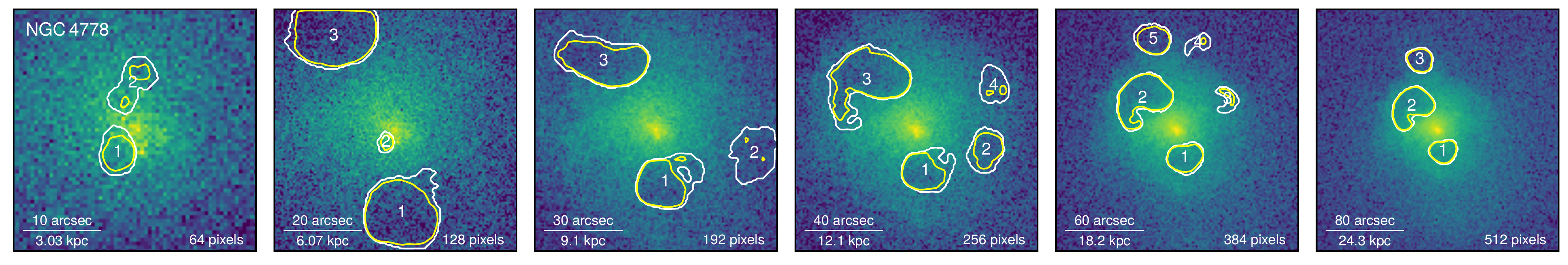}};
    \draw (0, -10.8) node {\includegraphics[height=85pt]{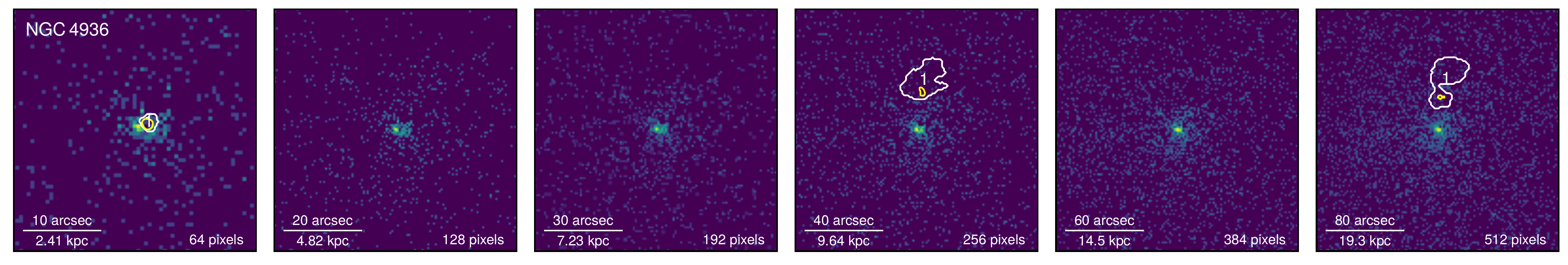}};
\end{tikzpicture}
\end{figure*}

\clearpage

\begin{figure*}
\begin{tikzpicture}[remember picture,overlay,shift={(current page.center)}]
    \draw (0, 10.2) node {\includegraphics[height=85pt]{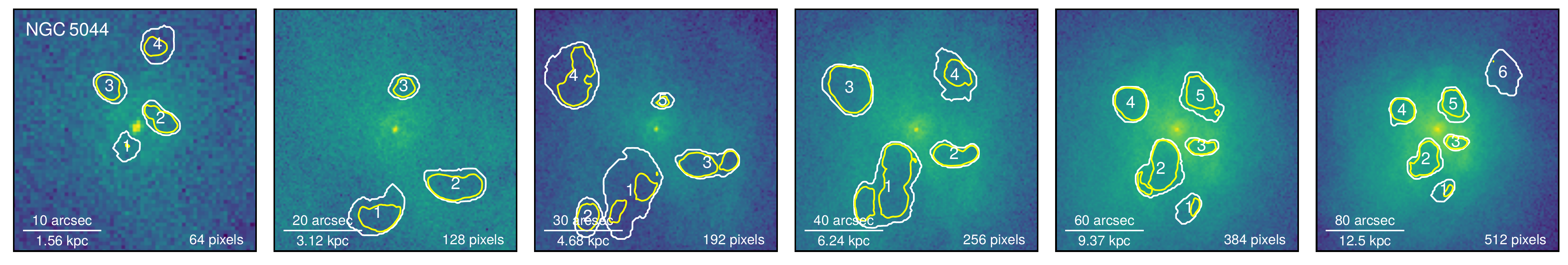}};
    \draw (0, 7.2) node {\includegraphics[height=85pt]{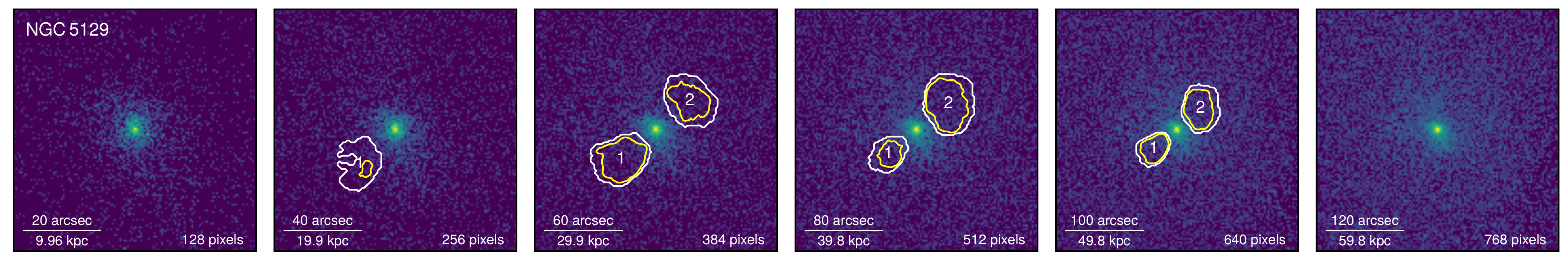}};
    \draw (0, 4.2) node {\includegraphics[height=85pt]{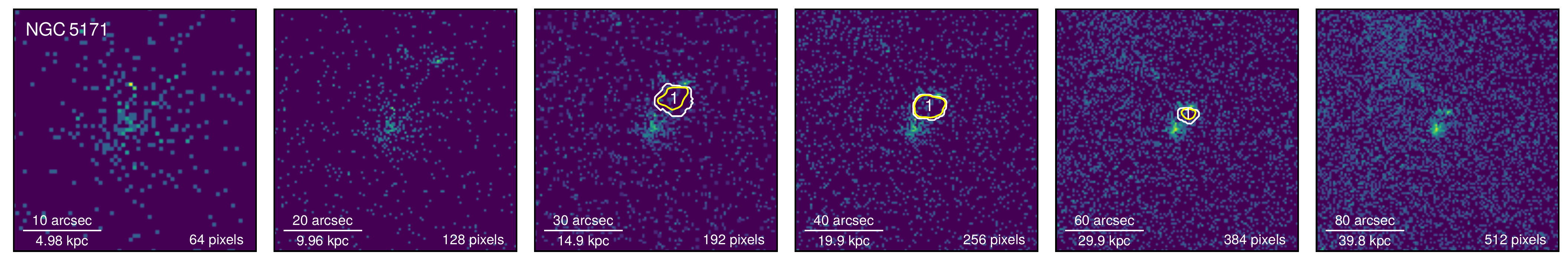}};
    \draw (0, 1.2) node {\includegraphics[height=85pt]{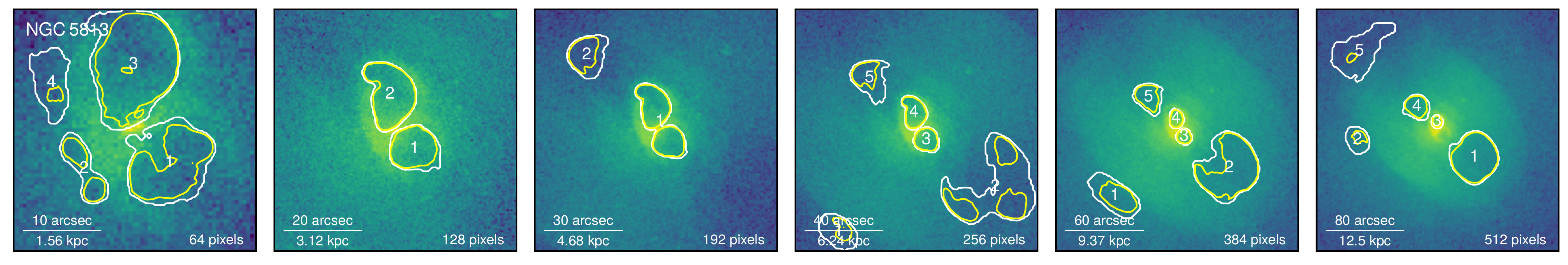}};
    \draw (0, -1.8) node {\includegraphics[height=85pt]{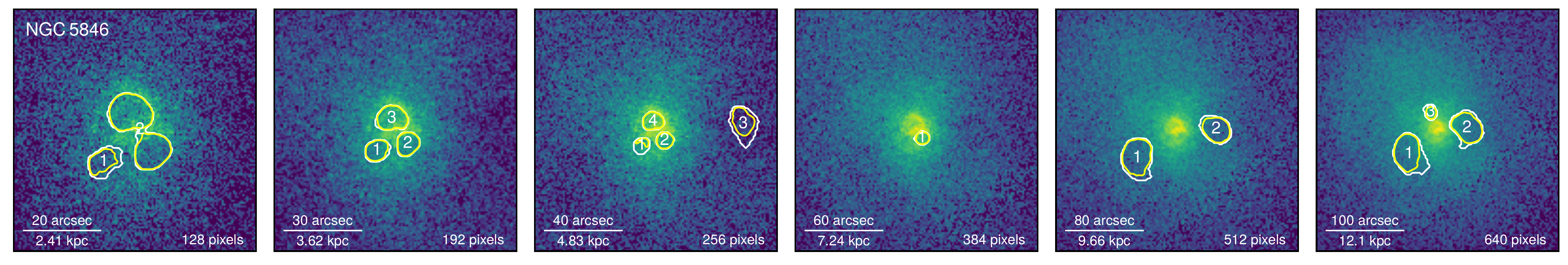}};
    \draw (0, -4.8) node {\includegraphics[height=85pt]{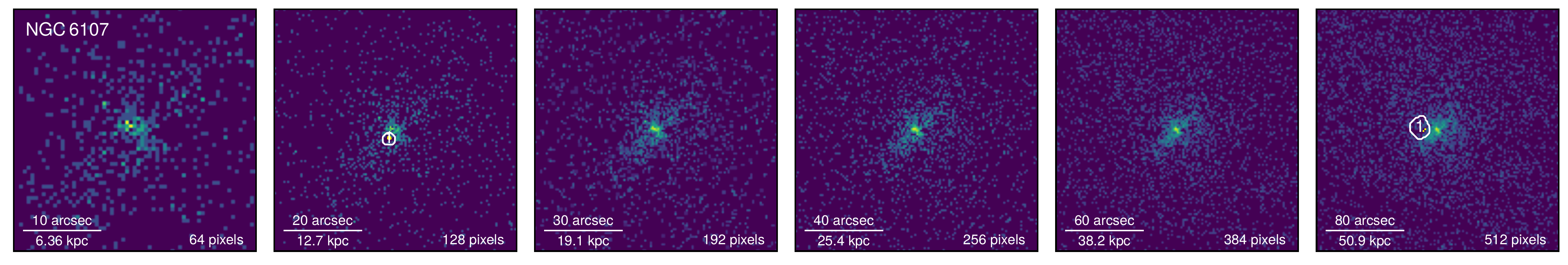}};
    \draw (0, -7.8) node {\includegraphics[height=85pt]{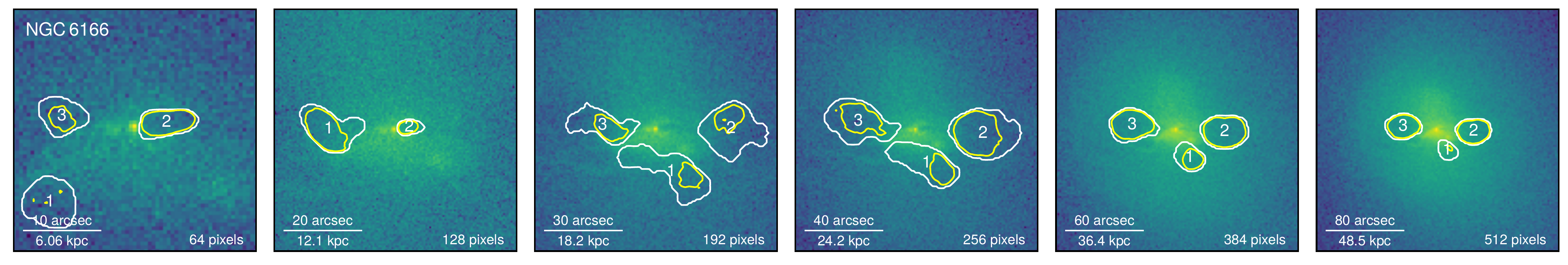}};
    \draw (0, -10.8) node {\includegraphics[height=85pt]{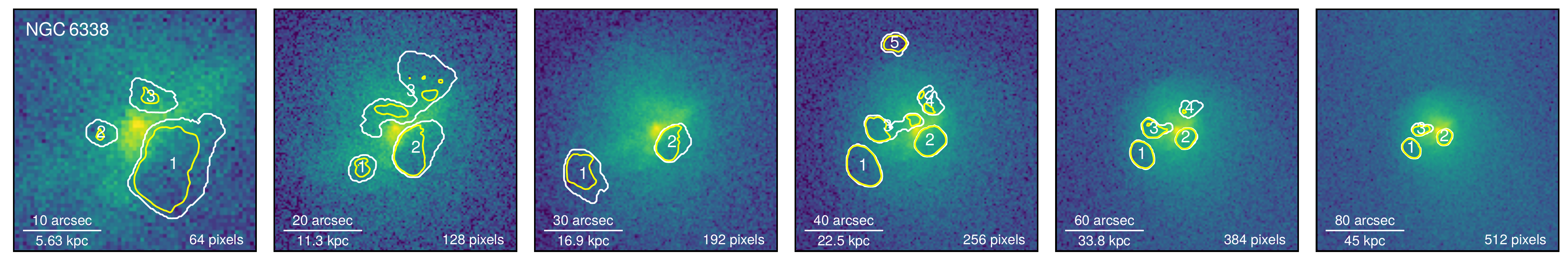}};
\end{tikzpicture}
\end{figure*}

\clearpage

\begin{figure*}
\begin{tikzpicture}[remember picture,overlay,shift={(current page.center)}]
    \draw (0, 10.2) node {\includegraphics[height=85pt]{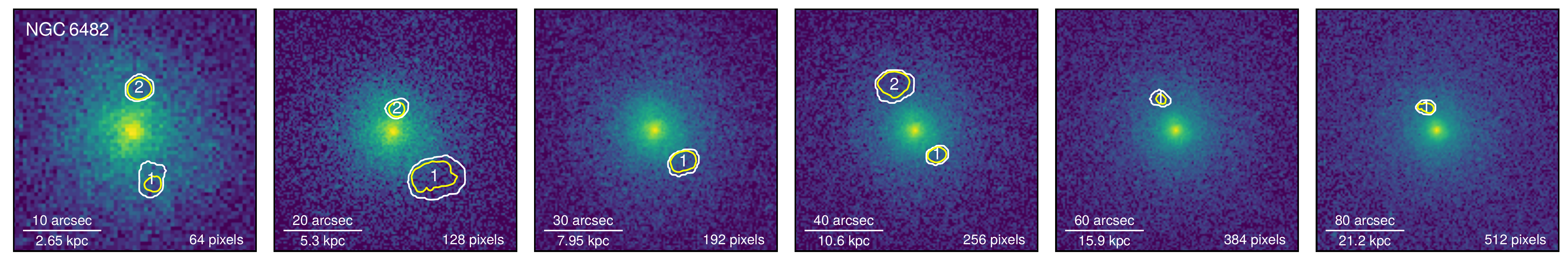}};
    \draw (0, 7.2) node {\includegraphics[height=85pt]{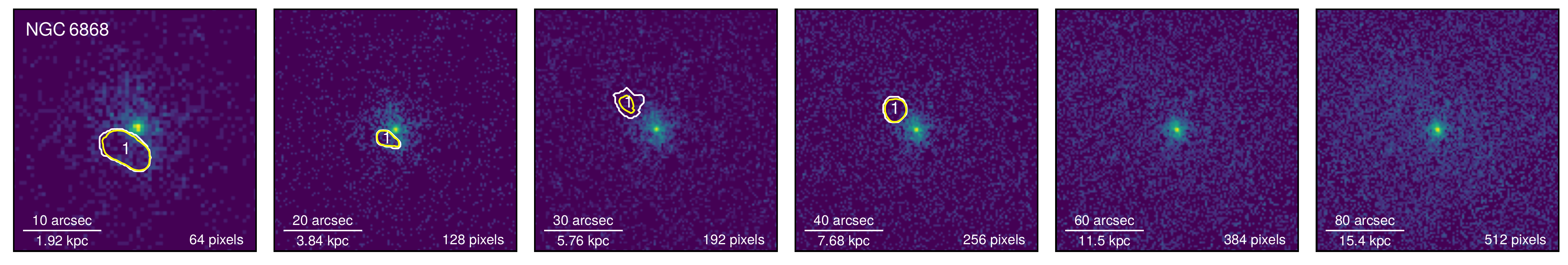}};
    \draw (0, 4.2) node {\includegraphics[height=85pt]{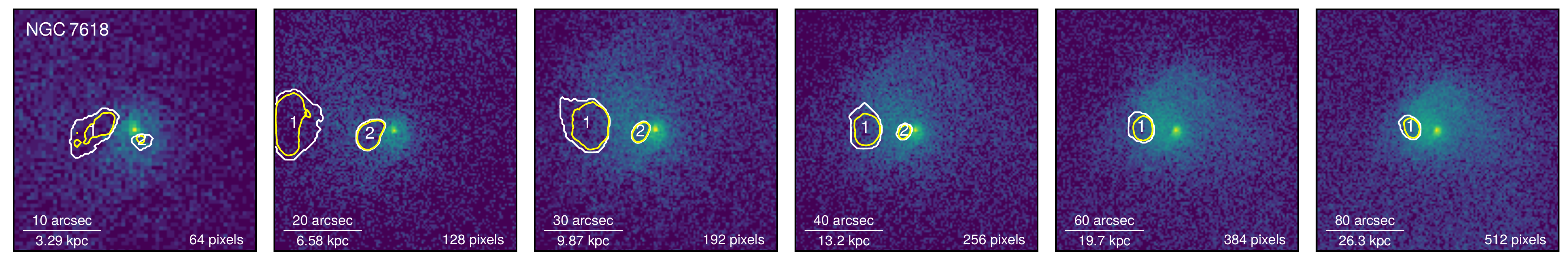}};
    \draw (0, 1.2) node {\includegraphics[height=85pt]{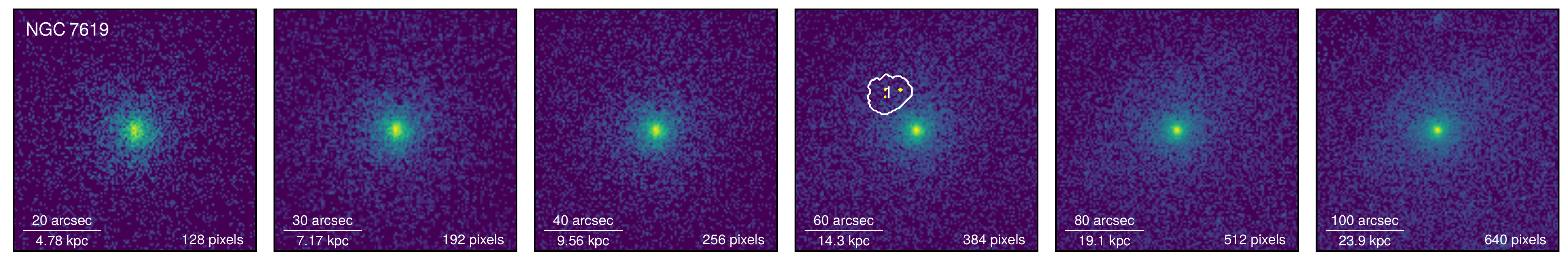}};
    \draw (0, -1.8) node {\includegraphics[height=85pt]{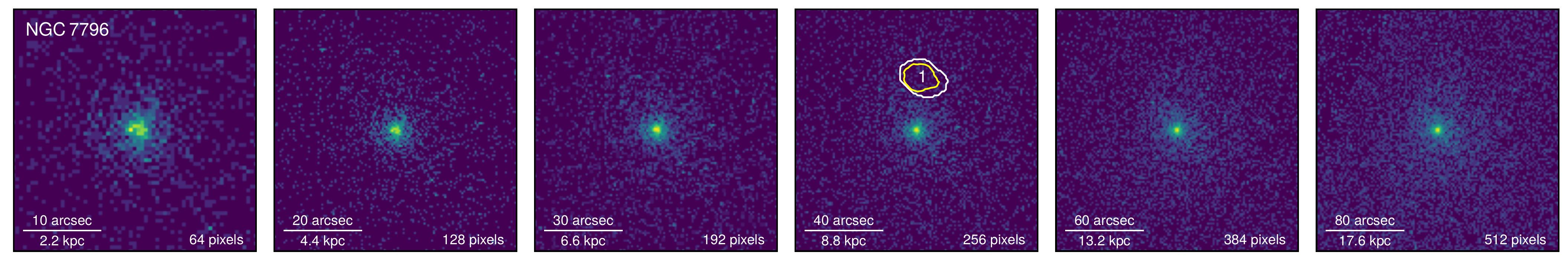}};
\end{tikzpicture}
\vspace{153mm}
\caption{Images of real galaxies overlaid by contours of CADET predictions. Contours correspond to discrimination threshold of 0.5 (white) and 0.9 (yellow). For each galaxy, the CADET pipeline was applied on 5 different scales, from left to right: 0.5 binning (64 pixels), 1 binning (128 pixels), 1.5 binning (192 pixels), 2 binning (256 pixels), 3 binning (394 pixels), and 4 binning (512 pixels). For very extended or nearby sources, we omitted the 0.5 binning and added one additional scale based on visual inspection of given images (typically binning 5, 6 or 7).}
\end{figure*}


\clearpage

\begin{figure*}
    \begin{tikzpicture}[remember picture,overlay,shift={(current page.center)}]
        \draw (0, 11.9) node {\large \bf Application of CADET on images of distant galaxy clusters};
        \draw (0, 9.9) node {\includegraphics[height=85pt]{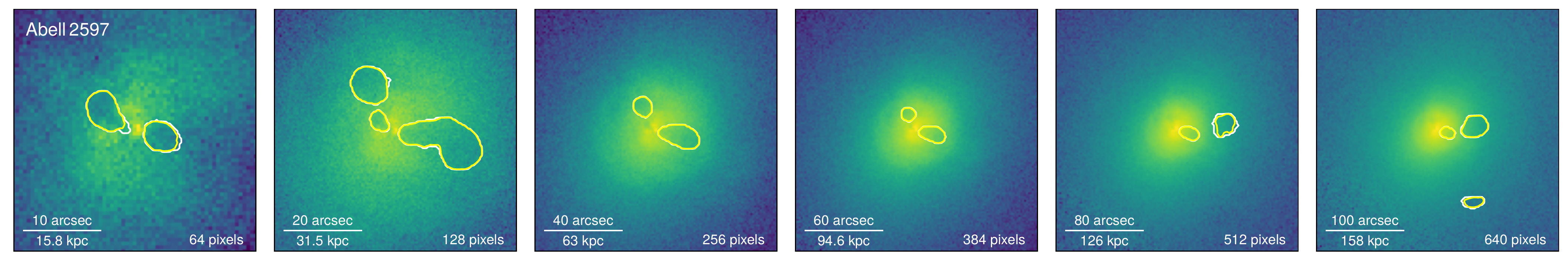}};
        \draw (0, 6.9) node {\includegraphics[height=85pt]{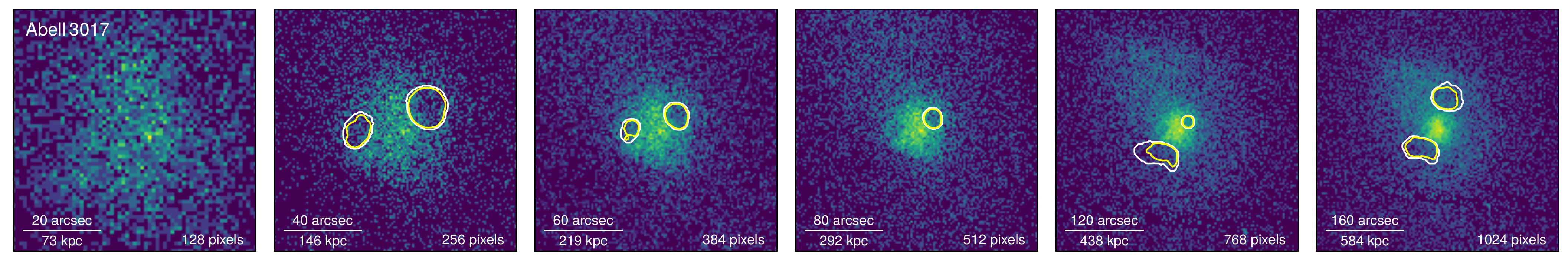}};
        \draw (0, 3.9) node {\includegraphics[height=85pt]{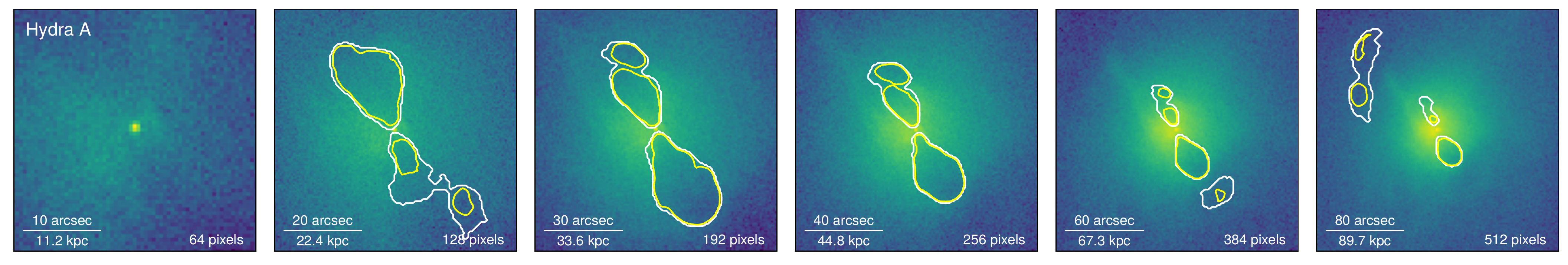}};
        \draw (0, 0.9) node {\includegraphics[height=85pt]{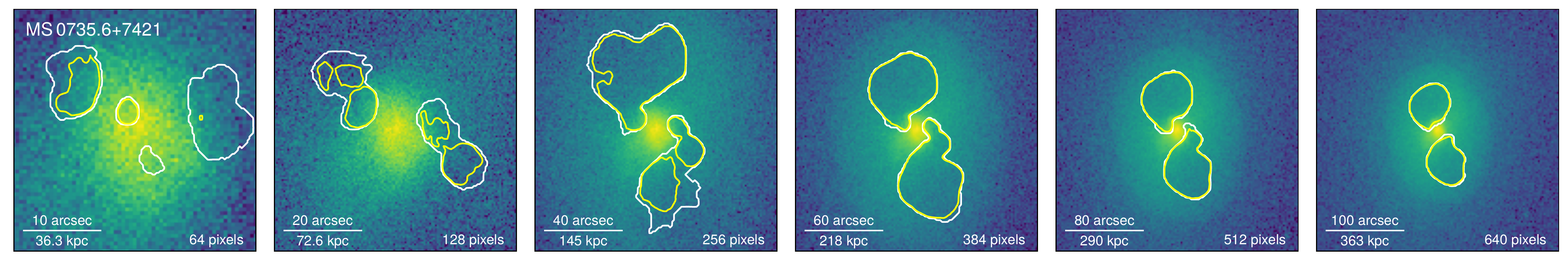}};
        \draw (0, -2.1) node {\includegraphics[height=85pt]{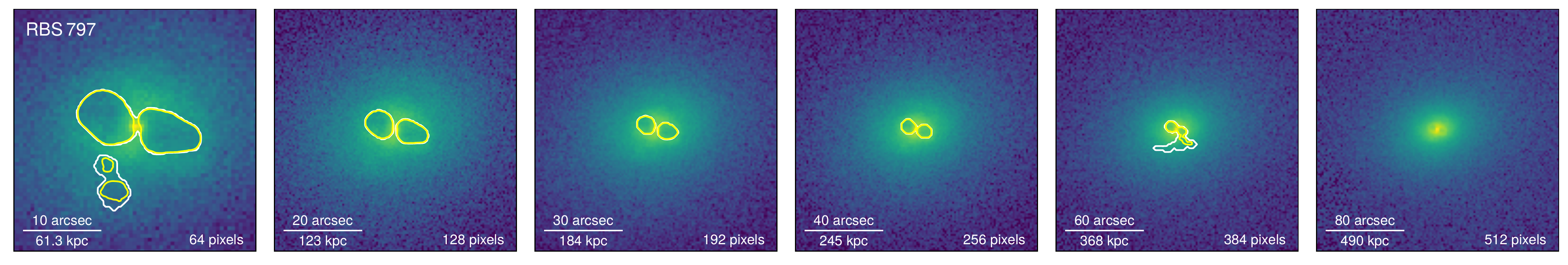}};
        \draw (0, -5.1) node {\includegraphics[height=85pt]{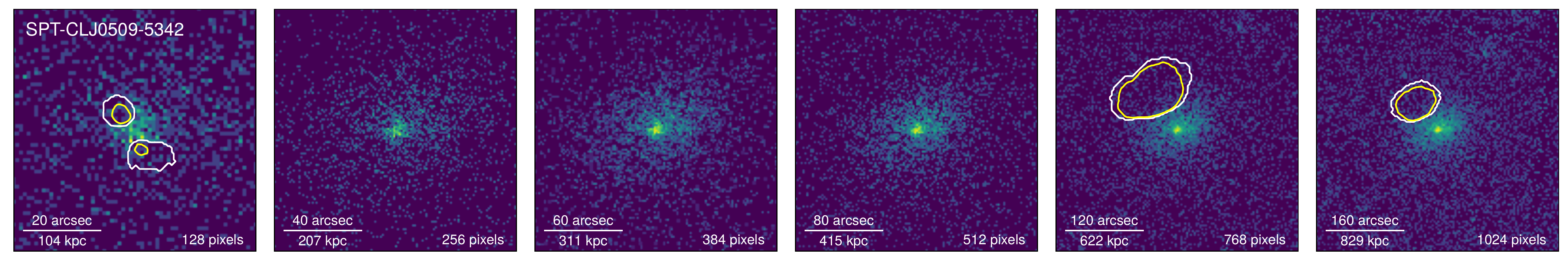}};
        \draw (0, -8.1) node {\includegraphics[height=85pt]{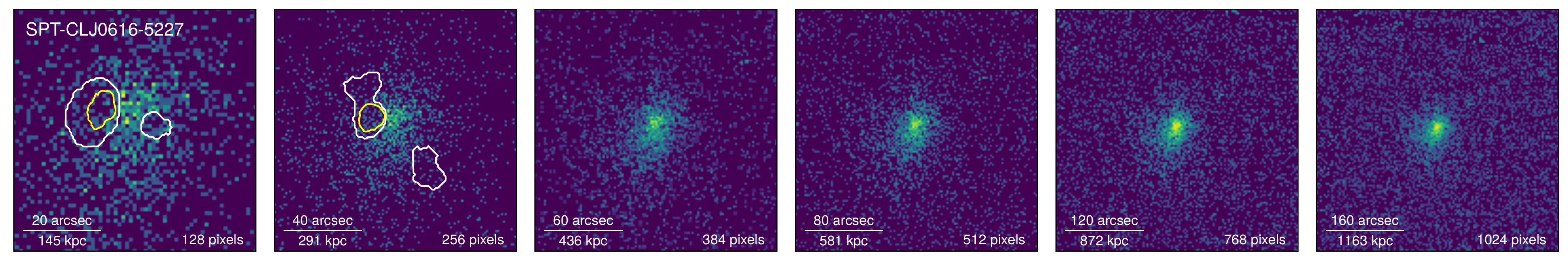}};
    \end{tikzpicture}
    \vspace{216mm}
    \caption{Application of CADET pipeline on more distant galaxy cluster systems. For each galaxy cluster, the CADET pipeline was applied on 5 different scales (specified in the bottom right corner of each subplot).}
\end{figure*}